\documentclass[twocolumn]{aastex62}

\usepackage{natbib}
\usepackage{multirow}
\usepackage{graphicx}
\usepackage{subfigure}
\usepackage{float}
\usepackage{footmisc}
\newcommand{\mum}{\ifmmode{\rm \mu m}\else{$\mu$m}\fi}

\newcommand{\ha}{{H$\alpha$\ }}

\newcommand{\msun}{{M$_{\sun}$}}
\newcommand{\rmn}[1]{{\mathrm{#1}}}
\newcommand{\chitwo}{$\chi^{2}\ $}

\begin{document}
\title{Elliptical Galaxy in the Making: The Dual Active Galactic Nuclei and Metal-enriched Halo of Mrk~273}

\author[0000-0003-3762-7344]{Weizhe Liu}
\altaffiliation{lwz@astro.umd.edu}
\affiliation{Department of Astronomy, University of Maryland,
  College Park, MD 20742, USA}

\author[0000-0002-3158-6820]{Sylvain Veilleux}
\altaffiliation{veilleux@astro.umd.edu}
\affiliation{Department of Astronomy, University of Maryland,
  College Park, MD 20742, USA}
\affiliation{Joint Space-Science Institute, University of Maryland,
  College Park, MD 20742, USA}

\author[0000-0002-4923-3281]{Kazushi Iwasawa}
\affiliation{Institut de Ci\`encies del Cosmos (ICCUB), Universitat de Barcelona (IEEC-UB), Mart\'i i Franqu\`es, 1, 08028 Barcelona, Spain}
\affiliation{ICREA, Pg. Llu\'is Companys 23, 08010 Barcelona, Spain}

\author[0000-0002-1608-7564]{David S. N. Rupke}
\affiliation{Department of Physics, Rhodes College, Memphis, TN 38112, USA}

\author{Stacy Teng}
\affiliation{Science and Technology Division, Institute for Defense Analyses, 4850 Mark Center Drive, Alexandria, VA 22311, USA}

\author[0000-0002-1912-0024]{Vivian U}
\affiliation{Department of Physics and Astronomy, 4129 Frederick Reines Hall, University of California, Irvine, CA 92697, USA}

\author[0000-0002-6562-8654]{Francesco Tombesi}
\affiliation{Department of Astronomy, University of Maryland,
  College Park, MD 20742, USA}
\affiliation{NASA/Goddard Space Flight Center, Code 662, Greenbelt, MD 20771, USA} 
\affiliation{Department of Physics, University of Rome ``Tor Vergata'', Via della Ricerca Scientifica 1, I-00133 Rome, Italy}

\author[0000-0002-1233-9998]{David Sanders}
\affiliation{Institute for Astronomy, 2680 Woodlawn Drive, University of Hawaii, Honolulu, HI 96822-1839, USA}

\author[0000-0003-0682-5436]{Claire E. Max}
\affiliation{Department of Astronomy and Astrophysics, University of California, Santa Cruz, CA 95064, USA}

\author[0000-0001-8485-0325]{Marcio Mel{\'e}ndez}       
\affiliation{Space Telescope Science Institute, 3700 San Martin Drive, Baltimore, MD 21218} 

\begin{abstract}
A systematic analysis of the X-ray emission from the nearby ultraluminous infrared galaxy Mrk~273 was carried out by combining new 200 ksec {\em Chandra} data with archived 44 ksec data. The active galactic nucleus (AGN) associated with the Southwest nucleus is confirmed by the new data, and a secondary hard X-ray (4-8 keV) point source is detected, coincident with the Northeast nucleus at a projected distance of 0.75 kpc from the Southwest nucleus. The hard X-ray spectrum of the Northeast nucleus is consistent with a heavily absorbed AGN, making Mrk~273 another example of a dual AGN in a nearby galaxy merger. Significant 1-3 keV emission is found along the ionization cones and outflowing gas detected in a previous study. The data also map the giant X-ray nebula south of the host galaxy with unprecedented detail. This nebula extends on a scale of $\sim$ 40 kpc $\times$ 40 kpc, and is not closely related to the well-known tidal tail seen in the optical. The X-ray emission of the nebula is best described by a single-temperature gas model, with a temperature of $\sim$ 7 million K and a super-solar $\alpha$/Fe ratio. Further analysis suggests that the southern nebula has most likely been heated and enriched by multiple galactic outflows generated by the AGN and/or circumnuclear starburst in the past, on a time scale of $\lesssim$0.1 Gyr, similar to the merger event itself. 
\end{abstract}


\section{Introduction}
Major mergers of gas-rich galaxies may lead to the formation of elliptical galaxies and the growth of supermassive black holes. {(Ultra)luminous infrared galaxies \citep[(U)LIRGs, log(L$_{IR}) \geq$ (12)11 L$\sun$, e.g.][]{Sanders1996}} are generally considered good examples of such merging process \citep[e.g.][]{Sanders1988,Veilleux2002,Hopkins2009}. Powerful outflows, driven by the central quasar and/or cicumnuclear starburst, have been invoked to quench or regulate star formation in the merger remnants, creating a population of ``red and dead'' ellipticals and setting up the observed tight black hole - galaxy relations \citep[e.g.][]{veilleuxaraa,Fabian2012araa,Kormendy2013araa}. There is growing observational support for these influential outflows, at both local and high-redshift universe \citep[e.g.][and references therein]{Cicone2015,Nadia2016,Harrison2017,Rupke2017,Veilleux2017}. 
 
Recent deep {\em Chandra} observations have revealed giant ($\simeq$ 50 kpc), X-ray-emitting gaseous halos in two of the nearest (U)LIRGs: NGC~6240 \citep[113 Mpc; 150 ksec; ][]{Nardini2013} and Mrk~231 \citep[200 Mpc; 500 ksec; ][]{Veilleux2014}. In both sources, super-solar $\alpha$/Fe abundance ratios are measured throughout the halo. In order to produce the amount of $\alpha$ elements detected in these halos, star formation activity over an extended period of time ($\lesssim$0.1 Gyr) is required if a star formation rate (SFR) at the current level is assumed. Repeated outflow events, like the ones currently seen in both objects, have been suggested as one plausible mechanism to help carry the $\alpha$ elements produced in the circumnuclear region all the way to the halo, on scales of several tens of kpc \citep[e.g.][]{Nardini2013,Veilleux2014}. Such outflow-driven metal transport is directly seen in the nearby starburst M82, though on a significantly smaller spatial scale {\citep[on the order of $\sim$ 10 kpc, e.g.][]{Konami2011}}. 

Another contentious issue in these galaxy mergers is the duty cycle of black-hole accretion activity. The inward-flowing gas induced by the merging activity has long be thought as one important fueling mechanism for Active Galactic Nucleus (AGN) \citep[e.g.][]{DiMatteo2005}, although the observational evidence in support of this scenario is still incomplete. Hard X-ray observations can penetrate the high column densities usually found in the central regions of those galaxies, and thus serve as a good probe for the AGN activity that might be otherwise hidden by dense clouds. Over the last decade, dozens of nearby kpc-scale dual AGN have been found serendipitously in interacting galaxies through X-ray observations {\citep[e.g.][]{Komossa2003,Koss2011a,DeRosa2018}} 

Being the second nearest ULIRG (176 Mpc\footnote{{Based on a redshift $z$ =
  0.0377 and a cosmology with $H_0$ = 69.6 km s$^{-1}$ Mpc$^{-1}$,
  $\Omega_{\rm matter}$ = 0.286, and $\Omega_{\rm vacuum}$ = 0.714. \citep{cosmology2014}}}; $\sim$0.74 kpc arcsec$^{-1}$), Mrk~273 is an excellent laboratory to explore the effect of feedback and dual AGN activity. {Mrk~273 is brighter in the X-ray (2-10 keV) than Arp 220, the nearest ULIRG, and is well known to harbor a large X-ray halo, based on an old 44-ksec {\em Chandra} exposure analyzed in \citet{Xia2002} and reanalyzed in a series of papers \citep{Ptak2003, Grimes2005, Teng2010, Iwasawa2011cgoal,Iwasawa2011}}. The halo of Mrk~273 appears to be unrelated to the brightest tidal features, as the southern tidal tail casts a shadow on the nebula emission; Super-solar $\alpha$ element abundance has been tentatively reported in this halo, but this is based on a spectrum with only 300 counts \citep{Iwasawa2011}. More recently, a remarkable AGN-driven, bipolar ionized outflow has been detected in this object, on a scale of $\sim$ 4 kpc \citep[][]{RupkeVeilleux2013b}, accompanied by warm and cold molecular outflows \citep{Vivian2013,Veilleuxherschel,Cicone2014}. Outflowing ionized gas has also been detected on a larger scale of $\sim$ 10 kpc \citep[][]{Zaurin2014}.

Mrk~273 is a late merger with dual nuclei in the mid-infrared\footnote{High-resolution radio observations have revealed a third component in the southeast of the central region of Mrk~273 \citep{Condon1991,Smith1998}, which is only weakly detected in the near-infrared \citep{Scoville2000}. The steep radio spectrum of this component perhaps points to a starburst origin \citep{Bondi2005}. Also, it is not detected in the X-ray ($>$2 keV) based on the {\em Chandra} data, and is thought to be a candidate star cluster \citep{Iwasawa2011}.}, located 0.75 kpc apart in projection, similar to that of NGC~6240 ($\sim$ 0.74 kpc), but larger than that in Arp 220 (0.33 kpc) and Mrk~231 {\citep[coalesced into a single nucleus;][]{Surace1998,Veilleux2002}}. While the existence of an AGN in the southwest nucleus (hereafter SW nucleus) has been demonstrated at multiple wavelengths \citep[e.g.][]{Sanders1988,Veilleux1999,Veilleux2009,Teng2009,Iwasawa2011}, the nature of the northeast nucleus (hereafter NE nucleus) is still controversial. The NE nucleus is seen in the near-infrared \citep{Armus1990,Surace2000,Scoville2000}. \citet{mrk273CO1998} suggests that the {NE nucleus} is an extreme compact starburst with a high luminosity density, similar to the western nucleus of Arp 220. High-resolution radio continuum imaging supports this hypothesis \citep{Carilli2000,Bondi2005}. Meanwhile, integral field spectroscopy of the nuclear region by \citet{Colina1999} shows characteristic of LINER for the NE nucleus. A point-like hard X-ray source has been identified with the NE nucleus \citep{Xia2002,Gonzalez2006}, suggesting the existence of a heavily absorbed AGN. In \citet{Iwasawa2011}, the tentative detection of Fe K$\alpha$ line emission associated with the NE nucleus supports this scenario. Nevertheless, the S/N of the data is insufficient to determine whether the Fe K$\alpha$ emission originates only from the NE nucleus or there is contamination from the SW nucleus. A fast-rotating molecular gas disk and coronal line [Si {\sc vi}] 1.964\mum\ emission flowing from the NE nucleus is revealed in \citet{Vivian2013}, which also favors the existence of a heavily absorbed AGN, although it is not a unique interpretation. Additionally, Mrk~273 was also detected, though the nuclei were unresolved, above 10 keV in the X-ray by {\em Suzaku} \citep{Teng2009} and {\em NuSTAR} \citep{Teng2015}. Spectral fitting and variability analysis of these data suggest the presence of a single, partially covered, and heavily obscured AGN.  A recent re-analysis of the {\em NuSTAR} data by {\citet{Iwasawa2017}} assumed a double nucleus model, suggesting that the spectrum above 10 keV can be modeled by two heavily absorbed AGN. {They also discussed the X-ray variability of Mrk~273 over more than a decade. The uncorrelated variability above and below 10 keV may suggest that two distinct sources are present in the respective bands.} 

In this paper, we analyze newly obtained 200 ksec {\em Chandra} data of Mrk~273, in combination with the old 44 ksec data, to explore the dual AGN activity,  the outflow, and the extended X-ray halo. The paper is organized as follows. In Section 2, the datasets and reduction procedures are described. In Section 3, the main results are presented. A discussion of the implications is presented in Section 4, and the conclusions are summarized in Section 5.

\section{Observation and Data Reduction}

\subsection{Chandra Observations}
Mrk~273 was aimed at the back-illuminated S3
detector of ACIS. The rationale behind the setup used for the new 200-ksec observation
(PID 17700440; PI Veilleux) was to match the observational parameters of the $\sim$ 44 ksec exposures obtained in 2000 and analyzed in \citet{Xia2002}, and thus to facilitate the task of combining both data sets into a single $\sim$ 244 ks exposure when appropriate.  

Due to scheduling constraints, the planned 200-ksec observation was divided into five segments of 61, 32, 35, 34 and 38 ksec, with the first one taken on 2016 Sep. 6, and the other four obtained on 2017 Feb. 14, 16, 18, and 26, respectively. All the observations were performed in 1/2 subarray mode in order to avoid pileup and take advantage of {\em Chandra}'s excellent angular resolution ($\sim$ 0$\farcs$5 ).

In this paper, effort has been taken to combine both the old $\sim$ 44 ks observation and new 200 ks observations together when appropriate. For convenience, the archival data analyzed by \citet{Xia2002} are denoted as the 2000 data, and all other data observed recently in year 2016-2017 are denoted as the 2016 data. 
 
Data reduction was carried out through standard {\em Chandra} data analysis package CIAO 4.8 and CALDB 4.7. All the data were reprocessed using the CIAO script \textit{chandra\_repro}. The \textit{deflare} routine was used to detect and discard any possible flare in the data, where data with background count rate exceeds 3 standard deviations from the mean of the distribution are removed. No attempt was taken to deconvolve the data using, e.g., Lucy or EMC2 algorithms \citep{Lucy1974,Esch2004,Karovska2005}. This strategy better preserves diffuse features and possible slight asymmetries in the point-spread function (PSF). 

The images were merged for analysis using the CIAO script \textit{merge\_obs}. In the hard X-ray band (2-8 keV), the response of ACIS-S has not changed to a noticeable extent, so a direct stack of the hard X-ray images in counts of all data was adopted to generate the images. In the soft X-ray band (0.4-2 keV), however, the response of ACIS-S has dropped significantly (factor of $\sim$ 2). Therefore, images of both datasets in flux units were produced after accounting for the energy dependence of the exposure maps at different epochs. 

All the spectral extractions were done by the CIAO script \textit{specextract}, and the combined spectrum was generated by the script \textit{combine\_spectrum}. {The spectra were binned to 15 counts bin$^{-1}$ when the total number of counts was high enough, while the rest of them were binned mildly to 1 counts bin$^{-1}$ in order to conserve a good energy resolution. The spectral fittings were done by XSPEC version 12.9.0 \citep{xspec}. The \chitwo statistic was used by default when the spectra were binned to 15 counts bin$^{-1}$. The CSTAT statistic \citep{Cash1979} was used when the spectra were binned to 1 counts bin$^{-1}$, and the default MCMC method in XSPEC was used to calculate the errors of the measurements.}{\footnote{When applying CSTAT statistic, unbinned spectra with bins of zero counts will bias the results of the fit. The spectra are thus mildly binned to 1 counts bin$^{-1}$ to avoid this problem. See section \textit{cstat} on ``http://cxc.harvard.edu/sherpa4.4/statistics/index.html'', as well as sections \textit{Poisson data (cstat)} and \textit{Poisson data with Poisson background (cstat)} in ``https://heasarc.gsfc.nasa.gov/xanadu/xsp ec/manual/XSappendixStatistics.html'' for more information.}}

\subsection{Ancillary Datasets}
In order to better understand the {\em Chandra} observation, multi-wavelength ancillary data were gathered from the literature and archives. 

The reduced Hubble Space Telescope (HST) I-band image taken with the F814W filter of ACS \citep{Armus2009GOALS,Kim2013ACS} and H-band image taken with the NIC2 F160W filter of NICMOS \citep{Scoville2000,Cresci2007} were downloaded from the Hubble Legacy Archive. The I-band image was used to trace the stellar component of the galaxy, and the H-band image was used to locate the two nuclei within Mrk~273. The continuum-subtracted, narrow-band [O {\sc iii}] $\lambda$5007 image from \citet{Zaurin2014} was used to trace the spatial distribution of the ionization cones and outflowing gas. The continuum-subtracted, narrow-band \ha image from \citet{Spence2016} was used to trace the large-scale H${\alpha}$ emission.

\begin{figure}[!htb]   
\epsscale{1.1}
\plotone{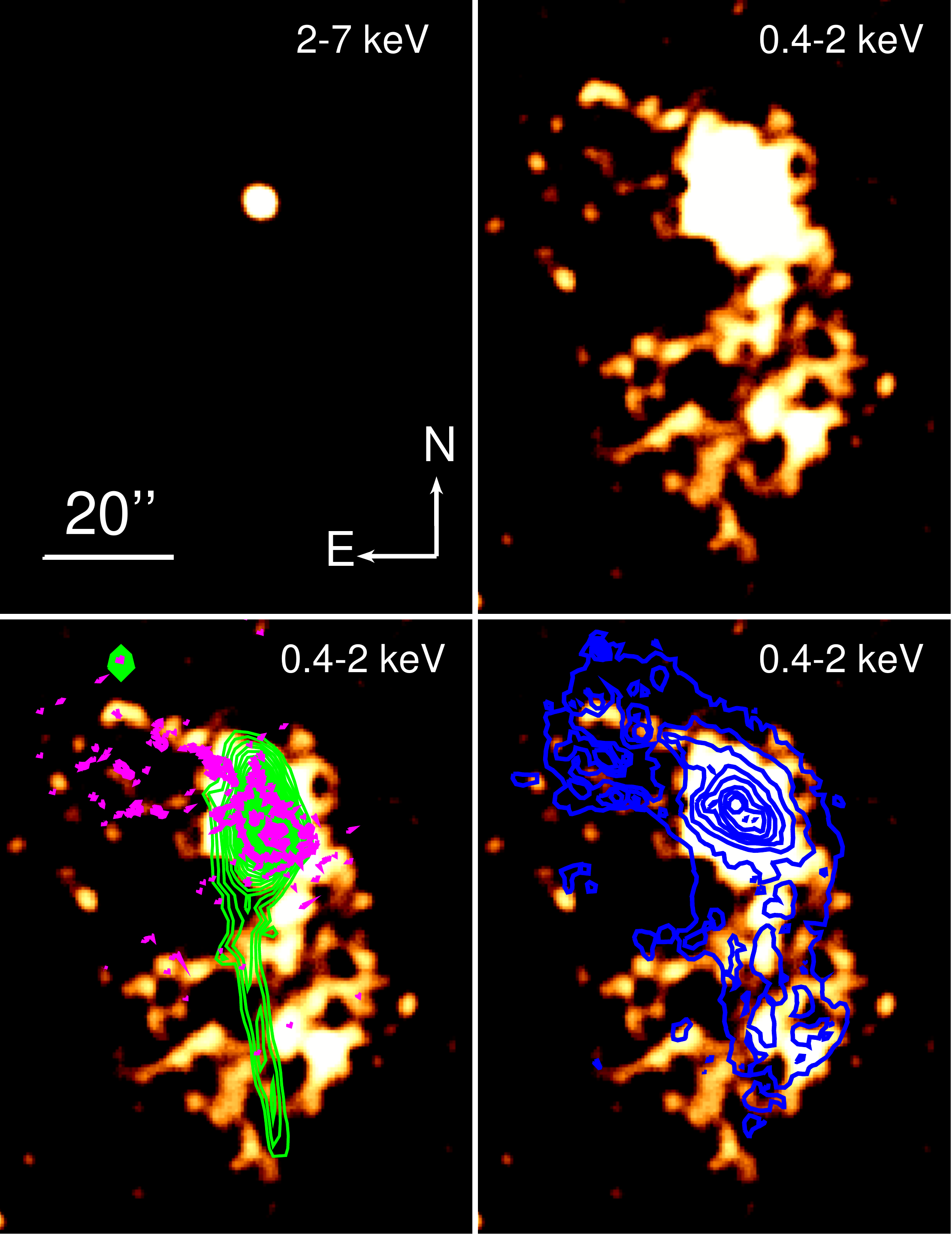}
\caption{Merged and smoothed X-ray images from all observations (244 ksec in total) in energy bands 2.0-7.0 keV ({\em Top Left}) and {0.4-2.0 keV} ({\em Top Right \& Bottom}). In the bottom left panel, the HST I-band image is over-plotted in magenta contours and the continuum-subtracted [O {\sc iii}] $\lambda$5007 image from \citet{Zaurin2014} is over-plotted in green contours. In the bottom right panel, the continuum-subtracted H$\alpha$ image from \citet{Spence2016} is over-plotted in blue contours. The images are on different logarithmic scales. North is up and East is to the left. The linear spatial scale is 0.74 kpc arcsec$^{-1}$.}
\label{fig:0470ct}
\end{figure}

\begin{figure}[!htb]
\epsscale{1.16}   
\plottwo{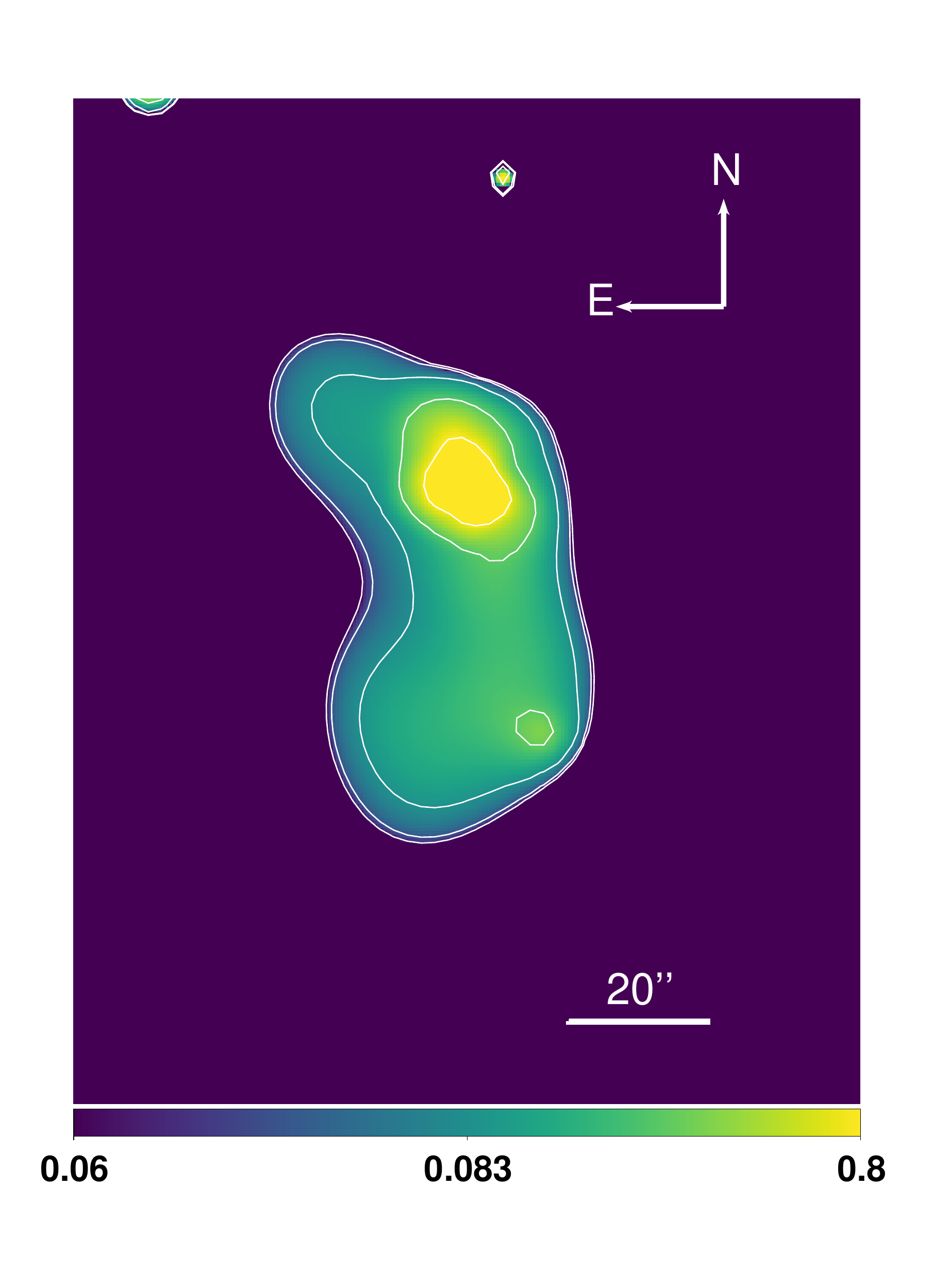}{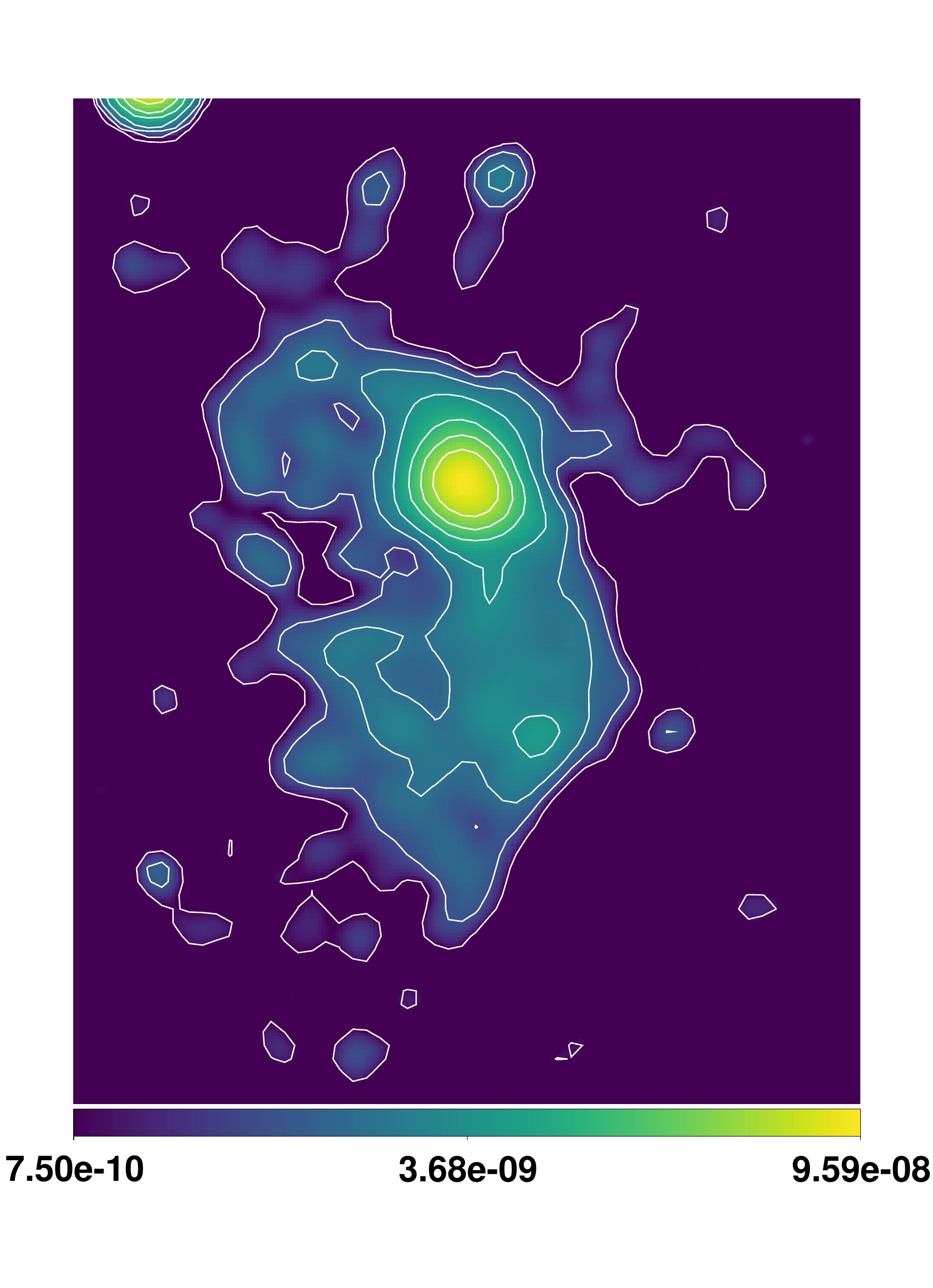}
\caption{Adaptively smoothed images from all combined observations (244 ksec in total), on logarithmic scales. North is up and East is to the left. {\em Left:} {Image in the 0.4-7.0 keV band} in counts. The CIAO script \textit{csmooth} was used for the smoothing, in order to emphasize the faint, extended emission in the image. The contours are at 2, 3, 5, 10 and 25$\sigma$. {\em Right:} Image {in the} 0.4-1.0 keV band, {in units of photons cm$^{-2}$ s$^{-1}$ .} The CIAO script \textit{dmimgadapt} was used for the smoothing. It provided a superior resolution of the bright small-scale structures when compared to the script \textit{csmooth}, while keeping the extended structures. The contours are at 2, 3, 5, 10, 25, 50, 100$\sigma$.
}
\label{fig:0470csmooth}
\end{figure}

\begin{figure}[!htb]
\epsscale{1.15}
\plotone{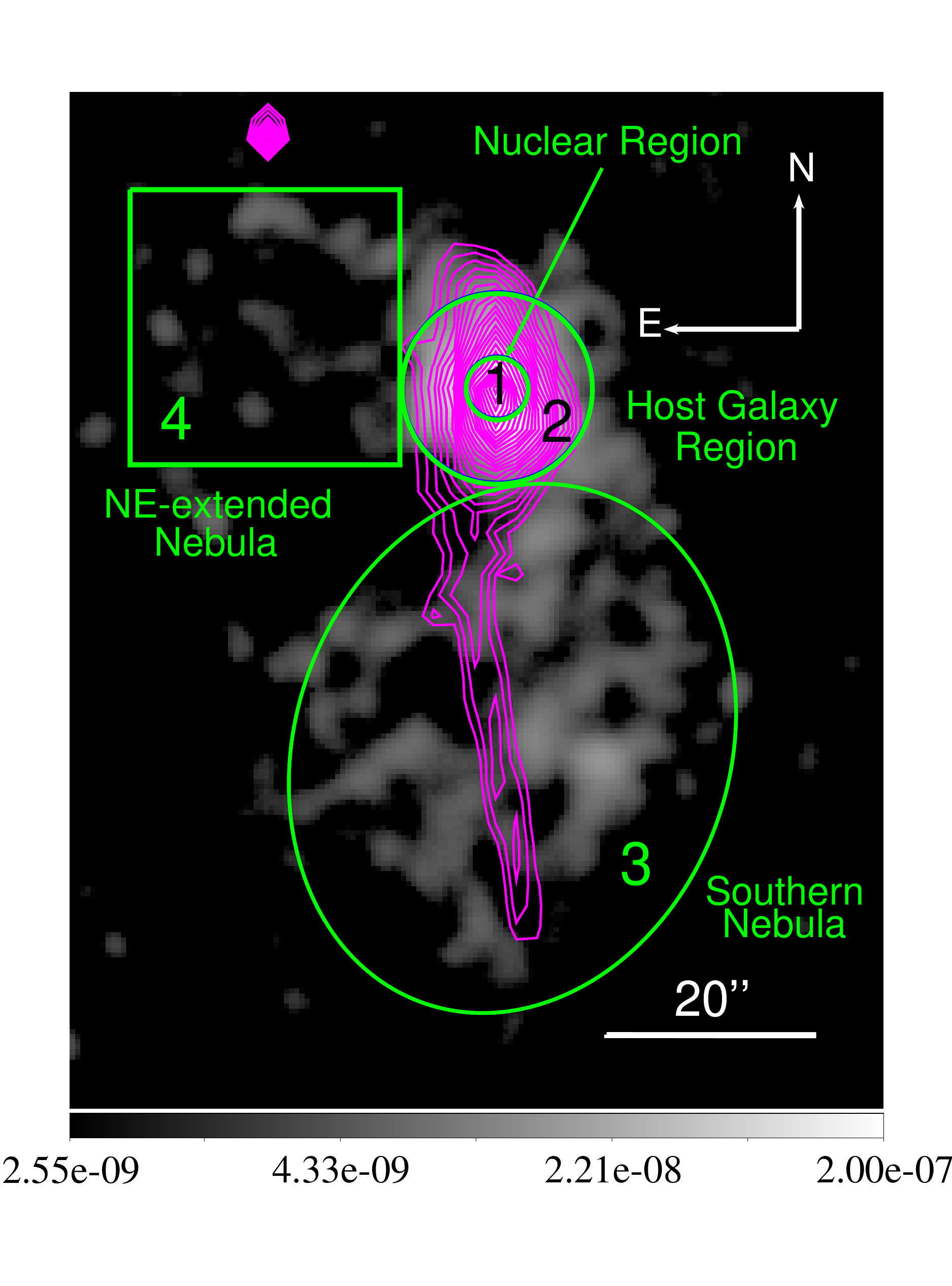}
\caption{The different spatial regions used for the analyses presented in Section 3. {Region 1}: Nuclear Region, denoted by the green aperture with 3$\arcsec$ in radius. {Region 2}: Host Galaxy Region, denoted by the green circular annulus with an inner radius of 3$\arcsec$ and an outer radius of 9$\arcsec$. {Region 3}: Southern Nebula region, denoted by the green ellipse in the south. {Region 4}: Extended emission in the northeast (i.e. NE-extended Nebula), denoted by the green box to the northeast. The {grayscale image is in} the 0.4-2.0 keV band, {in units of photons cm$^{-2}$ s$^{-1}$} and on a logarithm scale. The overlaid magenta contours are generated from the HST I-band image. 
\label{fig:regions}}
\end{figure}

\section{Results}

\subsection{Overview of the Chandra Data}
The stacked images of the 2000+2016 data are shown in Figure \ref{fig:0470ct}. In general, the hard X-ray emission (2-7 keV) is confined to the nuclear region. {Similarities between the soft X-ray emission (0.4-2 keV) and the HST I-band image are present for the brighter regions, but much less so for the fainter X-ray emission.}

Firstly, there is significant extended X-ray emission south of the galaxy, tracing the hot halo gas on a scale of $\sim$ 40 kpc $\times$ 40 kpc. Apparently the gas is not associated with the tidal tail, since it is much more extended than the tidal tail in the east-west direction. This nebula (hereafter called the Southern Nebula) has been reported in previous studies \citep[e.g.][]{Iwasawa2011}, but the structure of the nebula in these older data is not as clear due to fewer counts. In addition, northeast of the galaxy main body (i.e. the stellar component as seen in HST I-band, excluding the tidal tail in the south), there is also faint, extended X-ray emission (hereafter called NE-extended Nebula), which overlaps with the large-scale extended [O {\sc iii}] $\lambda$5007 emission seen in \citet{Zaurin2014}.   
 
The narrow-band H$\alpha$ image presented in \citet{Spence2016} shows extended nebulae south and northeast of the galaxy main body, as shown in blue contours in the bottom right panel of Figure \ref{fig:0470ct}. These H$\alpha$ nebulae resemble the Southern Nebula and the NE-extended Nebula in the X-ray, suggesting a possible connection between these nebulae. 

Adaptively smoothed images are presented in Figure \ref{fig:0470csmooth}. In the left panel, the stacked 0.4-7 keV image from the combined 2000+2016 data was adaptively smoothed with CIAO script \textit{csmooth}, which emphasized the faint extended structures in the image. In the right panel, the stacked, exposure-corrected 0.4-1.0 keV image from the same data was adaptively smoothed with another CIAO script \textit{dmimgadapt}. It provided a superior resolution of the bright small-scale structures when compared to the script \textit{csmooth}, while preserving the extended structures. The energy band of 0.4-1.0 keV was chosen for a best demonstration of the soft X-ray nebulae.

In order to further examine the X-ray emission in detail, the galaxy is divided into four different spatial regions for analysis, as shown and described in Figure \ref{fig:regions}. {The observed counts of those regions are summarized in Table \ref{tab:counts}, respectively.} The results of these analyses are discussed next.

{
\begin{deluxetable}{cccc}[!htb]
\tablecolumns{4}
\tablecaption{{Summary of Counts in the Data Sets}}
\tablehead{ \colhead{Region} & \multicolumn{2}{c}{Counts\tablenotemark{a}} \\
& 2000 data & 2016 data}
\startdata
Nuclear Region (1) & 1348 & 2681 \\
SW nucleus & 449 & 719 \\
NE nucleus & 204 & 647  \\
Host Galaxy Region (2)  & 553&  850  \\
Southern Nebula (3) & 376 & 680 \\
NE-extended Nebula (4) & 98 & 202  
\enddata
\tablenotetext{a}{For the Nuclear Region, SW nucleus and NE nucleus, the counts in 0.4-8 keV are shown; for the others, the counts in 0.4-2 keV are shown.}
\label{tab:counts}
\end{deluxetable}
}

\begin{figure*}[!htb]
\epsscale{1.15}
\plottwo{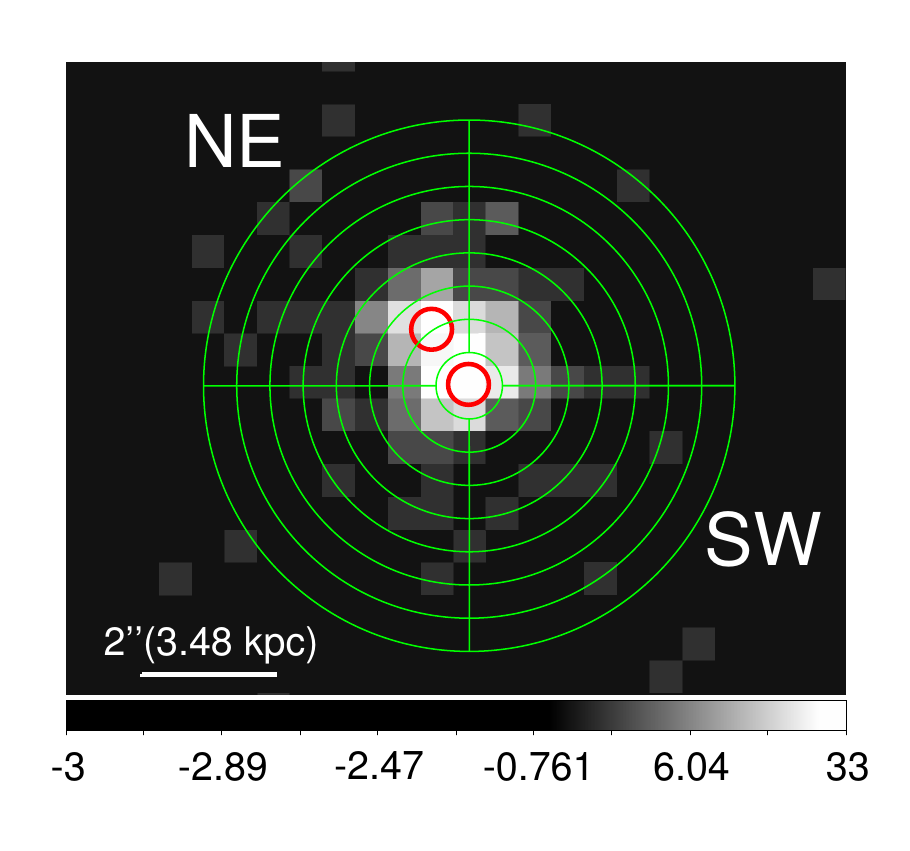}{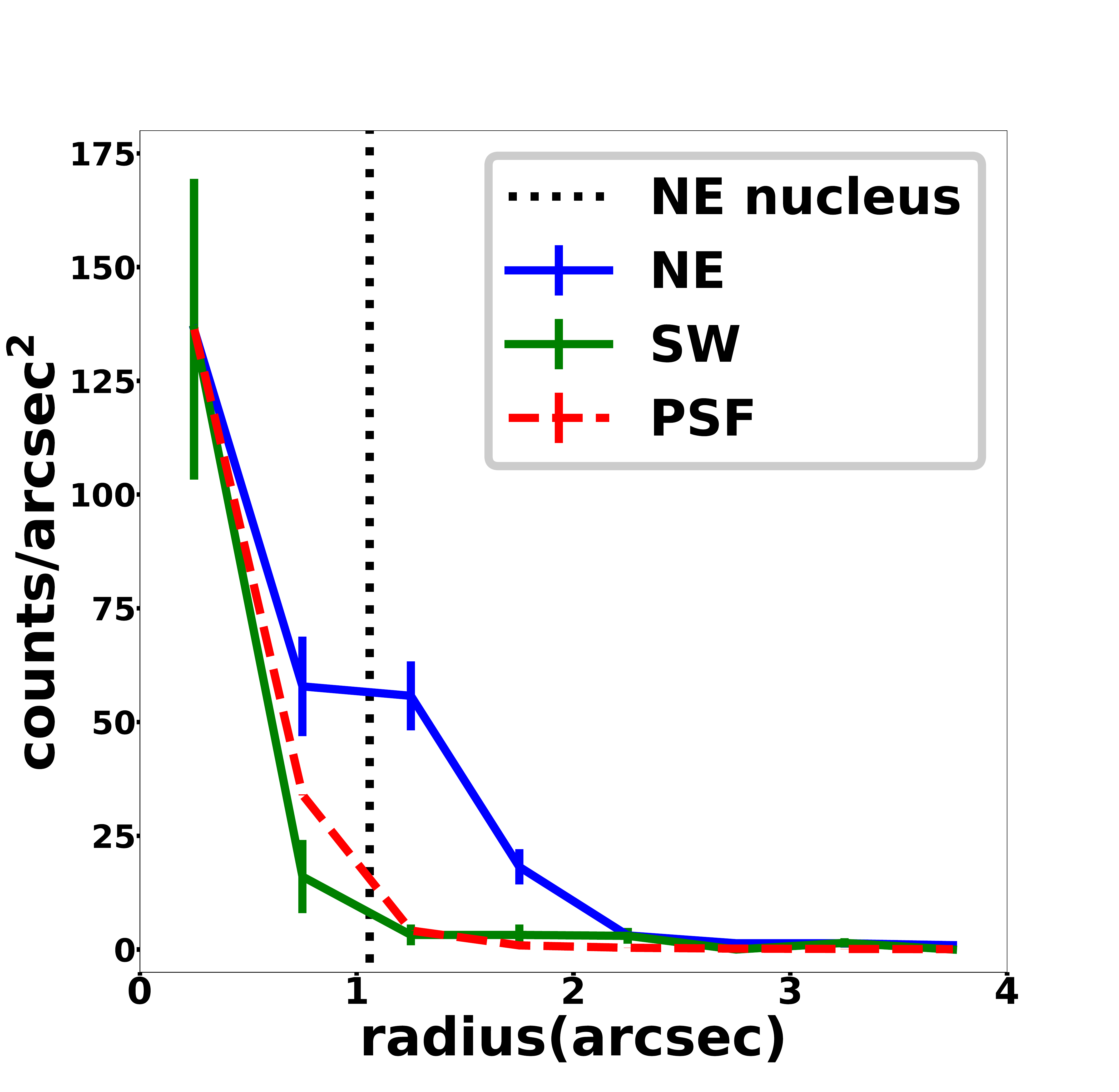}
\caption{{\em Left}: Image of the Nuclear Region in the 4-6 keV band, with sector annuli used to extract the radial profiles from the image. The spatial scale of the image is larger than the size of the Nuclear Region defined in Figure \ref{fig:regions} for better visualization. The annuli are centered on the peak of the emission (i.e. the SW nucleus) and are divided into four sectors with same area. The sectors in the northeast and southwest are denoted as NE and SW sectors separately. The image is shown in gray scale and in units of counts. It is scaled in logarithm and adjusted to show the full extent of the faint, hard X-ray emission. The two red circles denote the NE and SW nuclei seen in the HST H-band (F160W) image. The HST H-band (F160W) image and X-ray image are aligned following the same procedure described in Section 3.1 in \citet{Iwasawa2011}. {\em Right}: radial profiles of different sector annuli shown in the left panel, as well as the one from the PSF. The x-axis is the median radius for each sector annulus. A secondary maximum is obvious in the radial profile of NE sector (Blue). As a comparison, neither the radial profile of the SW sector (green) nor that of the PSF (red-dashed) shows a secondary maximum. {The PSF profile is simulated with the ray-tracing code ChaRT and MARX \citep{MARX}.} The black dotted line denotes the center of the NE nucleus. The pixel scale is {0$\farcs$49 (0.38 kpc)}. 
\label{fig:radial-hard}}
\end{figure*}

\begin{figure}[!htb]   
\epsscale{1.16}
\plotone{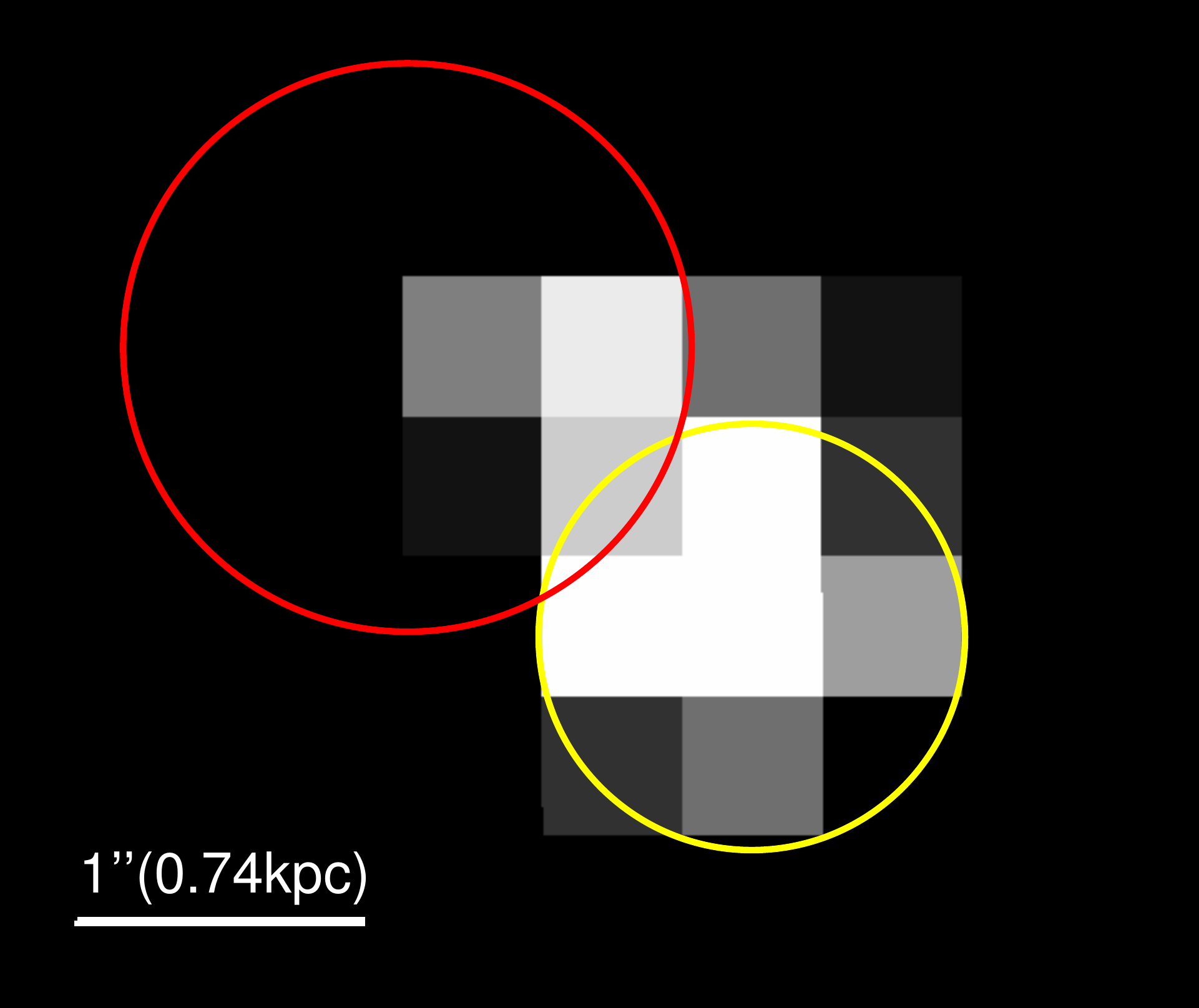}
\caption{The spectral extraction regions for the SW and NE nuclei. The yellow aperture denotes the extraction region for the SW nucleus, which is 0$\farcs$75 in radius. The red aperture denotes the extraction region for the NE nucleus, which is 1$\arcsec$ in radius. {The grayscale image is the stacked 4-6 keV image of the 2016 data.} The pixel scale is {0$\farcs$49 (0.38 kpc)}.
\label{fig:2nuclei}}
\end{figure}

\begin{figure}[!htb]   
  \centering
 \epsscale{1.15}
\plotone{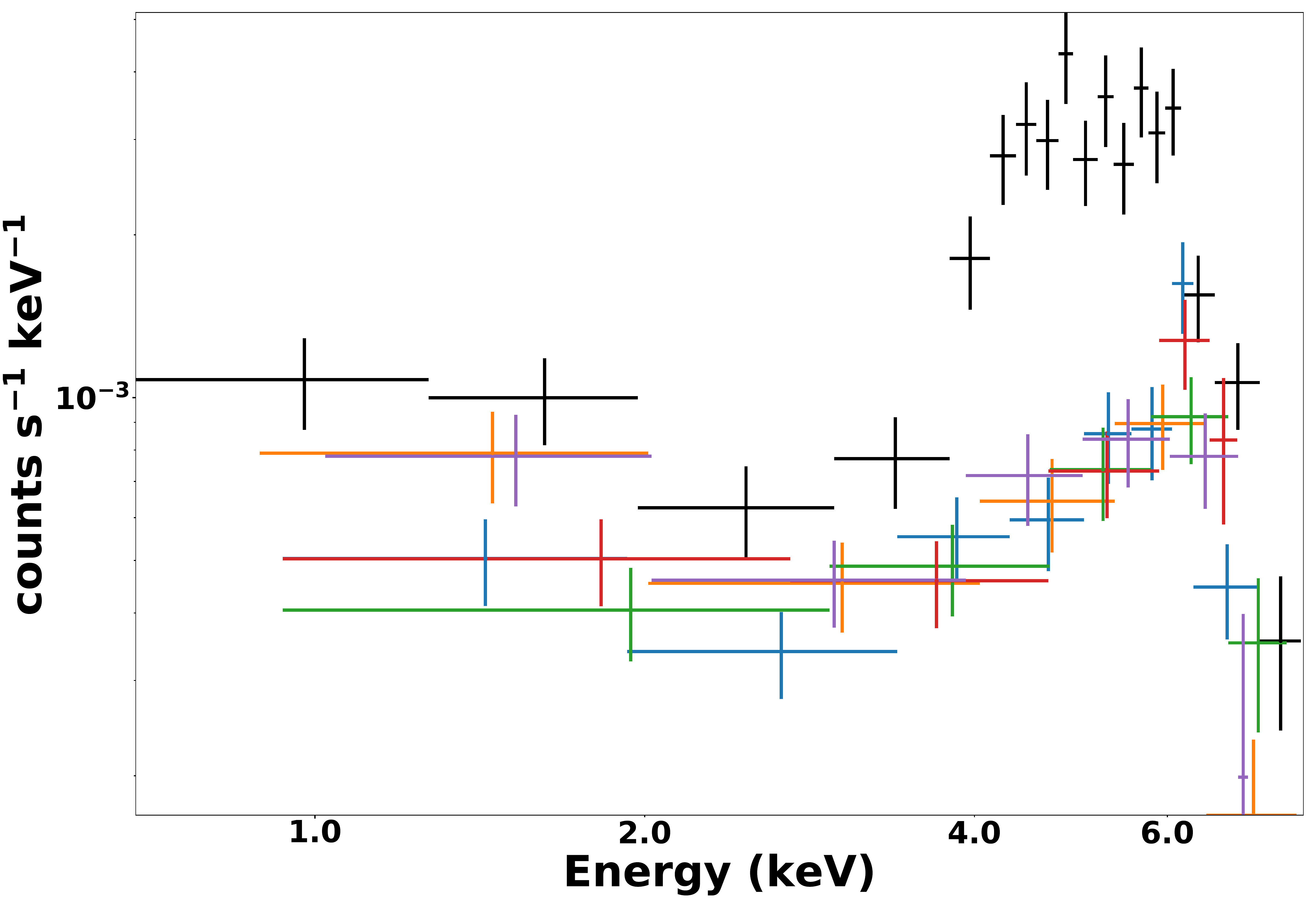} 
\caption{The original spectra of the SW nucleus, extracted from the yellow aperture defined in Figure \ref{fig:2nuclei}.  {\em Black spectrum}: the spectrum from the 2000 data. {\em Colored spectra}: spectra of {individual observations} from the 2016 data. All the spectra were binned to a minimum S/N of 3 bin$^{-1}$.
\label{fig:s-nuc-1-23456}}
\end{figure}

\subsection{Nuclear Region (Region 1)}

\subsubsection{Radial Profile of the Hard X-ray Emission}
The 4-6 keV image and the corresponding radial profiles in the Nuclear Region are shown in Figure \ref{fig:radial-hard}. The peak of the X-ray image overlaps with the SW nucleus seen in the HST NICMOS H-band (F160W) image. A possible secondary peak can also be seen by eye in the X-ray image, which is associated with the NE nucleus seen in the H-band image. In addition, the radial profiles of the surface brightness along the northeast (NE) and southwest (SW) directions are compared. The profiles are centered on the peak of the X-ray image. The radial profile along the northeast direction confirms the secondary peak identified by eye.

\begin{figure}[!htb]   
\epsscale{1.15}
\plotone{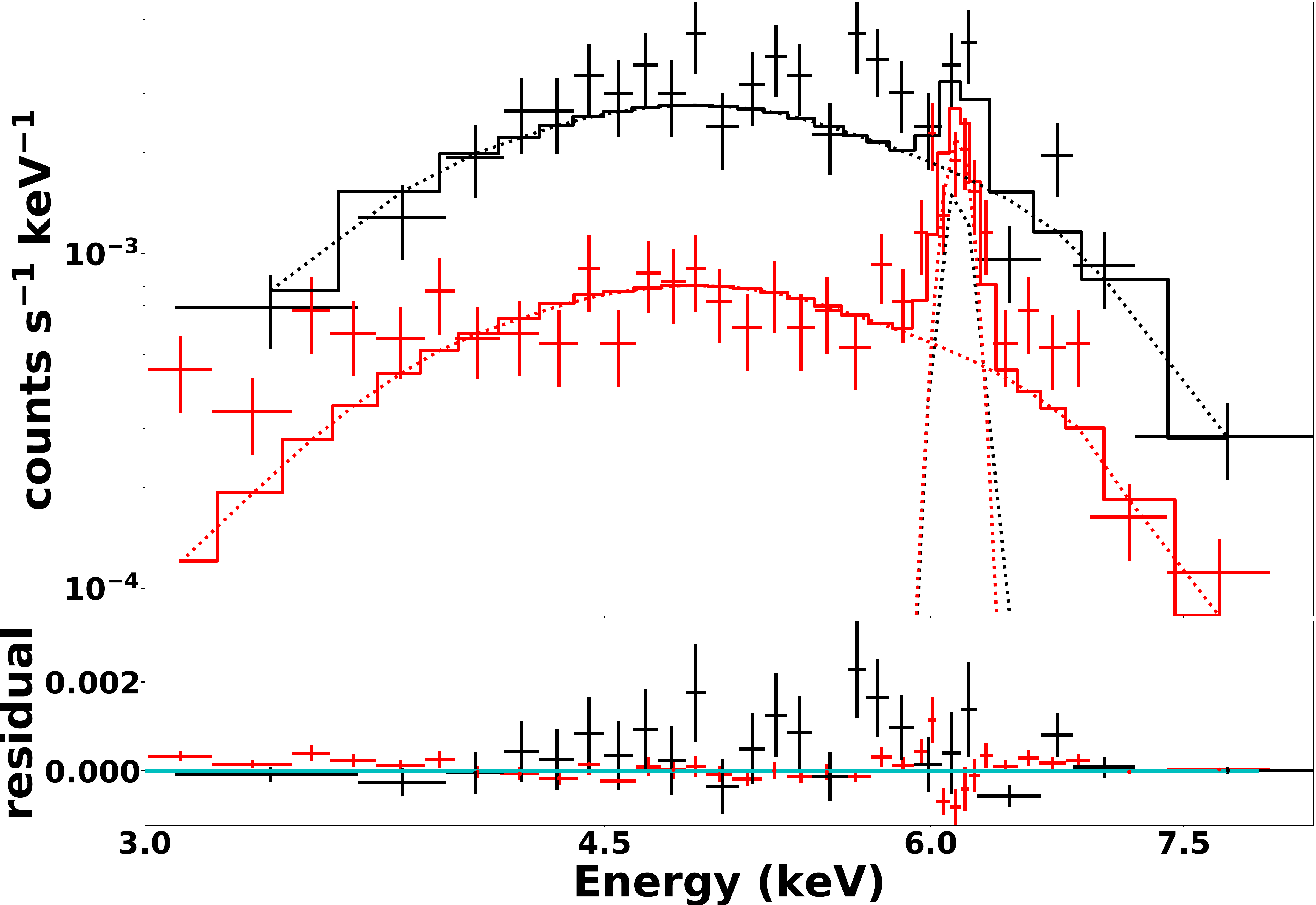}
\caption{Top: simultaneous fitting to the spectra of the SW nucleus from the 2000 data (black) and the 2016 data (red). Only data in the energy range of 3-8 keV is considered. The model is made up of one absorbed power-law component and an Fe K$\alpha$ line. All parameters for the two epochs are tied together except the normalizations for the power-law component and the Fe K$\alpha$ line. The best-fit parameters are summarized in Table \ref{tab:compare-nuclei}. The model is shown in solid lines and the components of it are shown in dotted lines. Bottom: residuals (data minus model).
\label{fig:SW-37kev}}
\end{figure}

\subsubsection{SW Nucleus}

Previous studies have revealed AGN activity in the central region of Mrk~273 \citep[e.g.][]{Colina1999,Scoville2000,Iwasawa2011}, and a possible dual AGN was suggested by \citet{Iwasawa2011}, \citet{Vivian2013}, and {\citet{Iwasawa2017}}. The analysis of hard X-ray image in Section 3.2.1 also suggests the possibility of dual AGN activity. In order to further explore the nature of the two nuclei, spectra of both SW nucleus and NE nucleus were extracted from apertures defined in Figure \ref{fig:2nuclei}. 

For the SW nucleus, the aperture was centered on the peak of the hard X-ray flux. It was chosen to include the majority of the emission from the SW nucleus revealed by the NICMOS/NIC2 F160W image, and the radius of the aperture was 0$\farcs$75. The original spectra of the SW nucleus extracted from the 2000 and 2016 data are shown in Figure \ref{fig:s-nuc-1-23456}. In general, the spectra from each individual observation of the 2016 data remains similar to each other. The spectrum from the 2000 data, however, shows clearly stronger emission in the hard X-ray band.

In order to improve the signal-to-noise ratio, all the spectra from the 2016 data were combined into one, while the spectrum from the 2000 data was left alone. A comparison of the two spectra in the energy range of 3-8 keV is shown in Figure \ref{fig:SW-37kev}. Again, it is clear that the hard X-ray emission has decreased significantly from the year 2000 to 2016-2017. By fitting the spectra in the 3-8 keV band from the 2000 and 2016 data separately, we find that both the absorption column density and the intrinsic luminosity of the AGN varies over the years (N$_{H}$ = 3.62$^{+1.83}_{-0.72}$ $\times$ 10$^{23}$ cm$^{-2}$ and 4-8 keV flux = 9.98$^{+2.73}_{-1.36} \times$ 10$^{-13}$ erg s$^{-1}$ cm$^{-2}$ for the 2000 data, and N$_{H}$ = 1.22$^{+0.43}_{-0.29}$ $\times$ 10$^{23}$ cm$^{-2}$ and 4-8 keV flux = 1.31$^{+0.12}_{-0.10}$ $\times$ 10$^{-13}$ erg s$^{-1}$ cm$^{-2}$ for the 2016 data, respectively\footnote{In this statement and those below, the errors from the spectral fits correspond to a confidence range of 90\%, or $\sim$$\pm{1.6}$$\sigma$}). In the following, we search for the origin of this drop of flux in the hard X-ray.

We first assume that only the intrinsic luminosity of the AGN has decreased over the years. Following this scenario, both spectra from the 2000 data and the 2016 data were fitted in 3-8 keV energy range simultaneously, with a model made up of one absorbed power-law component and an Fe K$\alpha$ line. All parameters for the two epochs were tied together except the normalizations for the power-law component and the Fe K$\alpha$ line. As a first trial, the photon index of the power-law component was set as a free parameter. {The fit is acceptable (reduced \chitwo = 1.50), } except that it gives a power-law photon index of $\sim$ 0.6, which is unreasonably small for an AGN. For example, this value is well below the measured lower limit \citep[$\Gamma$=1.4, see Figure 8 in][]{Ueda2014} for the {\em Swift/BAT} hard X-ray selected AGN. The photon indices derived in the same paper are $\Gamma$=1.94 with a standard deviation of 0.09 for Type 1 AGN and $\Gamma$=1.84 with a standard deviation of 0.15 for Type 2 AGN. Therefore, we fixed the power-law index to the standard value of $\Gamma$=1.9 \citep[e.g. see][ and references therein]{Piconcelli2005,Ishibashi2010,Ueda2014}. The best-fit model gave a reduced \chitwo of 1.51 (i.e. it is virtually the same as when $\Gamma$ was left as a free parameter). The results are shown in Figure \ref{fig:SW-37kev} and the best-fit parameters are summarized in Table \ref{tab:compare-nuclei}. {Although not statistically significant, there are possible residuals corresponding to other iron lines with energy higher than 6.4 keV in the rest frame.} From year 2000 to year 2016-2017, the total flux in 3-8 keV without absorption correction has decreased from 3.10$\times$10$^{-13}$ erg s$^{-1}$ cm$^{-2}$ to 1.21$\times$10$^{-13}$ erg s$^{-1}$ cm$^{-2}$. This is consistent with the variability on a scale of several years seen from observations at different epochs as discussed in \citet{Xia2002}, \citet{Teng2009,Teng2015} and {\citet{Iwasawa2017}}. 

Alternatively, an increase in the column density of the absorbing material in front of the central engine {alone} can also lead to the observed decrease of the hard X-ray flux. A model with tied power-law photon indices and normalizations but independent absorption column densities was fitted to the data. However, the {reduced \chitwo (2.7)} from this fit is poor.  Therefore, these results seem to favor the first scenario where the decrease of the hard X-ray luminosity is caused by the fading of the central engine. {However, we cannot formally rule out the possibility that both scenarios could be at work simultaneously, since we have to fix the photon indices of the power-laws in our fits due to the degeneracies among the parameters, which might bias our results.}  

{The measured hard X-ray (4-8 keV) continuum fluxes from the SW nucleus are $\sim$ 3.2$\times$10$^{-13}$ erg s$^{-1}$ cm$^{-2}$ in the 2013 {\em XMM-Newton} data \citep[with the contribution of the NE nucleus subtracted;][]{Iwasawa2017} and $\sim$ 7.2$\times$10$^{-14}$ erg s$^{-1}$ cm$^{-2}$ measured in the 2016 {\em Chandra} data, respectively (here we assumed that the continuum flux of the NE nucleus in the 2013 {\em XMM-Newton} data was the same as that measured in the 2016 {\em Chandra} data, given that the flux of the NE nucleus remained the same in 2000 and 2016-2017). The corresponding Fe K$\alpha$ line fluxes of the SW nucleus are $\sim$ 3.6$\times$10$^{-6}$ photons cm$^{-2}$ s$^{-1}$ for the 2013 {\em XMM-Newton} data \citep{Iwasawa2017} and $\sim$ 5.2$\times$10$^{-6}$ photons cm$^{-2}$ s$^{-1}$ for the 2016 {\em Chandra} data, respectively. Over these $\sim$ 3 years, the continuum flux has therefore dropped by a factor of $\sim$ 4.4, while the Fe K$\alpha$ line flux has increased by a factor of $\sim$ 1.4. Therefore the line emission has not followed the continuum precisely. This sets a lower limit on the physical scale of the inner edge of the torus ($\sim$ 1 pc) if the Fe K$\alpha$ line arises from that region. Note that this argument only relies on the fact that the fluxes of the Fe line and the continuum have changed differently over the years; the exact change in flux does not matter here.}

\begin{figure}[!htb]  
\centering
\epsscale{1.15}
\plotone{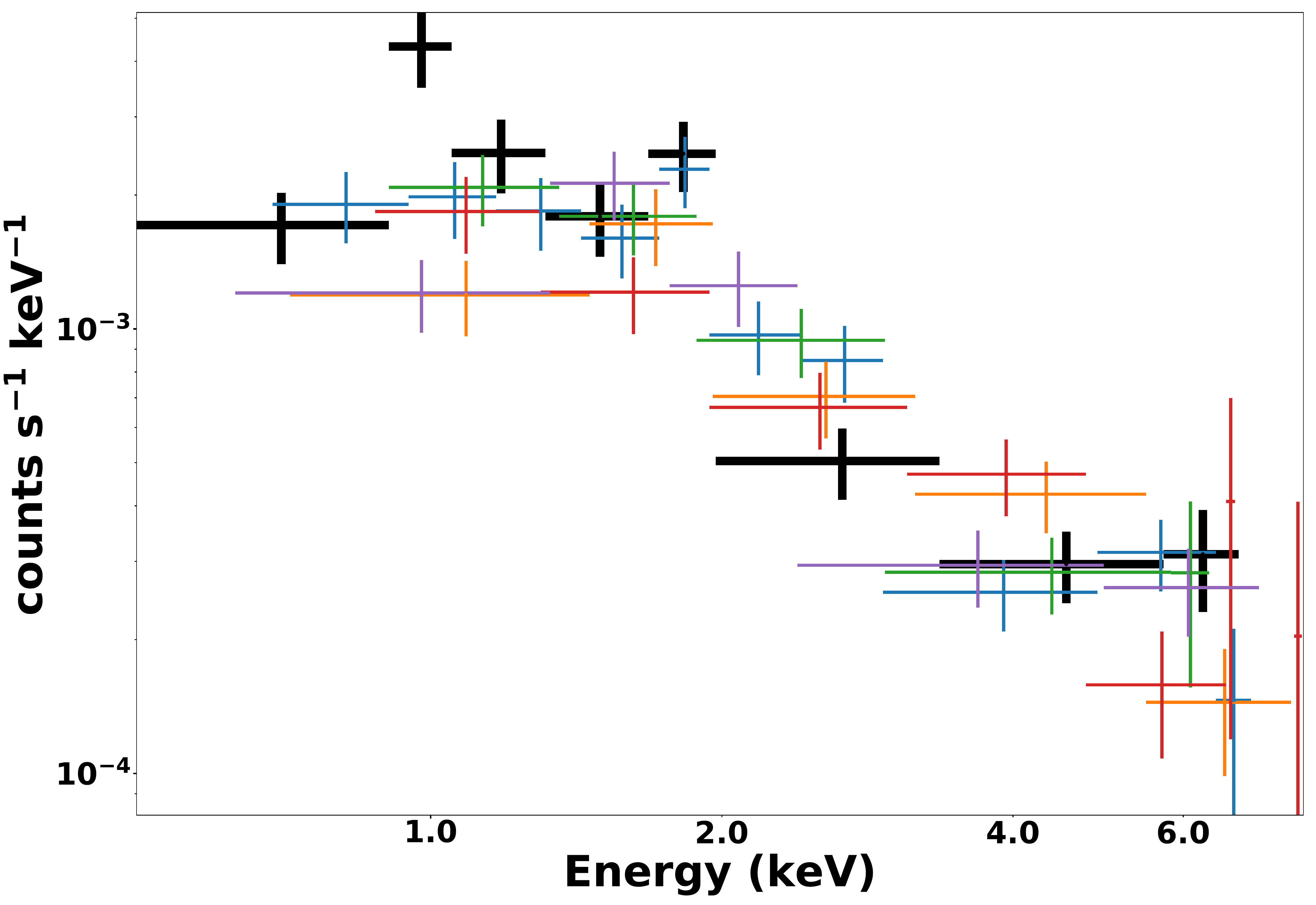}
\caption{The original spectra of the NE nucleus, extracted from the red aperture defined in Figure \ref{fig:2nuclei}.  {\em Black spectrum}: the spectrum from the 2000 data. {\em Colored spectra}: spectra of {each individual observation} from the 2016 data. All the spectra were binned to a minimum S/N of 3 bin$^{-1}$.
\label{fig:n-0501-raw}}
\end{figure}

\begin{figure}[!htb]   
\centering
\epsscale{1.15}
\plotone{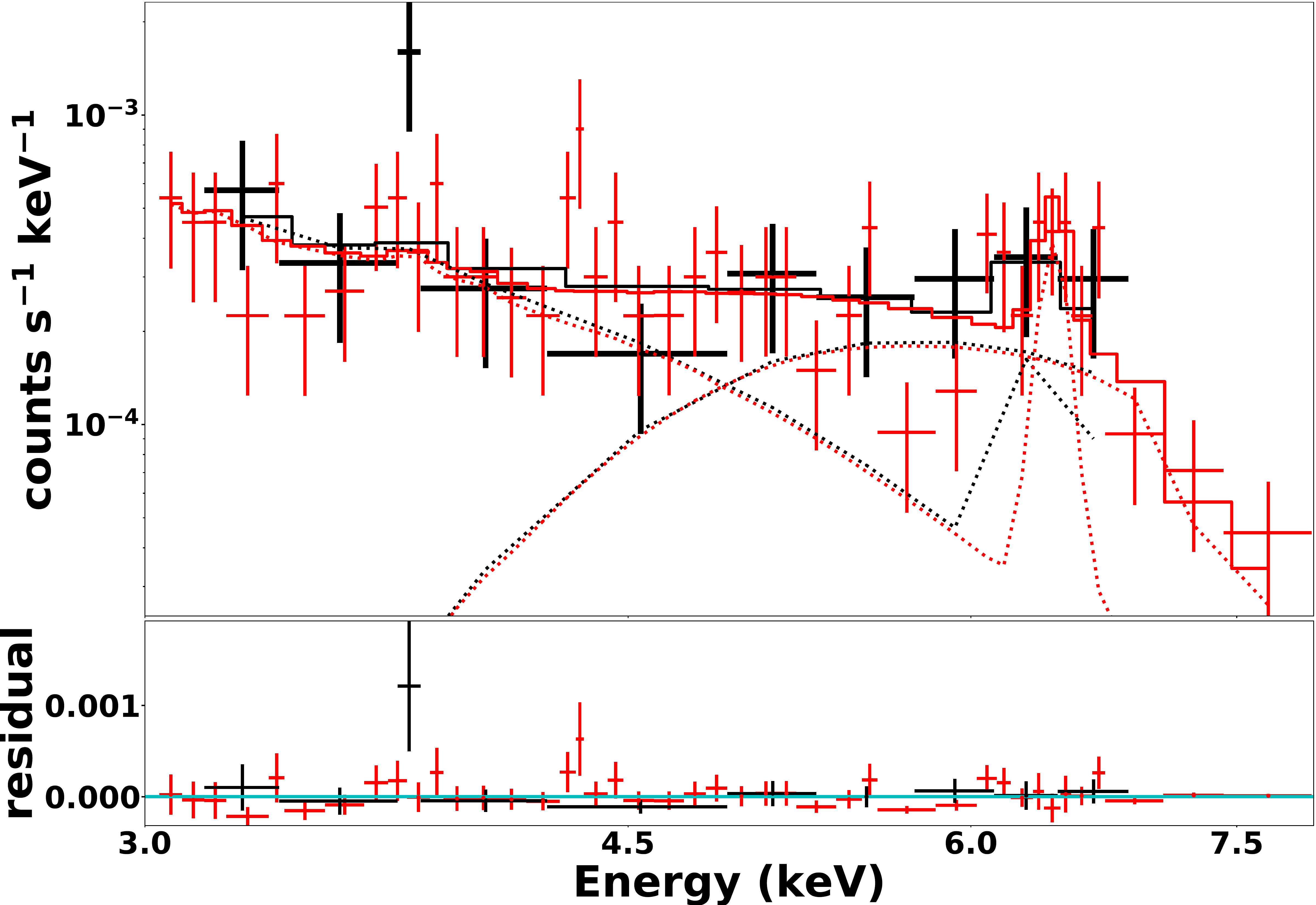}
\caption{Top: simultaneous fitting to the spectra of the NE nucleus from the 2000 data (black) and the 2016 data (red) in the 3-8 keV range. {The model consists of a thermal gas component (MEKAL) with variable Fe abundance, and a power-law component with absorption.} The best-fit parameters are summarized in Table \ref{tab:compare-nuclei}. {The spectra shown in this figure are binned to 5 counts bin$^{-1}$ only for the purpose of better visualization, while the spectra were binned to 1 counts bin$^{-1}$ in the fit.} The model is shown in solid lines and the components of it are shown in dotted lines. Bottom: residuals (data minus model).   
\label{fig:n-0501-38kev}}
\end{figure}

\subsubsection{NE Nucleus}
For the NE nucleus, the aperture was chosen to include all the hard X-ray emission in the vicinity of the secondary peak seen in the 4-6 keV image. The radius of the aperture was chosen to be 1$\arcsec$, in order to contain as much hard X-ray emission associated with the NE nucleus as possible, but minimize the overlap with the aperture used for the SW nucleus.

As shown in Figure \ref{fig:s-nuc-1-23456}, the hard X-ray flux of the SW nucleus has dropped by $\sim$ 60\% from 2000 to 2016, {making the hard X-ray emission from the NE nucleus less contaminated by emission from the SW nucleus}. As a result, the spectra of the NE nucleus from the 2016 data are less affected by the hard X-ray emission from the SW nucleus. All the spectra of the NE nucleus from the 2016 data were combined for this analysis.

The raw spectra are shown in Figure \ref{fig:n-0501-raw}, and the combined spectrum from the 2016 data is shown in Figure \ref{fig:n-0501-38kev}.  Due to the limited counts obtained, the spectral features critical for the spectral fitting are washed out if the spectra are binned to 15 counts bin$^{-1}$. Unbinned spectra, in this case, have bins with zero counts and will therefore bias the results (see footnote 3 on page 3). {Therefore, the spectra were binned mildly to 1 counts bin$^{-1}$, and the CSTAT statistic was used for the fits. The spectra were fitted in the energy range of 3-8 keV, with a model consisting of a thermal gas component with variable Fe abundance \citep[Mewe-Kaastra-Liedahl or MEKAL, see ][and references therein]{MEKAL} and Galactic absorption \citep[with column density of 9$\times$10$^{19}$ cm$^{-2}$, see][]{Kalberla2005}, as well as a power-law component with absorption. The photon index of the power-law component was fixed at 1.9, as adopted for the SW nucleus. The best-fit models are shown in Figure \ref{fig:n-0501-38kev}. The best-fit results are listed in Table \ref{tab:compare-nuclei}.} 

{We have also fitted the spectra of the NE nucleus in the full energy range (0.4-8 keV) with the same model, and the results of the best fit are consistent with those obtained from the fit in the energy range of 3-8 keV. Specifically, a temperature of kT=6.24$^{+3.86}_{-2.37}$ keV for the thermal component and an absorption column density of N$_{H}$=9.00$^{+4.98}_{-3.22}$$\times$10$^{23}$ cm$^{-2}$ for the power-law component were obtained. They are omitted in Table \ref{tab:compare-nuclei} to avoid redundancy.}  These results suggest the co-existence of a very hot, Fe {\sc xxv}-line-emitting gas and a heavily absorbed AGN.

{In order to gain a better knowledge of the Fe emission line from the NE nucleus, we also fitted the spectra with a model consisting of an absorbed power-law, a zero-metallicity thermal gas component, and an Fe line with Gaussian profile. The fit failed to converge when the absorption column density for the power-law and the temperature of the thermal gas component were set as free parameters. We have thus fixed those parameters to their best-fit values obtained from the fits with the power-law plus MEKAL model described in the last paragraph. The results of the new fit are summarized in Table \ref{tab:compare-nuclei}. While the Fe line seen in the SW nucleus is Fe K$\alpha$ with a rest frame energy of 6.4 keV, the dominant Fe line seen in the NE nucleus is instead Fe {\sc xxv} with a rest-frame energy of $\sim$6.7 keV. The broad width of the Fe {\sc xxv} line suggests that the line is blended with other highly-ionized Fe lines.} 

\begin{deluxetable*}{cc|cc|cc}[!htb]
\tablecolumns{6}
\tablecaption{Results from Fits of the Individual SW and NE Nuclei}
\tablehead{
\colhead{Model Component\tablenotemark{a}} & \colhead{Parameters} & \multicolumn{2}{c}{SW nucleus} & \multicolumn{2}{c}{NE nucleus}\\
&  &\colhead{2000 data\tablenotemark{b}} & \colhead{2016 data\tablenotemark{b}} & \multicolumn{2}{c}{2000+2016 data\tablenotemark{c}}
}
\startdata
\multirow{2}{*}{PL}      & N$_{H}$\tablenotemark{d}       & $2.79_{-0.22}^{+0.30}$ (t) & $2.79_{-0.22}^{+0.30}$ (t)     & $6.78^{+4.99}_{-3.31}$ &  6.78(fixed)         \\
                           & $\Gamma$                      & 1.9(fixed)                    &       1.9(fixed)      &  1.9(fixed)&       1.9(fixed)            \\ \hline
\multirow{4}{*}{Fe Line} & $E_{line}$\tablenotemark{e} & $6.36_{-0.04}^{+0.04}$ (t)  & $6.36_{-0.04}^{+0.04}$ (t)  &    N/A    & $6.64^{+0.11}_{-0.11}$                         \\
                           & width\tablenotemark{e}      & $0.11_{-0.02}^{+0.02}$ (t)      & $0.11_{-0.02}^{+0.02}$ (t)      & N/A    & $0.20_{-0.07}^{+0.07}$                           \\
                           & EW\tablenotemark{e}     & 0.23$_{-0.05}^{+0.06}$      & 0.95$_{-0.12}^{+0.05}$      &          N/A  &     0.70$_{-0.29}^{+0.99}$
                                       \\
                           & flux\tablenotemark{f}           & 3.67$^{+2.35}_{-2.35}$   & 5.18$_{-0.07}^{+0.07}$   &      N/A   & $0.98_{-0.58}^{+0.62}$                        \\ \hline
\multirow{3}{*}{MEKAL}   & N$_{H}$\tablenotemark{d}       & N/A                         & N/A                         &  $0.0009$(fixed)   &  $0.0009$(fixed)             \\
                           & kT\tablenotemark{g}                            & N/A                         & N/A                         &   7.21$^{+4.47}_{-2.29}$ & 7.21(fixed)         \\
                           & Z/Z$\sun$                    & N/A                         & N/A                  & 2.31$^{+1.16}_{-1.46}\tablenotemark{h}$  & 0(fixed) \\ \hline
{Flux (4-8 keV)\tablenotemark{i}}  & & 2.94$^{+0.50}_{-0.50}$ & 1.16$^{+0.24}_{-0.24}$ & \multicolumn{2}{c}{0.39$^{+0.01}_{-0.22}$\tablenotemark{k}} \\
{Flux (4-8 keV, absorption corrected)\tablenotemark{i}} && 5.74$^{+0.95}_{-0.95}$  & 2.29$^{+0.51}_{-0.51}$  & \multicolumn{2}{c}{1.60$^{+0.04}_{-0.85}$\tablenotemark{k}} \\
{Luminosity (4-8 keV, absorption corrected)\tablenotemark{j}} && 1.22$^{+0.20}_{-0.20}$  & 0.49$^{+0.11}_{-0.11}$   & \multicolumn{2}{c}{0.35$^{+0.01}_{-0.18}$\tablenotemark{k}}  \\
$\chi^{2}_{\nu}$ (DOF)\tablenotemark{l}        &                               & 1.5(52)                   & 1.5(52)       &- &-   \\              
cstat(DOF)\tablenotemark{m}   &       &                -        &         -                             &  158.2(201) & 156.7(201)
\enddata
\tablenotetext{a}{Different model components used in the fitting. They are the absorbed power-law component (PL), the Fe line with Gaussian profile, as well as the thermal, hot diffuse gas component (MEKAL). ``N/A'' means that the component is not included in the corresponding model. {All the errors listed correspond to a confidence range of 90\%, or $\sim$$\pm{1.6}$$\sigma$. For the data fitted with the CSTAT statistic (i.e., the results of the NE nucleus), the errors are derived from the default MCMC method implemented in XSPEC.}}
\tablenotetext{b}{The results of the simultaneous fitting to the spectra of the SW nucleus from the 2000 data and 2016 data, assuming that the change of hard X-ray flux is caused by the intrinsic variability of the central engine. Label (t) means the corresponding parameters are tied together in the fitting.}
{\tablenotetext{c}{The spectra from the 2000 data and 2016 data are fitted simultaneously.}
\tablenotetext{d}{Absorption column density. In units of 10$^{23}$ cm$^{-2}$.}
\tablenotetext{e}{In the rest frame and in units of keV.}
\tablenotetext{f}{In units of 10$^{-6}$ photons s$^{-1}$ cm$^{-2}$.}
\tablenotetext{g}{In units of keV.}
\tablenotetext{h}{Abundance of Fe in units of solar abundance. All other elements are fixed at solar abundance.}
\tablenotetext{i}{In units of $10^{-13}$ ergs cm$^{-2}$ s$^{-1}$.}
\tablenotetext{j}{In units of $10^{43}$ ergs s$^{-1}$.}
\tablenotetext{k}{The results of the four fits are combined together for simplicity, where the errors represent the full range of the fluxes and luminosities obtained for the four fits.} 
\tablenotetext{l}{The reduced \chitwo ($\chi^{2}_{\nu}$) from the fits and the corresponding degrees of freedom (DOF).}
\tablenotetext{m}{The CSTAT statistics from the fits and the corresponding degrees of freedom (DOF).}}
\label{tab:compare-nuclei}
\end{deluxetable*}

\begin{figure}[!htb]   
\epsscale{1.0}
\plottwo{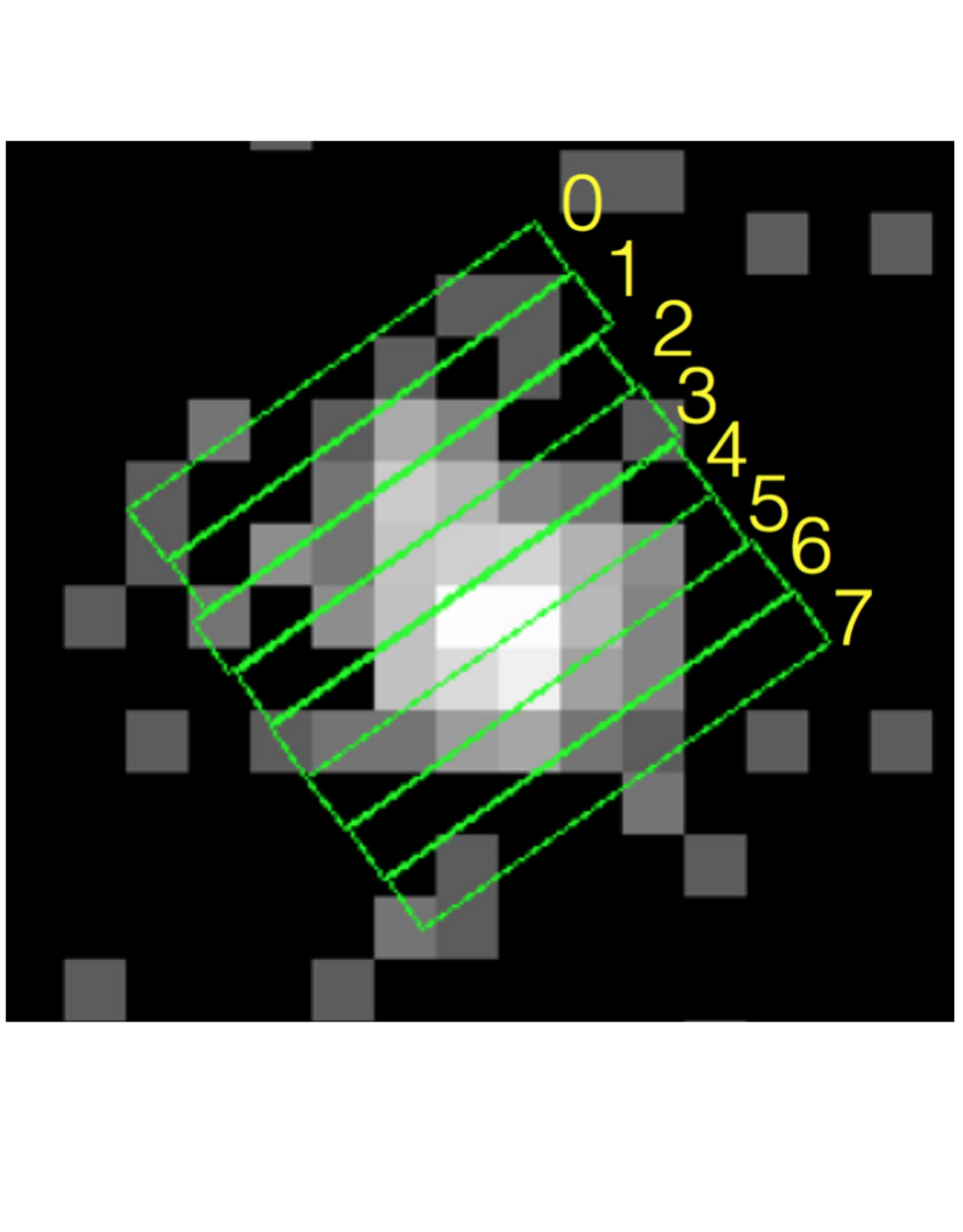}{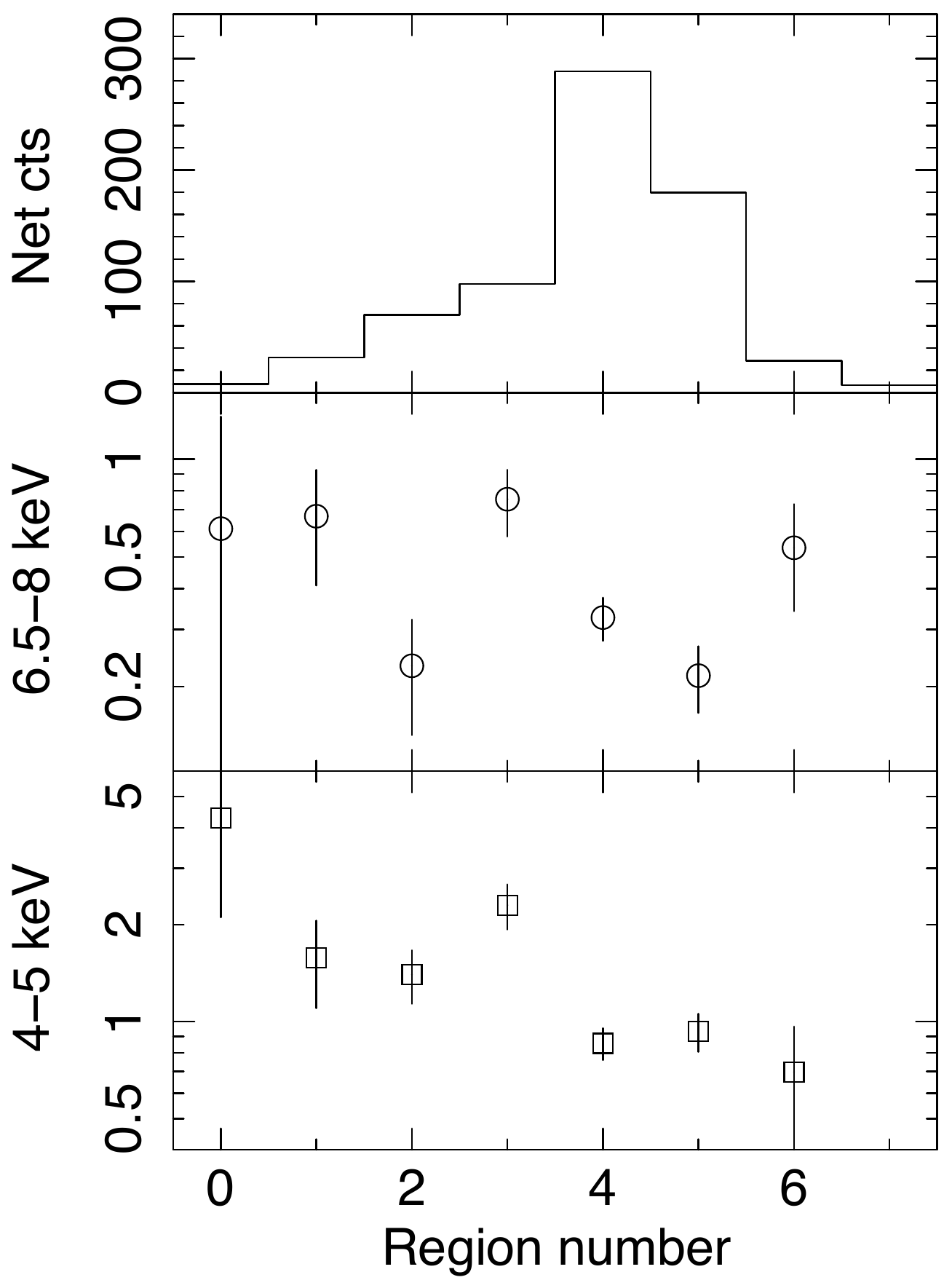}
\caption{{\em Left:} 8 pseudo slits used to extract spectra across the Nuclear Region. The grayscale image is at 6.5-8 keV in the observer's frame. {\em Right:} the histogram showing the total counts of different slits in 4-8 keV band (top), as well as the relative flux in the 6.5-8 keV band (middle) and 4-5 keV band (bottom) normalized to the flux in the 5-6 keV band. For slit 7, the 5-6 keV flux is formally negative and the corresponding data points are thus not shown in the middle and bottom panels. 
\label{fig:8slit}}
\end{figure}

\begin{figure}[!htb]   
\epsscale{1.2}
\plotone{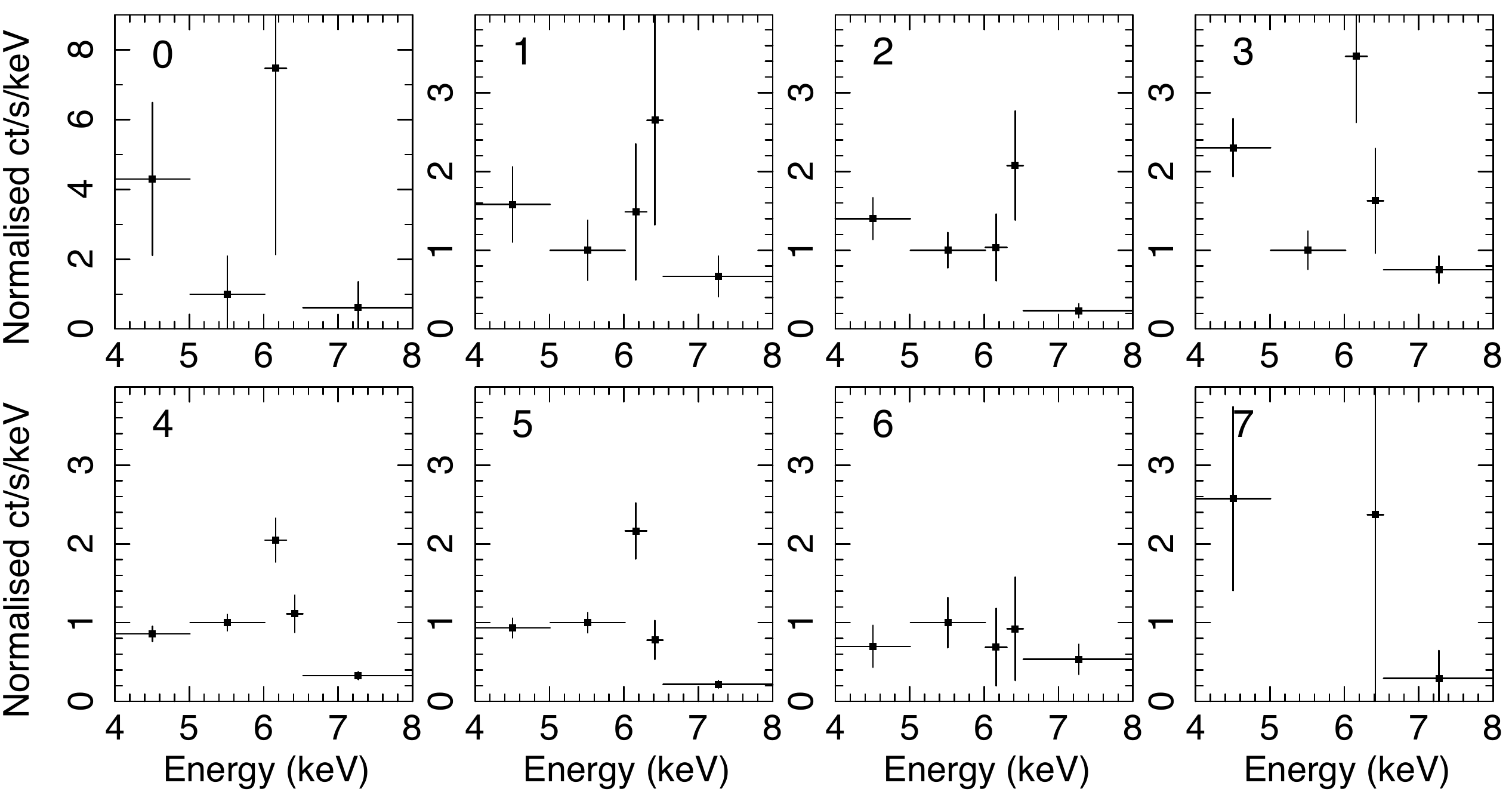}
\caption{Spectra extracted from the 8 pseudo slits, which are normalized at the 5-6 keV bin. The only exception is slit 7 where the 5-6 keV data point is formally negative, and the normalization of the spectrum is thus chosen arbitrarily. The slit names are shown in the top left corner of each panel. The 4-5 keV and 6.5-8 keV bins in the observer's frame were used to monitor the variation of the continuum across the pseudo slits. The Fe K$\alpha$ 6.4 keV line and highly ionized Fe {\sc xxv} 6.7 keV line (possibly blended with other highly ionized Fe lines) are located in 6-6.3 keV and 6.3-6.5 keV energy bins in the observer's frame respectively. 
\label{fig:8slit-spec}}
\end{figure}

\begin{figure*}[!htb]
\begin{minipage}[t]{0.33\textwidth}
\includegraphics[width=\textwidth]{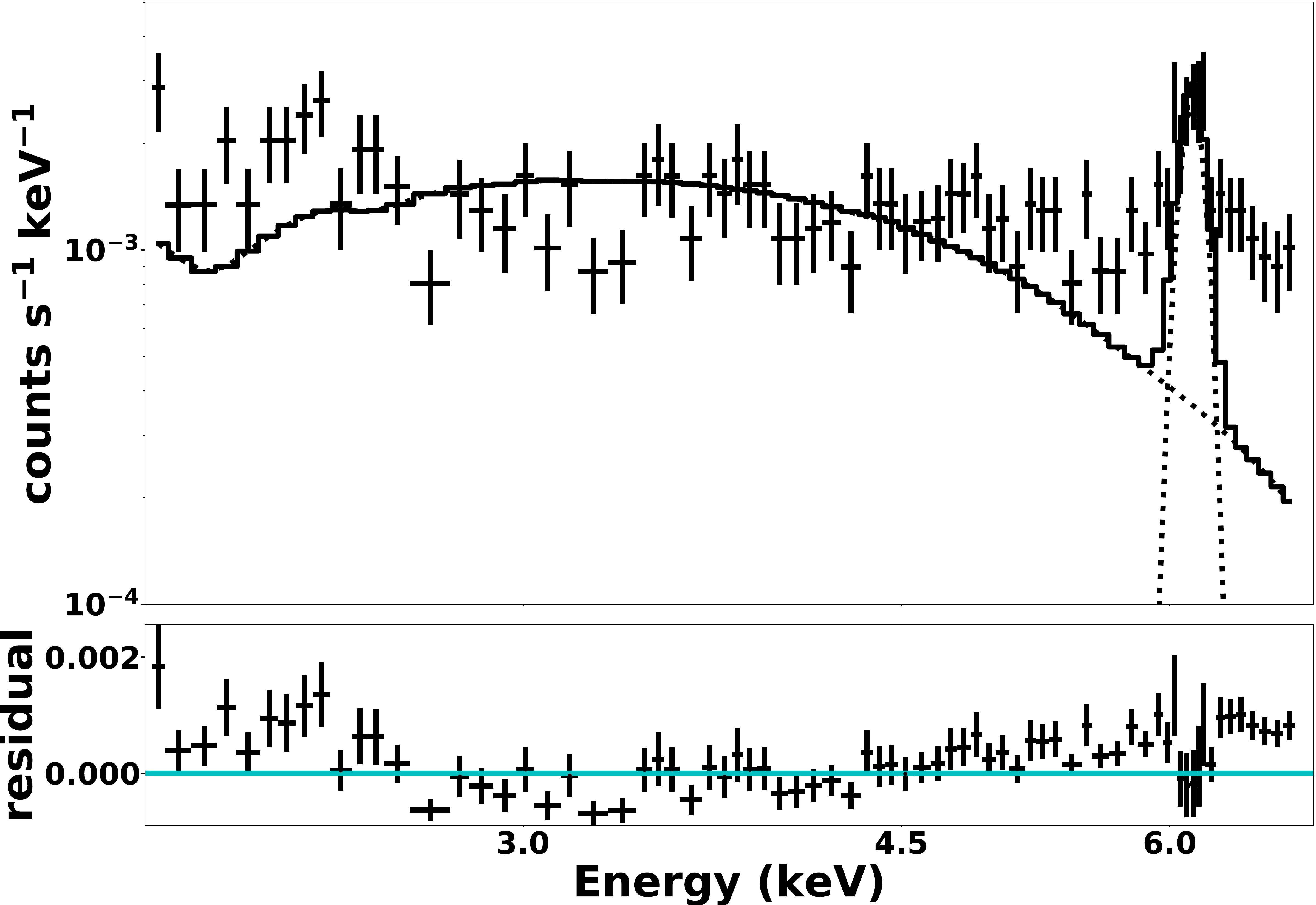}
\end{minipage}
 \begin{minipage}[t]{0.33\textwidth}
\includegraphics[width=\textwidth]{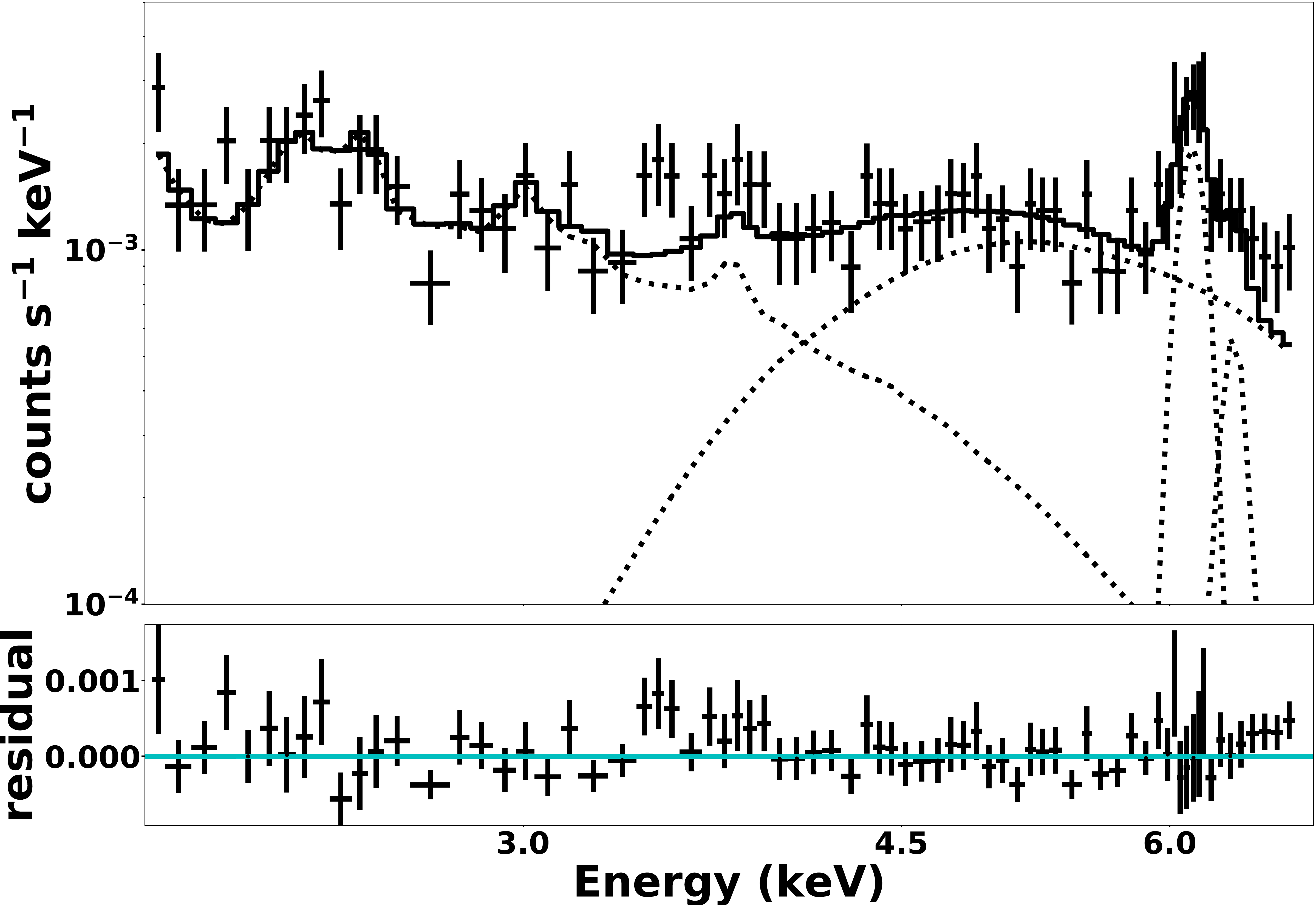}
\end{minipage}
 \begin{minipage}[t]{0.33\textwidth}
\includegraphics[width=\textwidth]{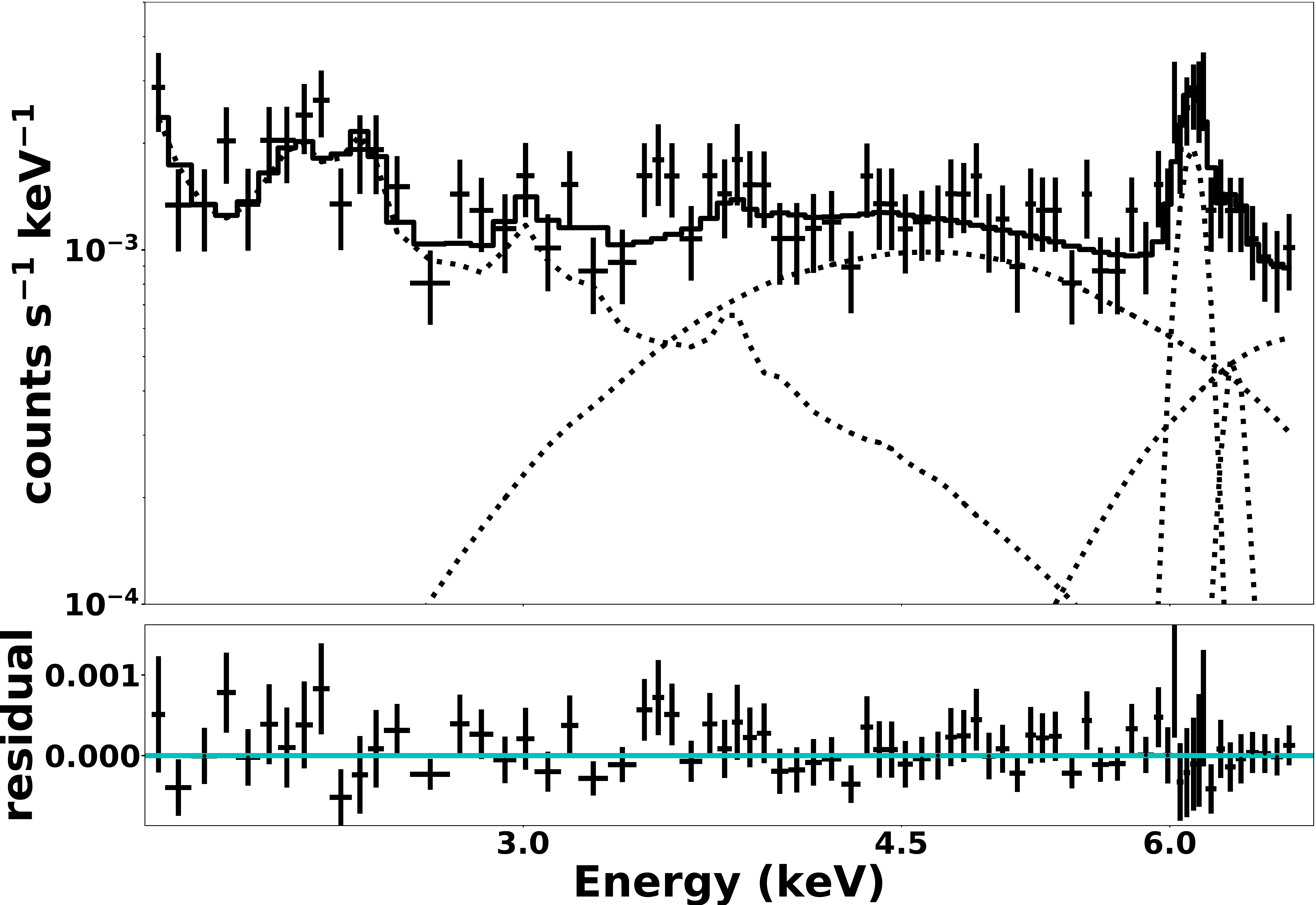}
\end{minipage}
\caption{Top: the fits of the spectrum of the Nuclear Region from the 2016 data at 2-7 keV. The model is made up of: {\em Left:} one absorbed power-law component with photon index fixed at 1.9 and an Fe line. {\em Middle:} one absorbed power-law component with photon index fixed at 1.9, an absorbed thermal gas component (MEKAL) and an Fe line. {\em Right:} two absorbed power-law components with photon index fixed at 1.9, an absorbed thermal gas component (MEKAL) and an Fe line. It is clear that only one power-law component cannot explain the spectrum (left panel). The models are shown in solid lines and the components of them are shown in dotted lines. Bottom: residuals (data minus model). 
\label{fig:nuclear-27kev}}
\end{figure*}

Another evidence of AGN activity in the NE nucleus comes from the shape of the hard X-ray continuum. As shown in Figure \ref{fig:8slit}, 8 parallel pseudo slits were used to extract spectra across the Nuclear Region. All the slits are perpendicular to the vector connecting the SW and NE nucleus. The spectra are shown in Figure \ref{fig:8slit-spec}.

The flux (in units of counts s$^{-1}$ keV$^{-1}$) in the two energy bins, 4-5 keV and 6.5-8 keV in the observer's frame, were used to monitor the variation of the continuum across the pseudo slits. The Fe K$\alpha$ 6.4 keV line and highly ionized Fe {\sc xxv} 6.7 keV line (possibly blended with other highly ionized Fe lines) are located in 6-6.3 keV and 6.3-6.5 keV energy bins in the observer's frame separately. Therefore, the line emission do not contribute to the 4-5 keV and 6.5-8 keV bins.  

At the position of the SW nucleus (slits 4 and 5), the 4-8 keV flux peaks, but the flux in both the 6.5-8 keV and 4-5 keV bins are relatively low. At the position of the NE nucleus (slits 1 and 2), the fluxes in both the 6.5-8 keV and 4-5 keV bins are larger than those in the SW nucleus. Firstly, the higher flux in the 6.5-8 keV bin of the NE nucleus suggests a harder continuum than that of the SW nucleus. This perhaps implies the existence of a more heavily-obscured AGN in the NE nucleus, compared to the one in the SW nucleus. Next, the higher flux in the 4-5 keV bin of the NE nucleus, together with the strong Fe {\sc xxv} 6.7 keV line seen at the same location, suggests a significant contribution from a hot gas component.

{The absorption-corrected flux of the NE nucleus in the 4-8 keV band is $\sim$ 1.6 $\times$ 10$^{-13}$ erg s$^{-1}$ cm$^{-2}$, and it is lower than the 5.6 $\times$ 10$^{-13}$ erg s$^{-1}$ cm$^{-2}$ estimated in \citet{Iwasawa2017}. This difference might be due partly by the fact that the former is calculated from an aperture of 1\arcsec\ in radius while the latter is estimated from the decomposition of the \textit{NuSTAR} spectrum (extracted from an aperture with a radius of 0\arcmin.8). The value estimated from the \textit{NuSTAR} data thus sets an upper limit to the flux of the NE nucleus.}

The heavily-absorbed nature of the AGN in the NE nucleus is also consistent with observations at {other wavelengths}. The high absorption column density is supported by the previous infrared and radio observations. The NE nucleus contains most of the molecular gas in the system within an extremely compact core \citep[radius $\sim$ 120 pc, see][]{mrk273CO1998}, and the NE nucleus is the source of most of the mid-infrared luminosity \citep{Soifer2000}. {Near-infrared integral field spectroscopy of the inner kiloparsec of Mrk~273 associates high-ionization coronal line emission ([Si {\sc vi}] 1.964 \mum) with the SE radio source (radio component to the south of the NE nucleus, see footnote 2 on page 2 for more information), which is likely caused by photoionization from the AGN in the NE nucleus \citep{Vivian2013}.} 

{In addition, current radio data suggest that the radio emission of the NE nucleus arises primarily from multiple radio supernovae and/or supernovae remnants, and it is thus dominated by the starburst \citep[e.g.][]{Carilli2000,Bondi2005}. The X-ray data agree with these radio observations that the starburst dominates here, as the X-ray emission in the soft band ($\textless$ 2 keV) is consistent with emission from hot, diffuse gas most likely heated by the starburst activity. However, possible radio emission from an obscured nucleus (if present) cannot be ruled out. Indeed, there is also tentative evidence that one of the compact radio components associated with the NE nucleus may be the radio counterpart of an AGN \citep{Bondi2005}. This is consistent with our result that a heavily obscured AGN exists in the NE nucleus based on the X-ray data.} 

{Overall, the multi-wavelength data suggest the coexistence of a heavily absorbed AGN and a hot gas component heated primarily by the starburst in the NE nucleus. }

\subsubsection{The Spectra of the Nuclear Region}

To get a global view of the Nuclear Region, we extracted spectra from Region 1 in Figure \ref{fig:regions} for analysis. All the spectra from the 2016 data were combined as a single spectrum for fitting, while the spectrum from the 2000 data was left alone. The AGN emission and thermal emission from the hot gas within the Nuclear Region should be treated separately due to their different physical origin. Therefore, spectra with energy range of 0.4 - 2 keV and those with energy range of 2-7 keV were fitted separately. 

{There is no sign of contribution from the scattering/reflection emission of the AGN in the 0.4 - 2 keV band. This is supported by the fact that no point-like peak is seen in the image in the soft X-ray band. We therefore conclude that a model with thermal gas component(s) is enough to describe the data.} In order to constrain the temperature and metal abundance of the thermal gas component in the Nuclear Region, simultaneous fitting of both the 2000 data and the 2016 data was carried over the 0.4 - 2.0 keV energy range. {A single temperature, hot diffuse gas (MEKAL) model could not describe the data ({$\chi^{2}_{\nu}$ = 3.1)}}. 

A model made up of two thermal gas components with variable metal abundance for individual elements (hereafter called VMEKAL) was then used in the simultaneous fitting. {As stated above, there is intense starburst activity happening in the Nuclear Region. It is therefore natural to expect that the metallicity pattern of the hot, X-ray emitting gas in this region is similar to that of the yield of SNe II (hereafter called SNe II metallicity pattern). On the other hand, a deviation from the SNe II metallicity pattern is also possible if there is significant amount of gas not enriched by the SNe II. Therefore, three metallicity patterns were tested in the fits for completeness.} For Pattern A, both gas components had the SNe II metallicity pattern used in \citet{Iwasawa2011cgoal}. For Pattern B, Z(Si), Z(O), and Z(Fe) of both gas components were set as free parameters, and Z(Ne)=Z(Si)=Z(Mg). Besides, all other elements were fixed at solar values. For Pattern C, the SNe II metallicity pattern used in Pattern A was adopted for the hotter gas component, and the metallicity pattern used in Pattern B was adopted for the cooler gas component. 

{The best-fit temperature of the two gas components are 0.75$^{+0.07}_{-0.08}$ keV and 2.21$^{+3.83}_{-0.73}$ keV, respectively. For the abundance ratios, all three fitting approaches adopted above agree that super-solar $\alpha$/Fe ratios are measured in both gas components, while the absolute values for individual element abundances are not exactly the same. The median values of the best-fit model parameters and their associated errors for the cooler gas component obtained from the fits above are adopted for further analysis (see Table \ref{tab:extendvarytable}). The $\alpha$/Fe ratio of the hotter gas component is SNe II-like, and is omitted in Table \ref{tab:extendvarytable} for simplicity.}

For the spectrum with energy range of 2-7 keV, the fit was carried out on 2016 data. As a first trial, a model made up of a single absorbed power-law component and an Fe line was used. As shown in the left panel of Figure \ref{fig:nuclear-27kev}, it can hardly describe the data {($\chi^{2}_{\nu}$ = 3.0)}. Secondly, a hot thermal gas component (MEKAL) was added to the model. As shown in the middle panel of Figure \ref{fig:nuclear-27kev}, the model matches the data better ($\chi^{2}_{\nu}$ = 0.95). {The best-fit temperature of the thermal gas component is 2.18$^{+1.80}_{-0.52}$ keV, which is consistent with that of the hotter gas component used in the fit of 0.4-2 keV spectra (2.21$^{+3.83}_{-0.73}$ keV)}. However, the flux of the best-fit model tends to be systematically lower than that of the data beyond 6.5 keV in the observer's frame, suggesting the possible existence of another hard X-ray component.

Another heavily absorbed power-law component was thus added to the model, as shown in the right panel of Figure \ref{fig:nuclear-27kev}. The best-fit model gives a reduced {\chitwo of 0.8}, and the absorption column densities of the two power-law component are $\sim$ 2.1$\times$10$^{23}$ cm$^{-2}$ and $\sim$ 1$\times$10$^{24}$ cm$^{-2}$ respectively. The error for the latter is relatively large, which is on the order of 5$\times$10$^{23}$ cm$^{-2}$. This is probably caused by the degeneracies among the fit parameters, most likely the degeneracy between the absorption column density and the temperature of the hot gas. {These results agree with those obtained in the fits of the SW and NE nuclei separately (Table \ref{tab:compare-nuclei}). They are also broadly consistent with the values listed in \citet[][]{Iwasawa2017} (2.6$^{+0.9}_{-1.1}$$\times$10$^{23}$ cm$^{-2}$ and 1.4$^{+0.7}_{-0.4}$$\times$10$^{24}$ cm$^{-2}$, respectively), which are obtained from the fit of {\em NuSTAR} data. }

{For the absorption-corrected flux of the two power-law components, the value of the less absorbed power-law component (1.97$^{+0.34}_{-0.09}\times$$10^{-13}$ ergs cm$^{-2}$ s$^{-1}$) agrees with that obtained from the 0$\farcs$75-aperture spectrum of the SW nucleus listed in Table \ref{tab:compare-nuclei}, while the value for the more absorbed one (corresponding to the NE nucleus) is not well constrained {($<$2$\times$$10^{-12}$ergs cm$^{-2}$ s$^{-1}$)}. Together with the results presented in Section 3.2.2 and 3.2.3, our results of the two heavily absorbed AGN are in general consistent with those obtained in \citet{Iwasawa2017}.}

\begin{deluxetable*}{ccccccc}[!htb]
\tablecolumns{7}
\tablecaption{Comparison among the Physical Properties of the Thermal Gas Component in Different Regions}
\tablehead{\colhead{Model Parameter\tablenotemark{a}} & \colhead{Nuclear Region\tablenotemark{b}} & \colhead{Host Galaxy Region} & \colhead{Ionization Cone} & \multicolumn{2}{c}{Southern Nebula} & \colhead{NE-extended Nebula}\\
                                         &                        &                          &             & 2016 data       &       2000 + 2016 data                 &                        
}
\startdata
N$_{H}$\tablenotemark{c}                  & $0.23_{-0.05}^{+0.07}$ & $0.03_{-0.006}^{+0.005}$ & 0.009(fix)             & 0.009(fix)             & 0.009(fix)             & 0.009(fix)             \\
kT (keV)                                 & $0.75_{-0.08}^{+0.07}$ & 0.73$_{-0.10}^{+0.08}$   & 0.82$_{-0.08}^{+0.09}$ & 0.60$_{-0.10}^{+0.10}$ & 0.57$_{-0.07}^{+0.06}$ & 0.31$_{-0.01}^{+0.07}$ \\
n$_{e}$\tablenotemark{d} & 24 & 21 & 16 & 3 & 3 & 2 \\
O/O$_\odot$                              & 0.59$^{+0.48}_{-0.34}$ & 0.48$^{+0.40}_{-0.21}$   & 0.44$_{-0.24}^{+0.45}$ & 0.79$^{+1.14}_{-0.40}$ & 0.83$^{+1.05}_{-0.22}$ & 0.16$_{-0.05}^{+0.63}$ \\
Si/Si$_\odot$                            & 0.53$^{+0.20}_{-0.22}$ & 0.17$^{+0.14}_{-0.15}$   & 0.37$_{-0.22}^{+0.14}$ & 1.00$^{+2.13}_{-0.04}$ & 1.32$^{+1.74}_{-0.16}$ & 0.64$_{-0.30}^{+1.25}$ \\
Fe/Fe$_\odot$                            & 0.08$^{+0.03}_{-0.03}$ & 0.10$^{+0.03}_{-0.05}$   & 0.11$_{-0.04}^{+0.04}$ & 0.22$^{+0.56}_{-0.06}$ & 0.30$^{+0.30}_{-0.05}$ & 0.31$_{-0.11}^{+0.14}$ \\
$\alpha$/Fe \tablenotemark{e}          & 6.6$^{+3.5}_{-3.7}$    & 1.7$^{+1.6}_{-1.6}$      & {3.3$^{+4.2}_{-2.6}$}    &     4.6$^{+7.1}_{-2.3}$       & 4.5$^{+3.1}_{-1.9}$   & {2.1$^{+5.9}_{-1.8}$}    \\
flux (0.4-2 keV) \tablenotemark{f}      & 7.96$^{+0.49}_{-0.22}$ & 3.75$^{+0.17}_{-0.18}$   & 1.00$_{-0.08}^{+0.06}$ &  1.88$^{+0.20}_{-0.16}$              & 1.78$^{+0.13}_{-0.12}$ & 0.54$^{+0.08}_{-0.08}$ \\
luminosity (0.4-2 keV)\tablenotemark{g} & 2.63$^{+0.16}_{-0.07}$ & 1.24$^{+0.06}_{-0.06}$   & 0.33$^{+0.02}_{-0.03}$ &  0.62$^{+0.07}_{-0.05}$   & 0.59$^{+0.04}_{-0.04}$ & 0.18$^{+0.03}_{-0.03}$ \\
$\chi^{2}_{\nu}$(DOF)\tablenotemark{h}        &   1.4(93)     & 0.9(67)      &  -    & - & - &-    \\                 
cstat(DOF)\tablenotemark{i}        &   - & -& 142.8(178) & 103.8(72) & 124.9(121) & 45.6(41)                       
\enddata
\tablenotetext{a}{Fitting Model: For the Nuclear Region, the model is two hot gas components with absorption (i.e. phabs(VMEKAL)+phabs(VMEKAL) in XSPEC). For all other regions, the model is just one hot gas component with absorption (i.e. phabs(VMEKAL) in XSPEC). The absorption column densities are fixed to the Galactic value of $9\times10^{19}$ cm$^{-2}$ when necessary. All other elemental abundance not included in the table are fixed at solar values. All the results are from the simultaneous fitting to both the 2000 data and 2016 data. {All the errors listed correspond to a confidence range of 90\%, or $\sim$$\pm{1.6}$$\sigma$. For the data fitted with the CSTAT statistic, the errors are derived from the default MCMC method implemented in XSPEC.}}
{\tablenotetext{b}{The median values obtained for the cooler gas component are listed. See Section 3.2.4 for more details. }}
\tablenotetext{c}{The absorption column density, in units of $10^{22}$ cm$^{-2}$.}
\tablenotetext{d}{Electron number density, in units of 10$^{-3}$ cm$^{-3}$. Spherical volumes with a filling factor of 1 are assumed in the calculation except for the Ionization Cone Region, where a bi-cone with an opening angle of 75\degr\ and a filling factor of 1 is assumed.}
\tablenotetext{e}{Abundance ratios of $\alpha$ elements to Fe, determined by the ratio of Si (row 4) and Fe (row 5) abundance. The values are rounded to one decimal. {For the data fitted with the \chitwo statistic, the errors were obtained assuming standard error propagation: $\sigma_{a/b} = \sqrt{(\sigma_{a}/b)^{2}+(\sigma_{b}a/b^{2})^{2}}$. For the data fitted with the CSTAT statistic, the errors were derived from the default MCMC method in XSPEC.}}
\tablenotetext{f}{In units of 10$^{-14}$ erg s$^{-1}$ cm$^{-2}$, absorption corrected.}
\tablenotetext{g}{In units of 10$^{41}$ erg s$^{-1}$, absorption corrected.}
\tablenotetext{h}{The reduced \chitwo ($\chi^{2}_{\nu}$) from the fits and the corresponding degrees of freedom (DOF).}
{\tablenotetext{i}{The CSTAT statistics and the corresponding degrees of freedom (DOF). The spectra were binned to 1 counts bin$^{-1}$.}}
\label{tab:extendvarytable}
\end{deluxetable*}

\subsection{Host Galaxy Region (Region 2)}

\begin{figure}[!htb]   
\centering
\epsscale{1.15}
\plotone{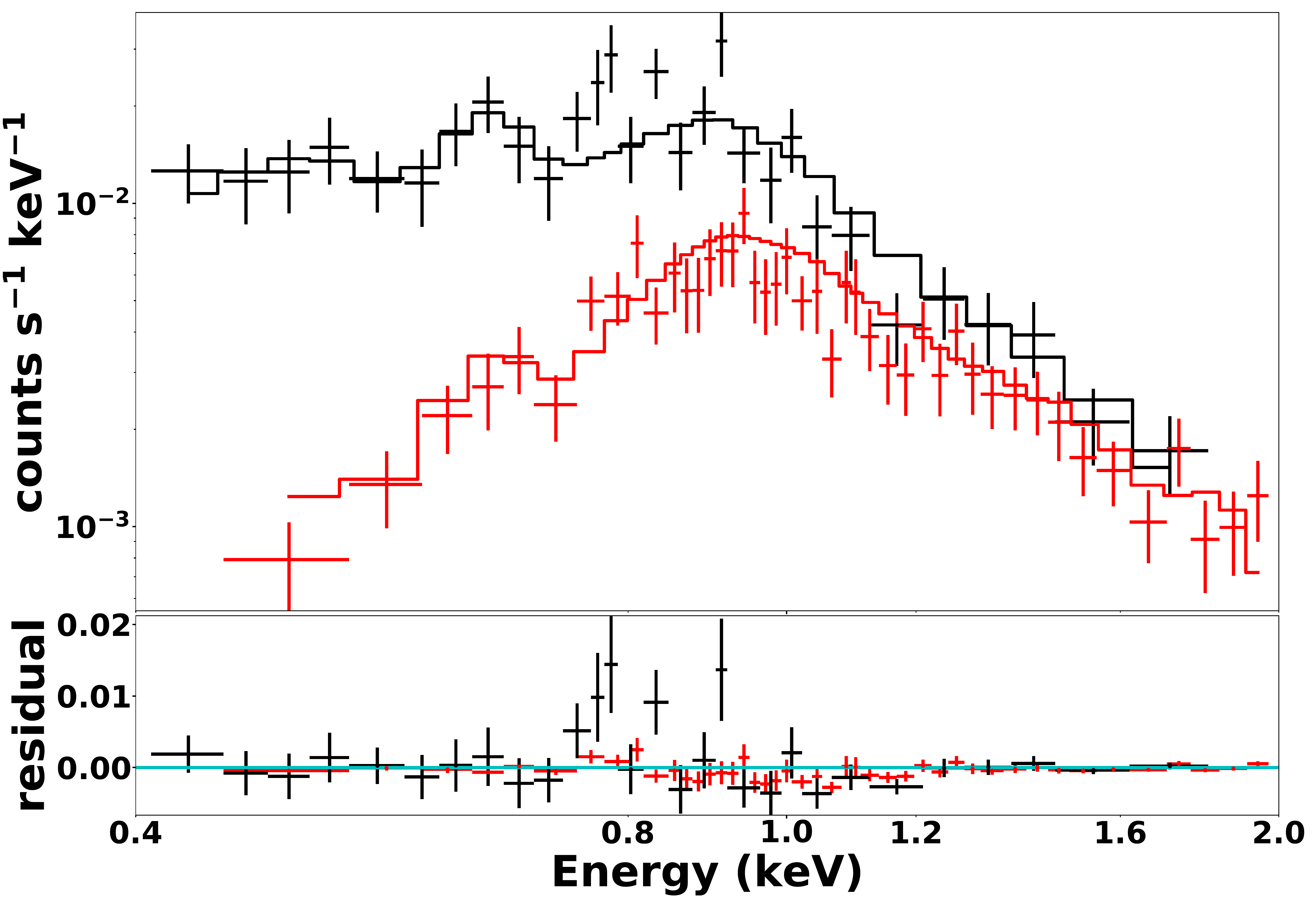}
\caption{Top: the results from the fits of spectra from the Host Galaxy Region (Region 2 as shown in Figure \ref{fig:regions}). Spectra from both the 2000 data (black) and the 2016 data (red) in the energy range of 0.4-2.0 keV were used. The model (solid lines) is only an absorbed thermal gas component with variable abundance of different elements(VMEKAL). The abundances of O, Si (Ne, Mg), and Fe are set as independent variables, while all other elements are fixed at solar values. The best-fit parameters are summarized in Table \ref{tab:extendvarytable}. Bottom: residuals (data minus model).
\label{fig:spec-galaxy}}
\end{figure}

The spectra of the host galaxy were extracted from Region 2 in Figure \ref{fig:regions} (i.e. excluding Region 1). To constrain the temperature and metallicity of the thermal gas in this region, simultaneous fitting to the spectra from both the 2000 data and the 2016 data was carried out. As the flux above 2 keV is negligible in the spectra, the model was only a thermal gas (VMEKAL) with Galactic absorption, and the fit was confined to 0.4-2 keV range. Again, the abundance of O, Si, and Fe were set as independent variables, and the abundances of Mg and Ne were tied to that of Si. The abundances of all other elements were fixed at solar values. The spectra and the best-fit model are shown in Figure \ref{fig:spec-galaxy}, and the results from the fits are summarized in Table \ref{tab:extendvarytable}. In general, the temperature of the thermal gas is similar to that of the cooler gas component in the Nuclear Region (kT $\sim$ 0.73 keV and kT $\sim$ 0.75 keV, respectively). {However, the Host Galaxy Region has a lower $\alpha$/Fe ratio (1.7$^{+1.6}_{-1.6}$) than that seen in the Nuclear Region ($\alpha$/Fe ratio$=$6.6$^{+3.5}_{-3.7}$) at the $\sim$2--$\sigma$ level}.  There is a non-negligible, intrinsic absorption in this region (see Table \ref{tab:extendvarytable}), which is not surprising given the dusty nature of Mrk~273 {\citep[A$_{V} \geqslant$ 15, derived from the ISO SWS with an aperture of 20\arcsec, which covers the whole galaxy except the tidal tail;][]{Genzel1998}.}

One thing to notice is that the absolute values of the abundances are quite low and they should be quoted with caution. When thermal spectra of different temperatures are summed together, emission-line features could be diluted and the best-fit abundance values might yield a falsely low value \citep[e.g.][]{Buote1999}. Therefore, there might actually be multi-temperature gas with higher metallicity in this region, although a two-temperature gas model does not improve the fit (i.e. the reduced \chitwo has only changed by $\sim$ 3\%). Nevertheless, the relative abundance ratios of different metals are much less affected by the effect described above, since all the measured metal abundances are affected in a similar way. 

\begin{figure*}[!htb]
  \centering
   \begin{minipage}{0.32\textwidth}
\includegraphics[width=\textwidth]{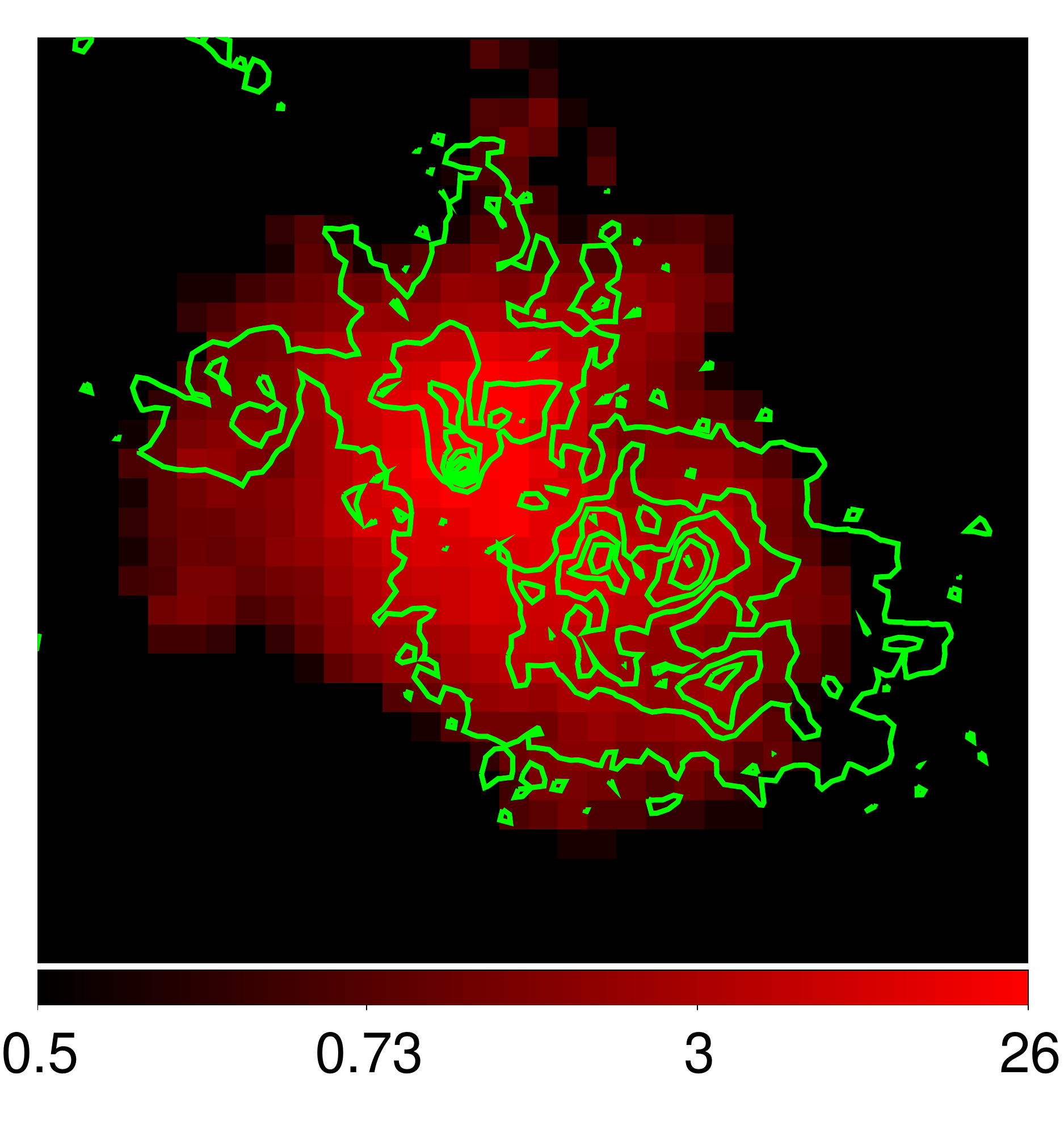}
  \end{minipage}
     \begin{minipage}{0.32\textwidth}
\includegraphics[width=\textwidth]{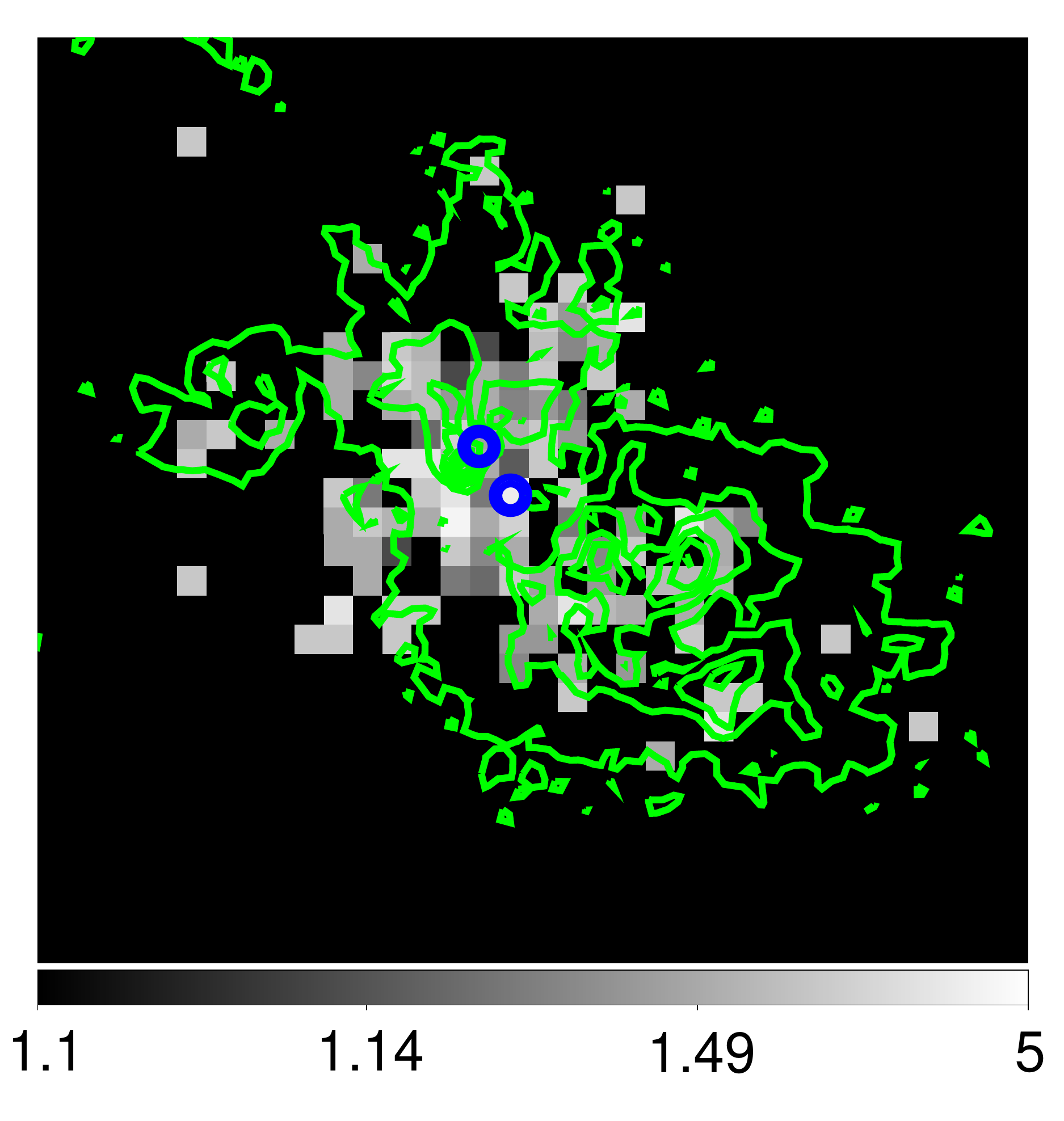}
  \end{minipage}
   \begin{minipage}{0.32\textwidth}
\includegraphics[width=\textwidth]{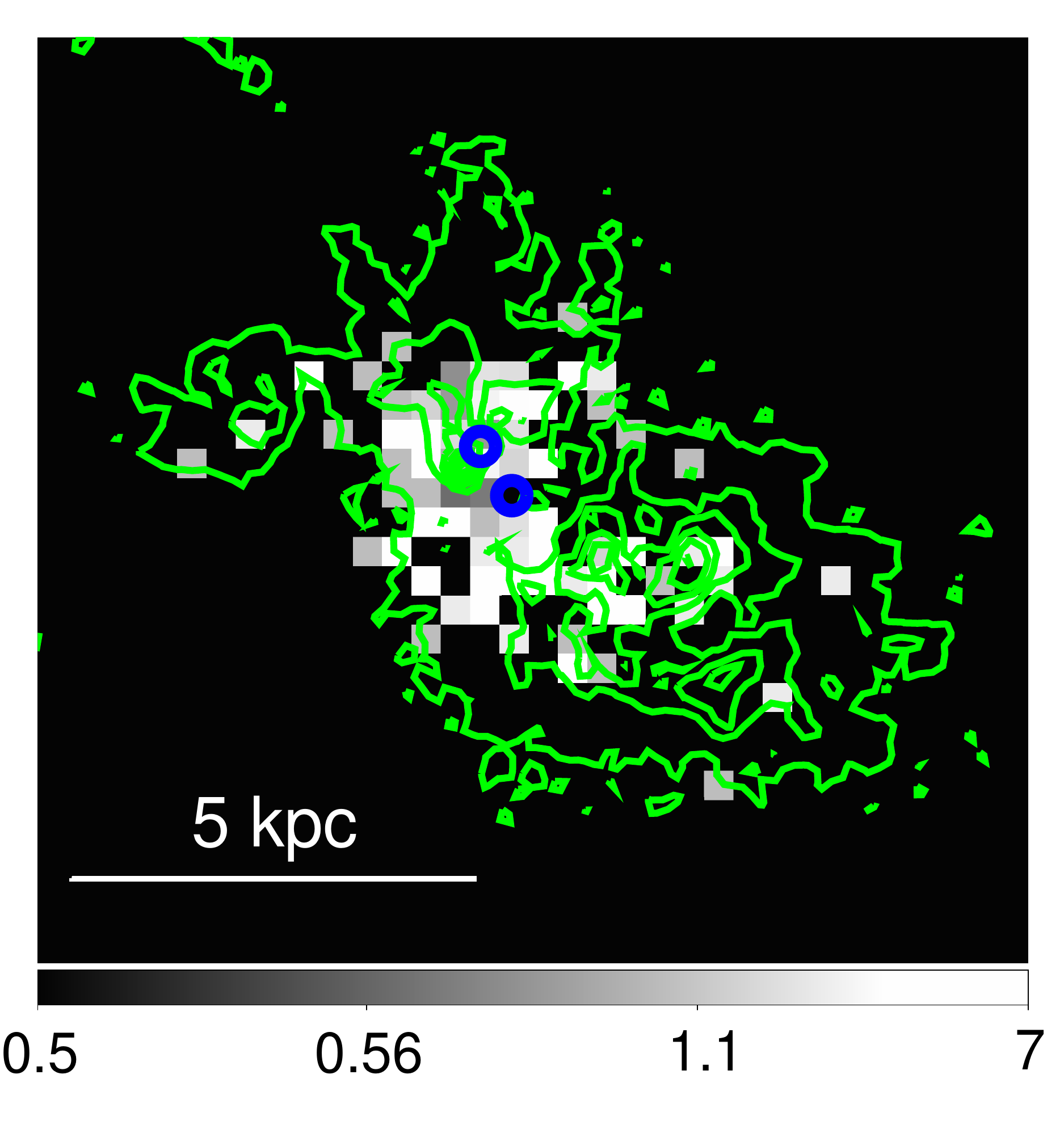}
  \end{minipage}
  \caption{{\em Left}: the image at 1-3 keV within Region 2. The image is on a logarithm scale and in units of counts. {\em Middle}: the map of 1-3 keV to 0.4-1 keV counts ratios. Pixels with counts fewer than 1 are excluded. {\em Right}: Si to Fe L line ratio maps, showing the significant Si {\sc xiii} 1.85 keV emission (without continuum subtraction) southwest and northeast the two nuclei. Pixels with counts fewer than 1 are excluded. In all the three panels, the overlaid green contours are the [O {\sc iii}] $\lambda$5007 emission, tracing the ionization cones and outflowing gas in optical and the two blue circles denote the location of the two nuclei.   
\label{fig:ouflow-sw-excess}}
\end{figure*}

\begin{figure*}[!htb]
\epsscale{1.15}
\plottwo{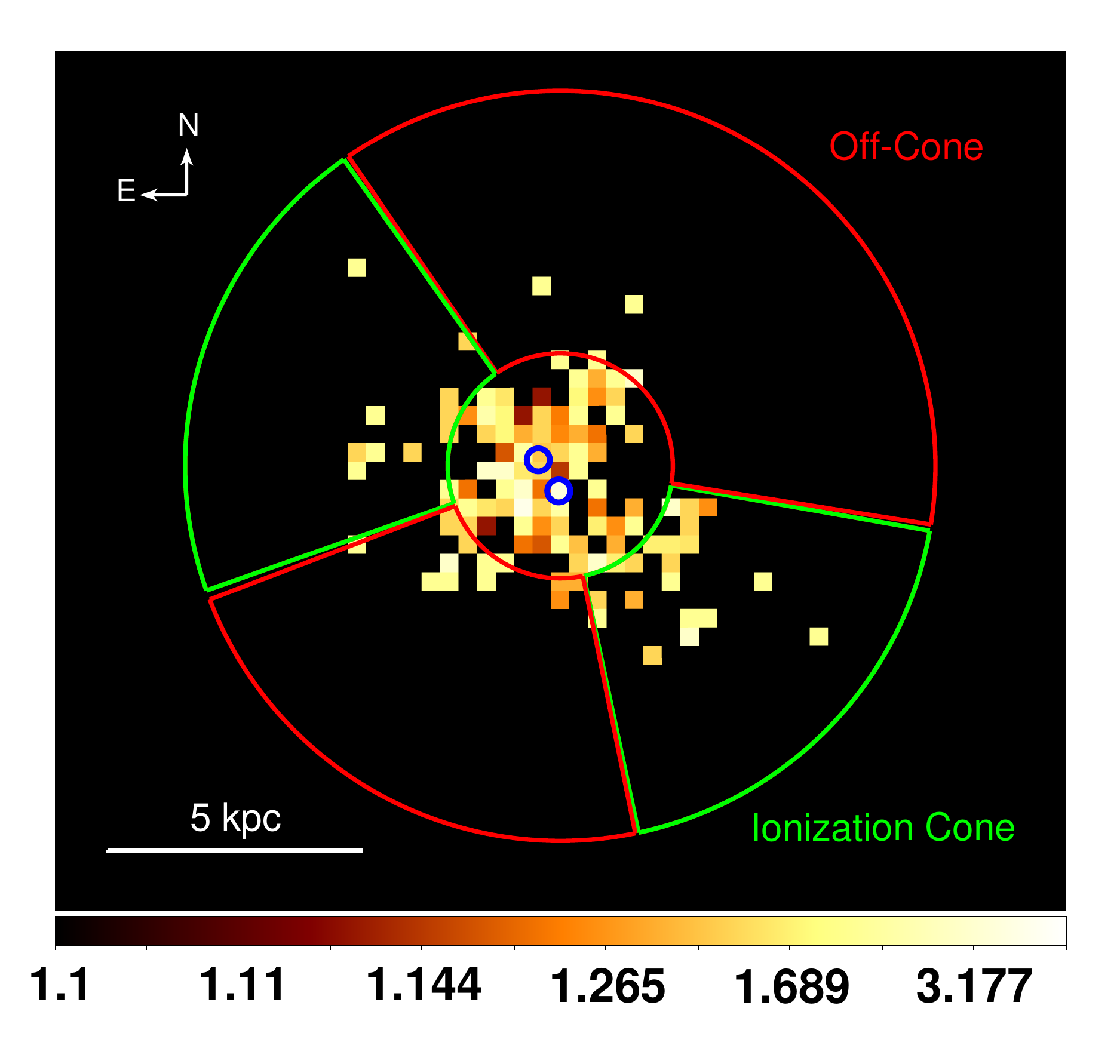}{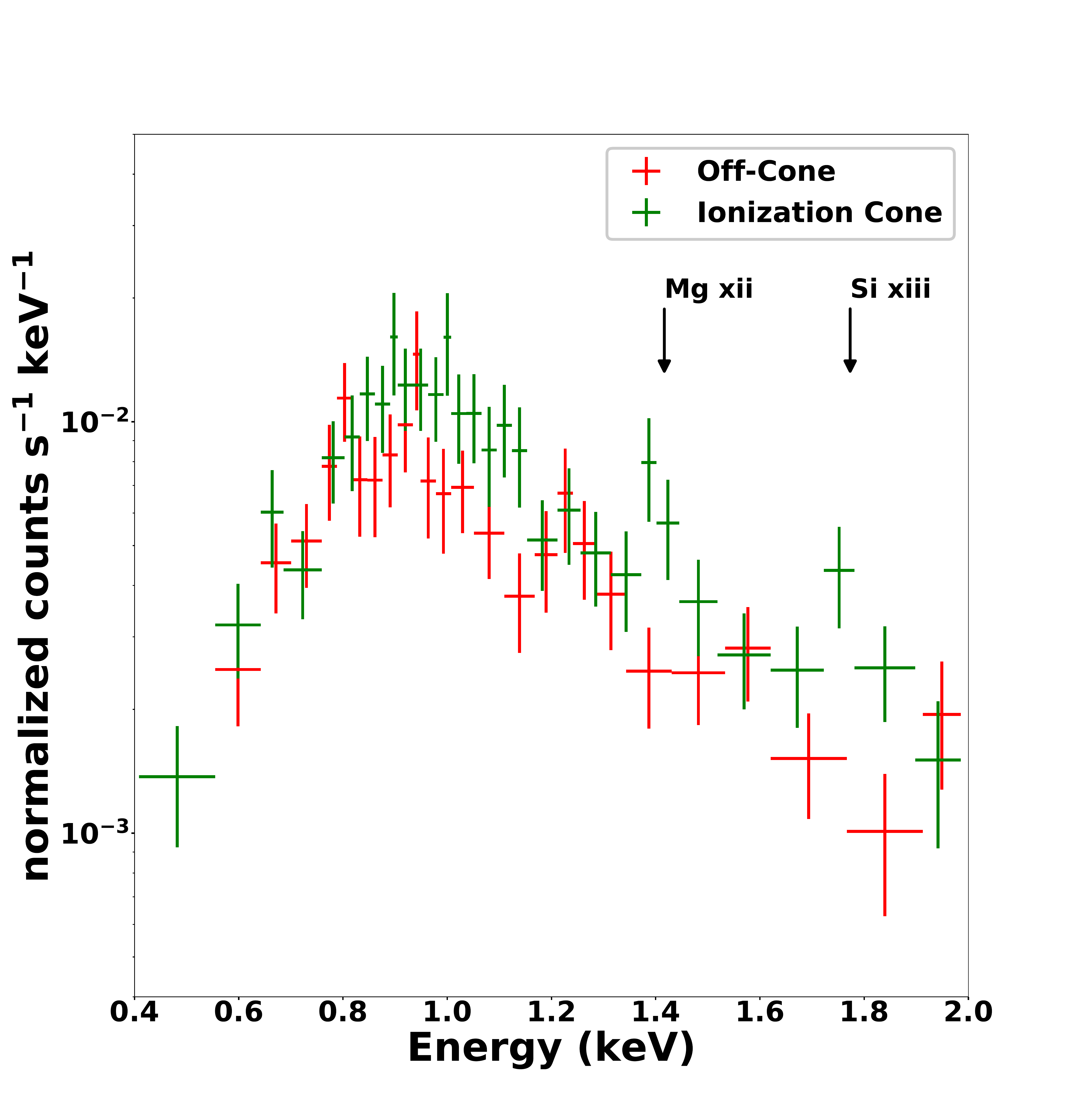}
\caption{{\em Left}: the same 1-3 keV to 0.4-1 keV ratio map as shown in the middle panel of Figure \ref{fig:ouflow-sw-excess} but on a different color scale, over-plotted with the extraction regions (annulus sectors) for Ionization Cone Region (green) and Off-Cone Region (red). The grayscale image is on a logarithm scale and in units of counts. {\em Right}: spectra of the 2016 data extracted from the Ionization Cone Region and Off-Cone Region shown in the left panel. The spectra are both binned to a minimum S/N of 3.5 bin$^{-1}$ and normalized at the first energy bin. 
\label{fig:onoffoutflow}}
\end{figure*}

 \subsection{Ionization Cones and Outflow}
Galactic outflows are usually multi-phase \citep[e.g.][]{veilleuxaraa}. 
The multi-phase nature of galactic outflows is not well understood. {Since the outflow in Mrk~273 has been detected in the ionized and
neutral gas phases \citep[$\sim$ 4 kpc;][]{RupkeVeilleux2013b,Zaurin2014}, as well as the molecular gas phase \citep[$\sim$ 550 pc;][]{Vivian2013,Veilleuxherschel,Cicone2014,Aladro2018}}, it is important to also explore the hot, X-ray-emitting phase of the outflow. For this, we examined  $\leq$ 3 keV X-ray emission where this hot phase may be seen.
  
The 1-3 keV image in counts, 1-3 keV to 0.4-1 keV ratio map and the Si to Fe-L line ratio map within the Host Galaxy Region and Nuclear Region (Region 1 \& 2) are shown in Figure \ref{fig:ouflow-sw-excess}. The Si to Fe-L line ratio map was generated from the division of the narrow-band images of the Si {\sc xiii} 1.85 keV line emission (1.78-1.95 keV) and Fe-L line emission (0.85-1.0 keV). The ratio map of Mg {\sc xii} 1.47 keV to Fe-L line was also obtained, which is similar to the Si to Fe-L line ratio map and is left out to avoid redundancy. Spatially extended 1-3 keV emission and Si (also Mg) line emission are clearly seen on a scale of $\sim$ 7 kpc $\times$ 7 kpc, suggesting the existence of highly ionized, hot gas. Within the Nuclear Region ($\lesssim$2 kpc), starburst, AGN and/or shocks caused by the outflowing gas impacting the ambient material can all be responsible for the ionization/heating of the gas, and it is hard to distinguish them based on the current {\em Chandra} data. However, excesses of 1-3 keV emission and Si {\sc xiii} 1.85 keV (also Mg {\sc xii} 1.47 keV) line emission are seen both southwest and northeast of the two nuclei, at a mean distance of $\sim$ 5 kpc in projection. Both of the excesses are coincident with the [O {\sc iii}] $\lambda$5007 ionization cones seen in \citet{Zaurin2014}, which also traces the outflowing gas along the same direction. This tentatively suggests that the excesses of 1-3 keV and Si {\sc xiii} 1.85 keV (also Mg {\sc xii} 1.47 keV) line emission are related to the outflowing, hot gas. 

The Host Galaxy Region was divided into Ionization Cone Region (green sector annuli) and Off-Cone Region (red sector annuli), based on the location of the [O {\sc iii}] $\lambda$5007 emission as well as the excesses in the 1-3 keV and Si {\sc xiii} 1.85 keV (also Mg {\sc xii} 1.47 keV) line emission, which is shown in the left panel of Figure \ref{fig:onoffoutflow}. Spectra were then extracted from the Ionization Cone Region and Off-Cone Region separately. The combined spectra from the 2016 data are shown in the right panel of Figure \ref{fig:onoffoutflow}. In the spectrum of the Ionization Cone Region, emission feature of Mg {\sc xii} 1.47 keV and Si {\sc xiii} 1.85 keV lines are seen, which are absent in the spectrum of the Off-Cone Region. Fitting the two emission lines with Gaussian profile gives equivalent widths of $\sim$ 0.04 keV and $\lesssim$0.01 keV respectively. These emission lines suggest the existence of highly ionized gas and/or high $\alpha$-element abundance in the Ionization Cone Region, and corresponding outflowing gas too.  

One thing to notice is that the outflow on this spatial scale is probably not confined only to the Ionization Cone Region. According to the integral field spectroscopy in \citet{Colina1999} and \citet{RupkeVeilleux2013b}, the outflows on smaller scales have different orientations (i.e. North-South for $\lesssim$1 kpc and Northwest-Southeast for $\lesssim$4 kpc). The outflow likely continues to move outwards in a similar direction but perhaps a wider opening angle, and the outflowing gas in the Ionization Cone Region may be part of this outflow. 

For the spectra from the Ionization Cone Region, a single-temperature gas model with variable abundance for elements O, Si (tied with Mg, Ne), Fe (VMEKAL) and Galactic absorption was used for the fitting. {The $\alpha$/Fe ratio in this region is 3.3$^{+4.2}_{-2.6}$. These results are summarized in Table \ref{tab:extendvarytable}.}

Additionally, the possible temperature difference of the gas in the Ionization Cone Region and Off-Cone Region was explored by fitting a single-temperature gas model with SNe II metallicity pattern adopted in \citet{Iwasawa2011cgoal} to the spectra. The temperatures derived from the fits are {kT=0.82$^{+0.09}_{-0.08}$ keV for the Ionization Cone Region and kT=0.51$^{+0.10}_{-0.12}$ keV for the Off-Cone Region.} The higher temperature in the Ionization Cone Region suggests possible heating from photoionization and/or shocks induced by the outflow. This result is consistent with the idea mentioned in Section 3.3 that the Host Galaxy Region is made up of multiple gas components of different temperatures.

\subsection{Southern Nebula (Region 3)}
\subsubsection{Soft X-ray Image}
In order to study the spatially extended halo emission, an image in flux unit is needed. To obtain such an image, the dependence of the collecting area on the energy and position as well as the effective exposure in different regions of the detector needs to be taken into account properly. The image of the Southern Nebula in flux unit was produced with the CIAO script \textit{flux\_image} for each observation. In order to obtain the spectral weights needed to produce the exposure map, preliminary spectra were extracted from an elliptical region (Region 3 in Figure \ref{fig:regions}) for each observation. All the spectra from the 2016 data were combined into one single spectrum. The combined 2016 spectrum and 2000 spectrum were then fitted with a single-temperature gas model (MEKAL) simultaneously, and the best-fit model spectra were used to generate the spectrum weights for exposure maps of each observation separately. Finally, all the images were combined.

The final product, the image of the Southern Nebula in flux unit, is shown in both the right panel of Figure \ref{fig:0470csmooth} and Figure \ref{fig:regions}. Soft X-ray emission is seen extending up to $\sim$ 40 kpc in the directions both perpendicular to and along the tidal tail. Apparently, the X-ray emission is suppressed in the tidal tail region seen in the optical image, where a decrease of $\sim$ 23\% in the surface brightness is measured, compared to that of the whole Southern Nebula. This suggests further that the X-ray emission is not exclusively caused by the tidal effects{, which is discussed further in Section 4 \citep[see also][]{Iwasawa2011}. }

\subsubsection{Spectra}
\begin{figure}[!htb]
\epsscale{1.15}
\plotone{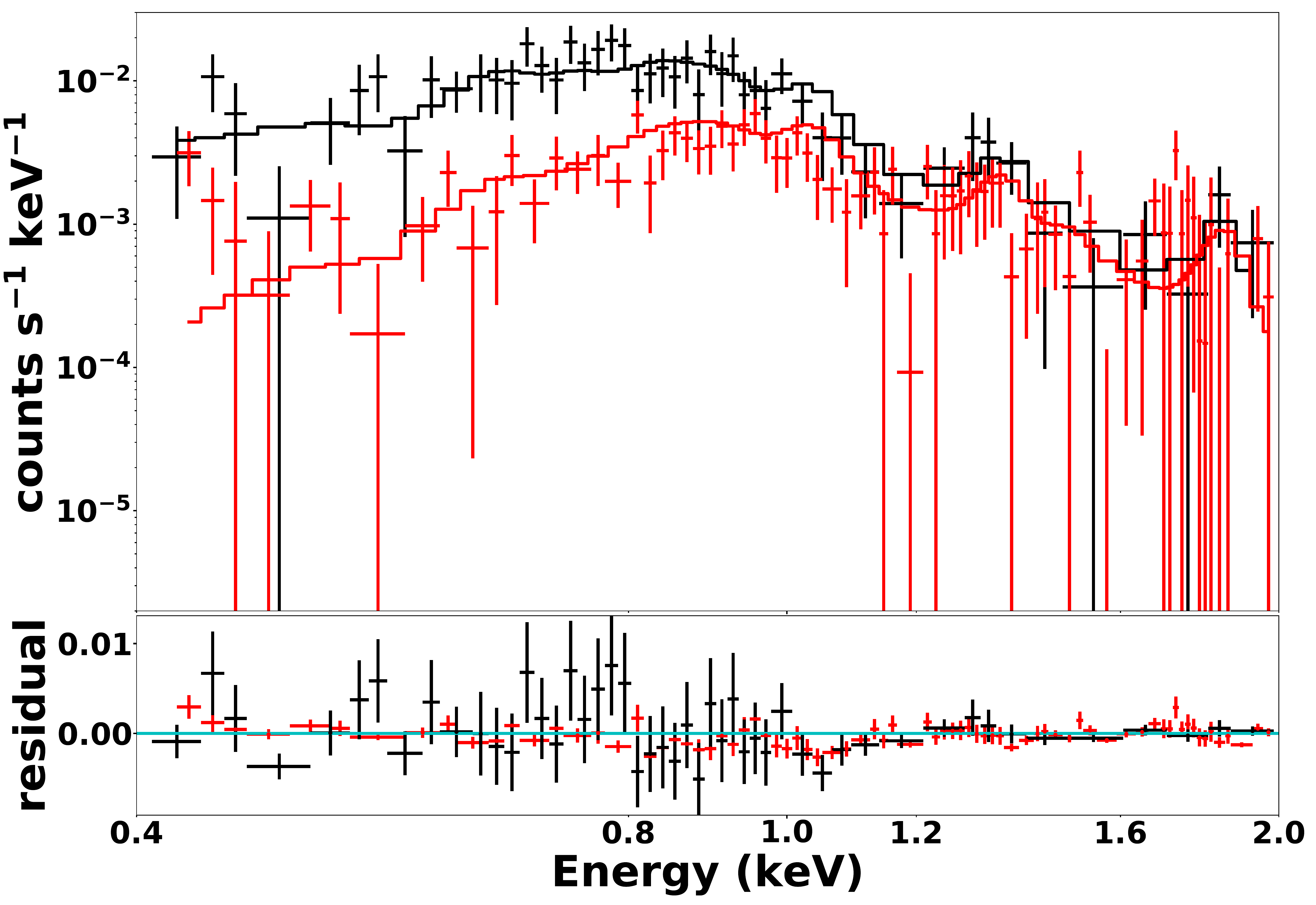}
\caption{Top: simultaneous fitting to the spectra of the Southern Nebula from 2000 data (black) and 2016 data (red). A thermal gas model (VMEKAL) is fitted to the spectra and shown in solid lines. The abundances of O, Si (Ne, Mg), and Fe are all set as independent variables. {The spectra shown are binned to 5 counts bin$^{-1}$ only for the purpose of better visualization only, while the spectra were binned to 1 counts bin$^{-1}$ in the fit.} The best-fit parameters are summarized in Table \ref{tab:extendvarytable}. Bottom: residuals (data minus model).
\label{fig:nebula-1-23456}}
\end{figure}

The spectra of both the 2000 and the 2016 data were extracted from Region 3 in Figure \ref{fig:regions}. As the 2000 data were taken 16 years before the 2016 data, the instrument response of ACIS-S in the soft X-ray has changed significantly. As shown in Table \ref{tab:counts}, only $\sim$ 680 counts were obtained in 2016-2017 for the Southern Nebula. Although the spectrum from the 2016 data seems more noisy, it does not mean that the data are of lower quality compared to the 2000 data. {In Figure \ref{fig:nebula-1-23456}, the 2000 and 2016 data were binned with the same method, i.e., at least 5 counts bin$^{-1}$ for visualization.} As more counts were obtained in the 2016 spectrum, the bin size of the 2016 spectrum (in units of keV) is much smaller than that of the 2000 spectrum (e.g., this can be seen most clearly in the 1.6-2.0 keV data in Fig 16). This is why the 2016 spectrum appear more ``noisy'' than they actually are. {In order to get robust measurements from the spectra of the Southern Nebula region, the spectral fitting was carried out as follows.}

{We firstly checked whether a model with fixed, solar abundance can fit the data. For this exercise, the spectra were binned to 15 counts bin$^{-1}$ so the \chitwo statistics and F-test could be applied. The fits gave a {\chitwo of 87.5} with a DOF of 55 in the case of the fixed solar abundance, and a {\chitwo of 49.2} with a DOF of 52 for the variable abundance model. The
p-value of the F-test is very low ($\sim$1.2$\times$10$^{-6}$), implying that the fixed abundance model is not sufficient to properly describe the data. }

{We then turned to the model with variable abundances.} Firstly, only spectra from the 2016 data were combined and fitted. In agreement with \citet{Iwasawa2011}, emission features of Mg {\sc xi} 1.34 keV and Si {\sc xiii} 1.85 keV are seen by eye in the spectrum. Such line emission suggests the existence of abundant $\alpha $ elements in the nebula. The spectrum was binned to {1 counts bin$^{-1}$} and {the CSTAT statistic was used for the fit. The spectrum was fitted with a single-temperature gas model (VMEKAL) with Galactic absorption.} Specifically, the abundances of Si, Mg, and Ne were tied together, while those of O and Fe were left as independent variables respectively. The abundance for all other elements not mentioned above were fixed at solar values. The corresponding results from the fits are summarized in Table \ref{tab:extendvarytable}. 

\begin{figure}[!htb]
\epsscale{1.1}
\plotone{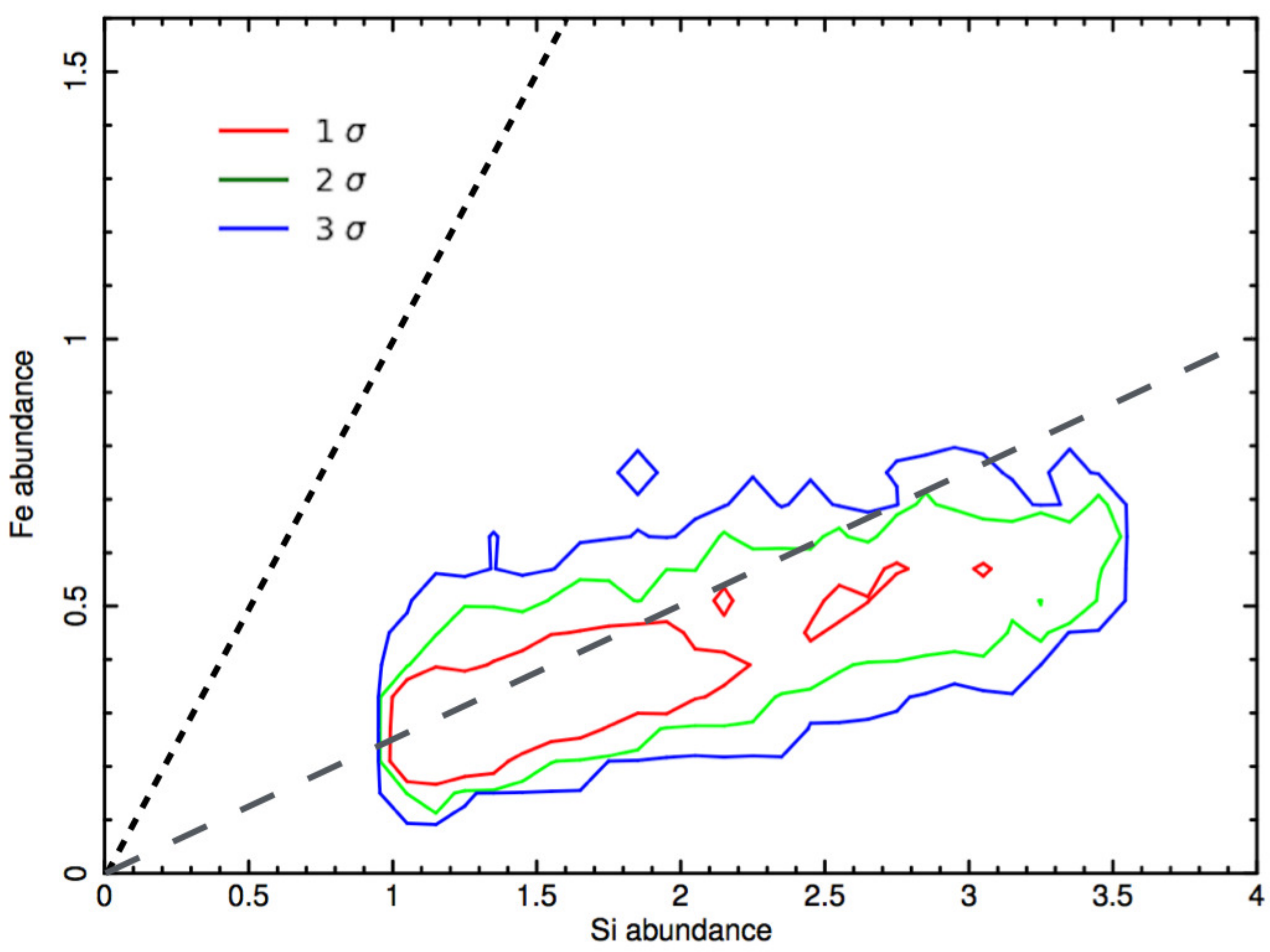}
\caption{{The probability contours of the Si (Ne,Mg) and Fe abundances (in solar units) for the Southern Nebula, using MCMC method and based on the results of the simultaneous fit to the spectra shown in Figure \ref{fig:nebula-1-23456}. The contours shown are for one, two and three sigma confidence levels. The black dotted line and gray dashed line denote the $\alpha$/Fe ratios of 1 and 4, respectively.} 
\label{fig:Si-Fe}}
\end{figure}

{The robustness of the results to the change of the abundance assumption was investigated. This was explored by setting the abundance of all other elements free in the fitting, but they had not changed the results from the fits to a noticeable extent. This is expected, since no line features of those elements are strong enough to affect the results of the fit in the spectral range of 0.4-2 keV. As a result, all those elements were fixed at solar values in the fits below.}

Secondly, simultaneous fitting to both the 2000 and the 2016 data was carried out. {Again, the spectra were binned to 1 counts bin$^{-1}$ and the CSTAT statistic was used for the fit.} The model and the parameter settings were the same as the one used when the 2016 data alone was fitted. The results are summarized in Table \ref{tab:extendvarytable}. {In general, the two fits give results that are consistent with each other. Taking into account the errors, both fits give the same temperature for the nebula. The largest difference of the two fits lies in the absolute metallicity, but both results still broadly agree with each other. }

{As stated at the end of the Section 2.1, the uncertainty of the $\alpha$/Fe ratio was obtained through the default Markov Chain Monte Carlo (MCMC) method implemented in XSPEC when the CSTAT statistic was used. The probability contours of Si (Ne, Mg) and Fe abundances are shown in Figure \ref{fig:Si-Fe}, based on the simultaneous fit of the 2000 and 2016 data. It can be seen that, although the absolute values of Si (Ne, Mg) and Fe abundance are uncertain, the ratio of $\alpha$/Fe is constrained relatively well by the data ($\alpha$/Fe = 4.5$^{+3.1}_{-1.9}$ for $\sim\pm{1.6}\sigma$, or $\alpha$/Fe = 4.5$^{+5.5}_{-2.4}$ for $\sim\pm{3}\sigma$). The detection of a super-solar $\alpha$/Fe ratio in the Southern Nebula is therefore robust ($>$3$\sigma$).}

\subsubsection{Search for Spatial Variations within the Nebula}

The spatial variations of physical properties within the Southern Nebula may help reveal the origin of the hot gas. The possible spatial variations of the temperature and metallicity of the hot gas were explored.

Spectra were extracted from regions on both the east and west sides of the tidal tail for inspection. The temperatures of the western and eastern parts of the Southern Nebula agree with each other and are similar to the value of the whole nebula (kT $\sim$ 0.57 keV), taking the {errors} into consideration. The relative metal abundance ratios of the two parts also show similar values to those of the whole nebula {($\alpha$/Fe $=$ 4.5$^{+3.1}_{-1.9}$)}. There is a hint that the absolute metal abundance in the western part of the nebula is systematically larger than that in the eastern part and that in the whole nebula. However, the uncertainties in the abundance are quite large (a factor of $\gtrsim$2) and this result is thus not conclusive. Moreover, the difference between the north and south parts of the Southern Nebula was also explored. Again, no clear difference was seen in either the temperature or the metallicity pattern.

\begin{figure}[!htb]   
\epsscale{1.15}
\plotone{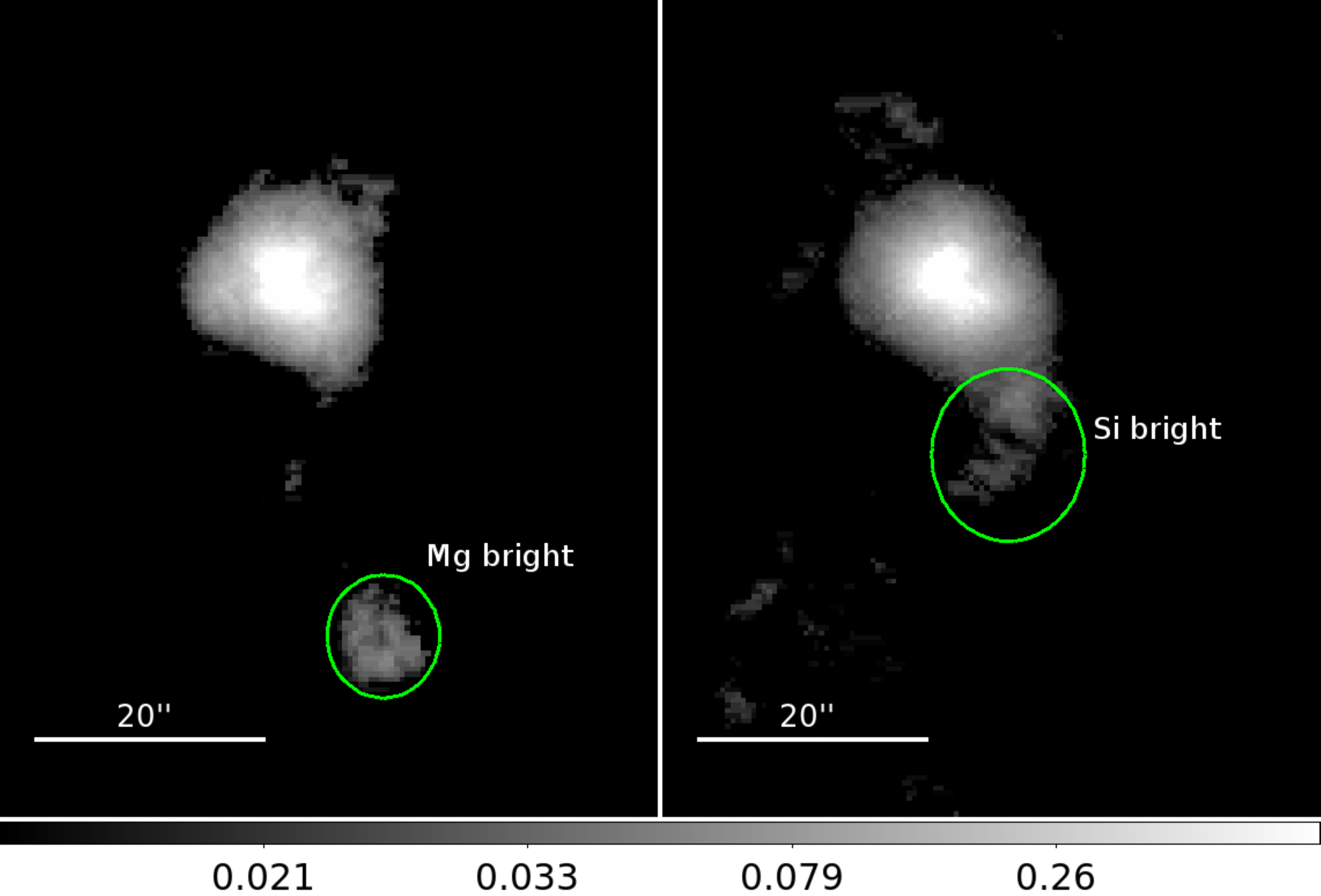}
\caption{{\em Left}: image of Mg {\sc xi} 1.34 keV emission (1.25-1.45 keV) from stacked 2000 and 2016 data. {\em Right}: image of Si {\sc xiii} 1.85 keV emission (1.78-1.95 keV) from the same data. On the left panel, an emission region dominated by Mg {\sc xi} 1.34 keV emission is shown, while on the right panel, another region dominated by Si {\sc xiii} 1.85 keV emission is shown. {Neither images are continuum subtracted. They are adaptively smoothed by FTOOL \textit{fadapt} \citep[http://heasarc.gsfc.nasa.gov/ftools/; see][]{ftool} with the same parameter settings, and are on a logarithm scale and in units of counts.}
\label{fig:Mg-Si}}
\end{figure}

\begin{figure}[!htb]   
\epsscale{1.25}
\plotone{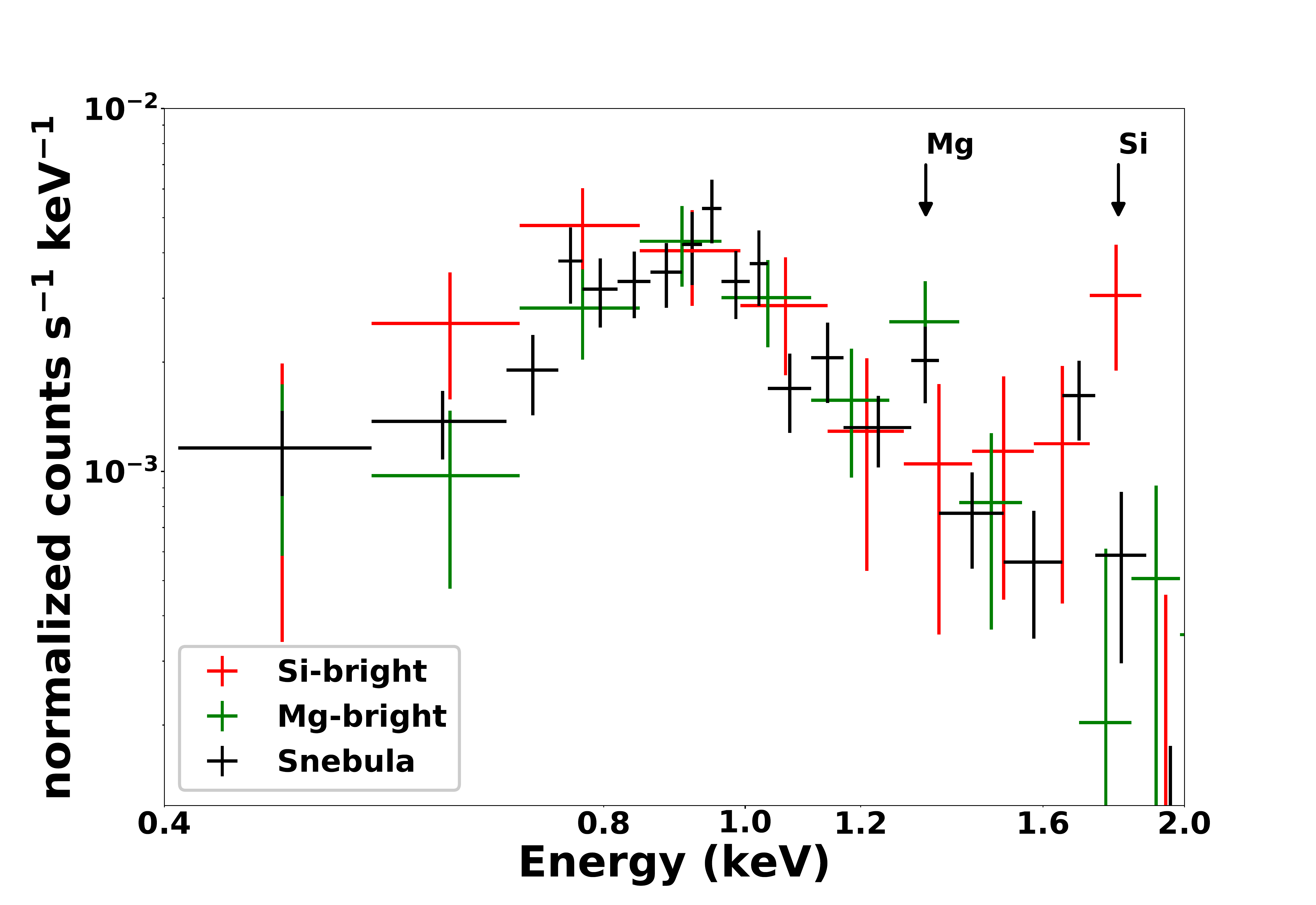}
\caption{The spectra from Si-bright region (red) and Mg-bright (green) region as shown in Figure \ref{fig:Mg-Si}, as well as the whole Southern Nebula (black). All spectra are normalized at the first energy bin. Locations of Mg and Si emission features are denoted with arrows.
\label{fig:Mg-Si-spec}}
\end{figure}

However, a detailed analysis of the narrow-band images centered on the Mg {\sc xi} 1.34 keV line and Si {\sc xiii} 1.85 keV line have revealed tentative spatial variations within the Southern Nebula. As shown in Figure \ref{fig:Mg-Si}, there are separate Mg {\sc xi} 1.34 keV bright region (Mg-bright region) and Si {\sc xiii} 1.85 keV bright region (Si-bright region) within the Southern Nebula. The spectra of both regions as well as the spectrum for the whole Southern Nebula are shown in Figure \ref{fig:Mg-Si-spec}, and similar temperatures (kT $\sim$ 0.57 keV) were obtained for all spectra by fitting a thermal gas model to them. The spectra demonstrates the stronger Mg {\sc xi} 1.34 keV line  and Si {\sc xiii} 1.85 keV line emission in corresponding regions respectively. As Si and Mg are both $\alpha$ elements, the relative abundance of these two elements should not vary much in different regions.  {One possible explanation for this variation is that the ionization states in the two regions are different, where the Si-bright region is more highly ionized than the Mg-bright region. The low S/N ($\lesssim$3) of these emission features prevent us from carrying out a more quantitative modeling of this apparent variation.} 

\subsection{NE-extended Nebula (Region 4)}

\begin{figure}[!htb]    
\centering
\epsscale{1.15}
\plotone{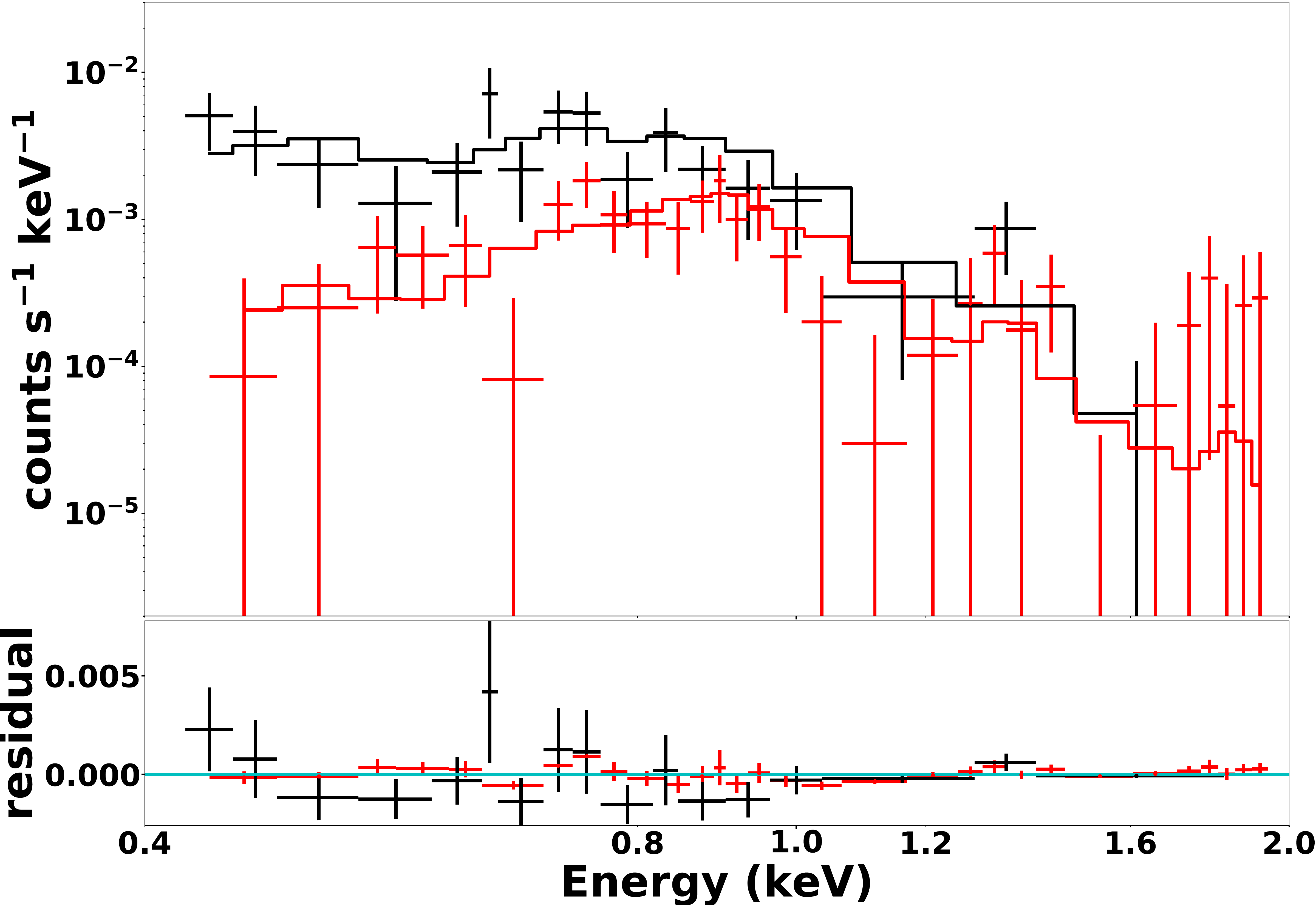}
\caption{Top: simultaneous fitting to the spectra of the NE-extended Nebula (Region 4 in Figure \ref{fig:regions}) from the 2000 data (black) and 2016 data (red). A thermal gas model (VMEKAL) is fitted to the spectra and shown in solid lines. The abundances of O, Si (Ne, Mg), and Fe are all set as independent variables. The best-fit parameters are summarized in Table \ref{tab:extendvarytable}. {The spectra shown are binned to 5 counts bin$^{-1}$ only for the purpose of better visualization, while the spectra were binned to 1 counts bin$^{-1}$ in the fit.} Bottom: residuals (data minus model).
\label{fig:ne-extend-1-23456}}
\end{figure}

Denoted as Region 4 in Figure \ref{fig:regions}, there is extended emission to the northeast of the galaxy (i.e. NE-extended Nebula), which is spatially associated with the extended, [O {\sc iii}] $\lambda$5007 emission studied in \citet{Zaurin2014}, most likely photoionized by the central AGN. The spectra were fitted in the same way that has been used for the Southern Nebula. The spectra for this region were fitted with a single-temperature gas model with Galactic absorption. {The spectra were mildly binned to 1 counts bin$^{-1}$ and the CSTAT statistic was used.} The results from the fits are summarized in Table \ref{tab:extendvarytable}, and the spectra are shown in Figure \ref{fig:ne-extend-1-23456}. Compared to the Southern Nebula, the temperature of this extended emission is significantly lower, perhaps $\sim$ 1/2 of that in the Southern Nebula. {The $\alpha$/Fe ratio in this region (2.1$^{+5.9}_{-1.8}$) seems slightly lower than that in the Southern Nebula (4.5$^{+3.1}_{-1.9}$), but only at the $\sim$1--$\sigma$ level so it is not significant.} The differences in temperature and perhaps in $\alpha$/Fe ratio between the NE-extended Nebula and the Southern Nebula probably point to their different origins. {See Sections 4.1.1-4.1.3 for a discussion of these results.}

\section{Discussion}
\subsection{Origin of the Southern Nebula}
An order-of-magnitude estimation of the total thermal energy and gas mass of the Southern Nebula was carried out. Despite the apparent irregular shape of the soft X-ray emission, it is assumed that the emission is generated in a sphere with a radius of 20 kpc and a filling factor of 1. Based on this simplification and the mean number density of 3$\times$10$^{-3}$ cm$^{-3}$ derived from the fits (see Table \ref{tab:extendvarytable}), a total energy of $\sim$ 1.2$\times$10$^{58}$ ergs and a total mass of $\sim$ 1.5$\times$10$^9$ \msun\ are obtained. The amount of kinetic energy deposited during a merger of two identical progenitors is of the order $M_g v_c^2/8$, where $M_g$ is the mass of the X-ray-emitting gas and $v_c$ is the relative speed during the collision \citep[see][]{Nardini2013}. In the case of Mrk~273, $v_c \sim$ 1250 km s$^{-1}$ is needed, which is significantly faster than the characteristic speed of head-on collisions in non-cluster environments {\citep[e.g., $\sim$ 800 km s$^{-1}$ in Taffy Galaxies; ][]{Braine2003}.} This implies the existence of additional energy source for the nebula.

\subsubsection{Extra Heating in the Southern Nebula}

As shown in Table \ref{tab:extendvarytable}, the measured temperature of the nebula is {0.57$^{+0.06}_{-0.07}$ keV}. Assuming the virialization of the hot gas and an NFW density profile for the dark matter halo \citep{Navarro1997}, the gas virial temperature can be written as follows:
\small
\[
kT_\rmn{vir} = \frac{1}{2} \mu m_p \frac{GM_\rmn{vir}}{r_\rmn{vir}}\simeq 0.135 \left(\frac{M_\rmn{vir}}{10^{12}M_{\sun}}\right) \left(\frac{r_\rmn{vir}}{100~\rmn{kpc}}\right)^{-1}~\rmn{keV},
\]
\normalsize
By further assuming the relations between the virial radius and both the halo mass and redshift presented in \citet{Mo2002}, the dark matter halo mass can be obtained through the equation $kT_{vir} \sim 0.042~M_{12}^{2/3} (1+z)$ keV, where $M_{12}$ is the virial mass expressed in units of 10$^{12}$ \msun. Similar to what is suggested for NGC~6240 in \citet{Nardini2013}, a possible evolutionary scenario consistent with the temperature ($\sim$ 0.57 keV) of the nebula is that of a cold dark matter halo with a total virial mass of $\sim$ 10$^{13}$ \msun\ formed at z $\sim$ 2.

This is almost an order of magnitude more massive than that of the Milky Way ($\sim$1.3$\times$10$^{12}$ \msun, see \citet[][]{McMillan2011} for details.) In addition, this value distinguishes group-scale halos from galaxy-scale halos \citep[see][]{Humphrey2006b}. The mass of the dark matter halo estimated here is unreasonably high, given the stellar mass of Mrk~273 \citep[$\sim$ 10$^{11}$ \msun; see][]{Veilleux2002,Zaurin2009,vivian2012} and its dynamical mass {\citep[$\sim$ 10$^{12}$ \msun, based on the velocity dispersion of the CO absorption features in the near-infrared, and assuming a King model with tidal radius to core radius of $\sim$ 50;][]{Tacconi2002} of Mrk~273.} Therefore, extra heating mechanism for the hot gas is at work.  

On the contrary, {the NE-extended Nebula shows a lower temperature of 0.31$^{+0.07}_{-0.01}$ keV}. Following the same procedure used for the Southern Nebula, a dark matter halo mass of $\sim$ 3.5$\times$10$^{12}$ \msun\ is obtained, which is a more reasonable value for the halo mass of Mrk~273. Compared to the Southern Nebula, the NE-extended Nebula is thus more likely to be the pre-existing, virialized gaseous halo. 

\subsubsection{Heating Source of the Gas}
Both galactic outflows generated from the nuclear/circumnuclear region and the merging activity can heat the gas in the nebula. It is thus necessary to discuss their relative importance.

One way to estimate it is to look at the required shock velocity. Assuming that the gas is heated up after being swept by an adiabatic shock, the temperature of the gas can be derived as $kT_\rmn{s} = 3 \mu m_p v_\rmn{s}^2/16$, where $v_\rmn{s}$ is the speed of the shock front. Adopting kT=0.57 keV (see Table \ref{tab:extendvarytable}), the $v_\rmn{s}$ will be 710 km s$^{-1}$, which is too high for the relative orbit speeds within a merging system like Mrk~273 (e.g. the inferred orbital velocity of the two nuclei of NGC 6240 is just 155 km s$^{-1}$; see \citet{Tecza2000} for details). In addition, the maximum speed of the non-outflowing ionized gas only reaches a maximum of 350 km s$^{-1}$, as is measured through the narrow component of \ha emission in \citet{RupkeVeilleux2013b}. Therefore, the merging activity alone seems not capable of causing shocks with required $v_\rmn{s}$.

As is presented in Section 3.4, hot gas (kT $\sim$ 0.8 keV, see Table \ref{tab:extendvarytable}) associated with the ionization cones and outflowing gas extends to a scale of $\gtrsim$5 kpc $\times$ 5 kpc. This outflowing hot gas might be the heating source for the Southern Nebula after it moves outward to a larger scale. Similar cases of hot gas heated by the AGN and/or starburst triggered outflow have been observed in several other galaxies \citep[e.g.][]{Wang2014,Tombesi2016,Tombesi2017}. {In addition, high velocity outflow in Mrk~273 has been detected in the ionized and
neutral gas phases \citep[$\sim$ 4 kpc;][]{RupkeVeilleux2013b,Zaurin2014}, as well as the molecular gas phase \citep[$\sim$ 550 pc;][]{Vivian2013,Veilleuxherschel,Cicone2014,Aladro2018}}. The ionized outflow has been observed to reach a speed as high as $\sim$ 1500 km s$^{-1}$ in projection, while the molecular outflow reaches a maximum speed of $\sim$ 900 km s$^{-1}$ in projection. These high-speed outflows may be capable of causing adiabatic shocks with $v_\rmn{s} \sim$ 710 km s$^{-1}$ in the nebula, if there is no significant decrease in the outflow speed before they reach the nebula. 

\subsubsection{The $\alpha$-elements Enrichment through Outflow}

Another independent piece of evidence regarding the origin of the nebulae comes from the $\alpha$/Fe abundance ratio of the hot gas. Iron is nearly entirely produced from Type Ia supernovae (SNe; i.e., exploded white dwarfs in close binary systems), while the majority of $\alpha$ elements originate from Type II SNe (i.e., core-collapsed massive stars). Based on the prediction of nucleosynthesis models for Type II SNe \citep[e.g,][]{Heger2010,Nomoto2006,Nucleosynthesis2013}, Si/Fe ratios can reach as high as $\sim$ 3-5 $\times$ solar value, while ratios of $\sim$ 0.5 solar value are expected from Type Ia SNe \citep[e.g,][]{Nagataki1998,Seitenzahl2013}. {Therefore, the $\alpha$/Fe ratios can provide information about how the gas was enriched.}

{The high $\alpha$/Fe ratios in both the Nuclear Region (Region 1; 6.6$^{+3.5}_{-3.7}$) and the Southern Nebula (Region 3; 4.5$^{+3.1}_{-1.9}$) might indicate a physical connection between the two, perhaps through the outflow generated in the Nuclear Region. Similar evidence for the transport of $\alpha$-enhanced material out to a scale $\lesssim$10 kpc by galactic winds has been seen in other nearby galaxies} \citep[e.g., M82;][and references therein]{Tsuru2007,Ranalli2008,Konami2011}.

\subsubsection{The Need for Multiple Outflow Events}
{If indeed the Southern Nebula is heated and enriched through the outflow, there are several reasons to believe that multiple outflow events are involved.}

First of all, the maximum silicon yield of a Type II SNe is $\sim$ 0.1 -- 0.3 \msun\ for a massive-star progenitor with $Z \le 0.02$ \citep[e.g,][]{Nucleosynthesis2013}. The total amount of silicon within the Southern Nebula is estimated to be $\sim$ 2$\times$10$^6$ \msun, adopting a derived gas mass of $\sim$ 1.5$\times$10$^9$ \msun\ and assuming a 1.5 $\times$ solar abundance for silicon estimated from the spectra. This can be translated into 1$\times10^7$ Type II SNe. Given the current star formation rate\footnote{{The SFR is quoted from Veilleux et al. 2009, and the contamination from the AGN is subtracted in the calculation. They have used six different methods to calculate the AGN fraction, which are based on the mid-infrared line ratios, PAH equivalent width, mid-infrared colors, and mid-to-far infrared continuum ratios. The mean AGN fraction of the six methods is used in the calculation, and the uncertainty in the AGN fraction is $\sim$15\%. See \citet{Veilleux2009} for more details.}} of $\sim$160 M$_{\sun}$ yr$^{-1}$, a continuous star formation activity of $\gtrsim$10$^7$ yr is required. It is thus very unlikely that the current star formation contributes solely to the enrichment. Past starforming and/or starburst activities over a much longer time scale are required. Detailed spectral analyses of the circum-nuclear region have indeed revealed stellar populations both young (3-60 Myr) and old (0.7-12.5 Gyr) \citep[see][for details]{stellarage2003,Zaurin2009}.

Secondly, the characteristic time scale needed for the outflow to affect the Southern Nebula on a spatial scale of $\sim$ 40 kpc can be estimated. The average outflow speed of warm ionized gas in projection is $\sim$ 1000 km s$^{-1}$  \citep[measured from the average V$_{98\%}$ of the H$\alpha$ emission,  see column (7) of Table 3 in][]{RupkeVeilleux2013b}. Assuming that the outflow travels with a constant speed of 1000 km s$^{-1}$, a time scale of 40 Myr is needed for it to reach the full extent of the Southern Nebula. 

Moreover, as no significant spatial variations of the $\alpha$/Fe ratio are seen within the Nebula, the whole duration of the enrichment might be as long as the dynamical time scale of $\sim$ 0.1 Gyr, which is measured from the sound-crossing time, $D/c_s = D (5 kT/3 \mu m_p)^{-1/2}$. The age of the current outflow in warm ionized phase is $\lesssim$10 Myr, based on its velocity and spatial extent \citep{RupkeVeilleux2013b,Zaurin2014}. As a result, the current outflow alone cannot be responsible for the $\alpha$-elements enrichment already observed in the Southern Nebula. 

In all, multiple outflow events in the past, on a time scale of $\lesssim$0.1 Gyr, are required to explain the super-solar $\alpha$/Fe ratio in the Southern Nebula. The merger event might have helped further erase spatial fluctuations in the $\alpha$/Fe ratio within the nebula {\citep[similar to the case of Mrk~231,][]{Veilleux2014}}. 

\begin{figure}[!htb]   
\centering
\epsscale{1.27}
\plotone{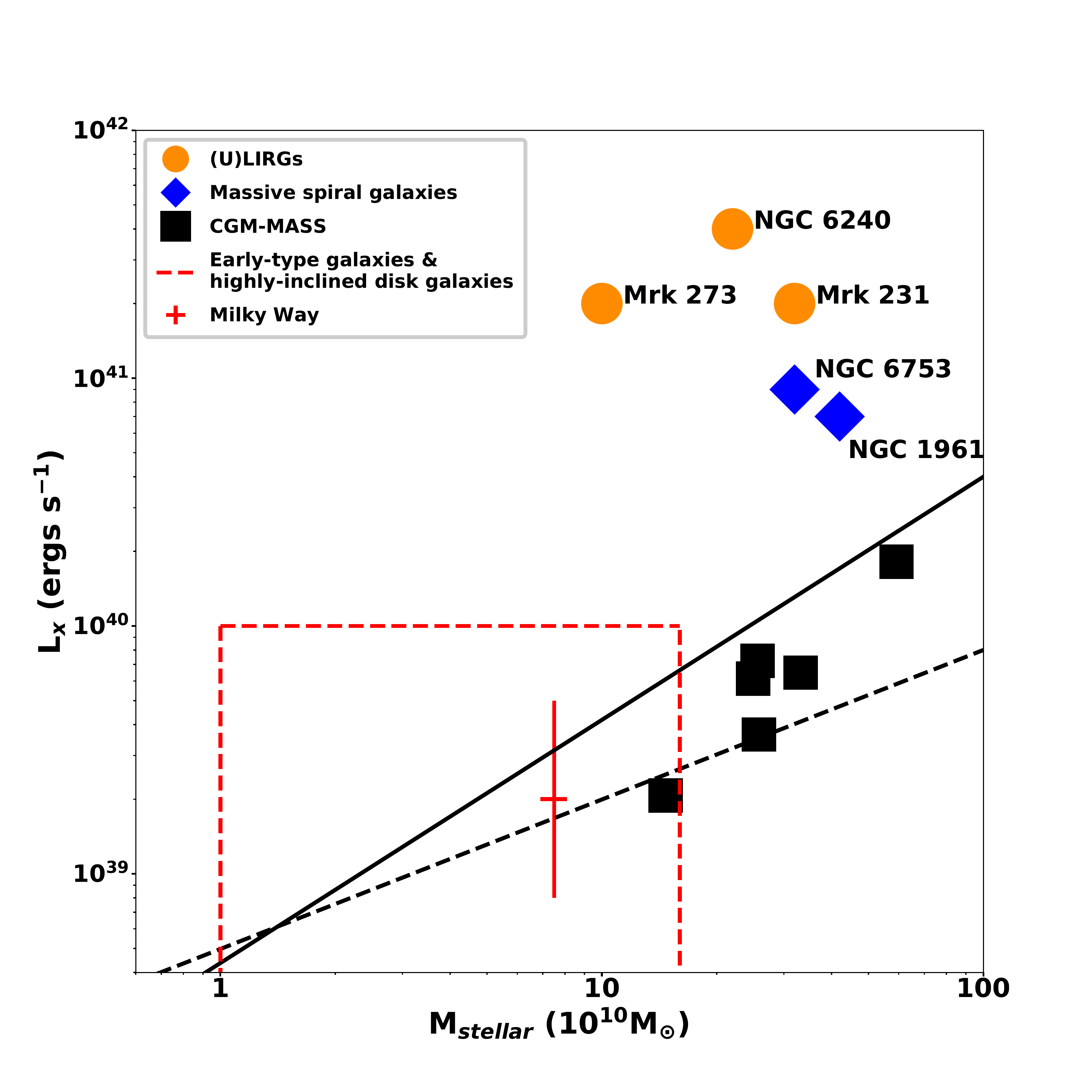}
\caption{Comparison of the nebula emission from (U)LIRGs and those from other galaxies with similar stellar mass. The x, y axis are the stellar mass and soft X-ray ($\sim$0.5-2 keV) luminosity. The filled circles, diamonds, squares and red cross denote (U)LIRGs (this work), massive spiral galaxies \citep{NGC1961,NGC6753} as well as CGM-MASS galaxies and the Milky Way \citep{Li2017}. The sizes of the labels represent the typical errors of the measurements, except that the error bars are drawn for the Milky Way. The region enclosed by the red dashed line represents the rough location of early-type galaxies from the ATLAS$^{3D}$ \citep{Kim2015} and MASSIVE \citep{Goulding2016} surveys in the corresponding mass range, as well as nearby highly-inclined disk galaxies \citep{Li2013a, Li2013b, Li2014}. The black solid line and black dashed line are linear and non-linear fits to the non-starburst field galaxies respectively \citep[see][for more details]{Li2017}. 
\label{fig:compare-Li}}
\end{figure}

\subsubsection{Comparison with the H$\alpha$ Nebula}
{As shown in the bottom right panel of Figure \ref{fig:0470ct}, the brighter portion of the extended H$\alpha$ emission south of the galaxy (T $\sim$ 10$^4$ K) resembles the Southern Nebula in the X-ray (T $\sim$ 7$\times$10$^6$ K).} The eastern part of the Southern Nebula in the X-ray seems to be more luminous and extended than the H$\alpha$ nebula, although this may be caused by the different sensitivity of the X-ray and H$\alpha$ observations. \citet{Spence2016} argue that the southern H$\alpha$ nebula is not related to the large scale outflow, based on the moderate velocity shifts ($|\Delta V| \lesssim$ 250 km s$^{-1}$) and line width (FWHM $\lesssim$ 350 km s$^{-1}$) measured in the emission lines. However, these results do not formally rule out the possibility that the H$\alpha$ nebula represents the accumulated reservoir of the outflowing gas if the turbulence caused by the outflows dissipated away as the material cooled down from $\sim$ 7$\times$ 10$^6$ K to $\sim$ 10$^4$ K. The high temperature and high $\alpha$/Fe ratio of the Southern Nebula favor an outflow-related origin. 
 
\subsection{X-ray Nebulae at Different Merger Stages}

The $\alpha$-enriched, luminous X-ray nebulae have been discovered in mergers at different stages: Mrk~273 (close-pair), NGC~6240 (close-pair),  and Mrk~231 (coalesced). The total luminosity of the nebula within Mrk~273 in 0.4-2 keV band is 2$\times 10^{41}$ erg s$^{-1}$ (adding up nebula emission in Region 2, 3 and 4), which is comparable to those of the nebulae within Mrk~231 \citep[2$\times 10^{41}$ erg s$^{-1}$ in 0.5-2.0 keV band, ][]{Veilleux2014} and NGC~6240 \citep[4$\times 10^{41}$ erg s$^{-1}$ in 0.4-2.5 keV band, ][]{Nardini2013}. As shown in Figure \ref{fig:compare-Li}, the soft X-ray emission of these nebulae is more luminous than that of galaxies with similar stellar mass, including massive spiral galaxies \citep{NGC1961,NGC6753}, CGM-MASS galaxies \citep{Li2017}, highly-inclined disk galaxies \citep{Li2013a, Li2013b, Li2014} as well as early-type galaxies from the ATLAS$^{3D}$ \citep{Kim2015} and MASSIVE \citep{Goulding2016} surveys in the corresponding mass range. 

Assuming that the X-ray nebulae in (U)LIRGs are in a quasi-equilibrium state, the heating of the gas should be balanced by the radiative cooling. If this assumption is valid, the brighter nebula emission suggests significant excess energy input from the outflow events and merging process. 

The basic properties of the X-ray nebulae in the three (U)LIRGs are summarized in Table \ref{tab:3ulirg}. In general, the three nebulae are quite similar to each other, in terms of luminosity, temperature and $\alpha$/Fe ratio. The existence of these similar X-ray nebulae in mergers at different stages reinforces the picture that multiple outflow events over a time scale of $\lesssim$0.1 Gyr must be at work during the whole merging process. 
\\

\begin{deluxetable}{cccc}[!htb]
\tablecolumns{4}
\tablecaption{Properties of the X-ray Nebulae within the three (U)LIRGs}
\tablehead{\colhead{Physical Property} & \colhead{Mrk~273} & \colhead{NGC~6240}\tablenotemark{a}  &\colhead{Mrk~231}\tablenotemark{b}
}
\startdata
merger stage & close pair & close pair & coalesced \\
nuclear separation &  0.75 kpc &  0.74 kpc & $-$ \\
L$_{nebula}$\tablenotemark{c} & $\sim$$2 \times 10^{41}$ & $\sim$$4 \times 10^{41}$ & $\sim$$2 \times 10^{41}$ \\
T (keV) & $\sim$0.57 & $\sim$0.65 & $\sim$0.67, 0.27 \\
n$_{e}$\tablenotemark{d} & $\sim$3 & $\sim$2.5 & $\sim$0.97, 0.92 \\
M$_{gas}$\tablenotemark{e} & $\sim$6 & $\sim$10  & $\sim$7 \\
$\alpha$/Fe& 4.5$^{+3.1}_{-1.9}$ & 4.0$^{+2.1}_{-1.8}$\tablenotemark{f} & 3.3$^{+2.2}_{-1.7}$, - \tablenotemark{f}
\enddata
\tablenotetext{a}{\citet{Nardini2013}.}
\tablenotetext{b}{\citet{Veilleux2014}. The best-fit model is made up of two gas components.}
\tablenotetext{c}{Soft X-ray luminosity of the nebulae, in units of erg s$^{-1}$. The luminosities are in 0.4-2.0 keV band for Mrk~273, in 0.4-2.5 keV band for NGC~6240 and in 0.5-2.0 keV band for Mrk~231.}
\tablenotetext{d}{Electron number densities, in units of 10$^{-3}$ cm$^{-3}$.}
\tablenotetext{e}{Total mass of the hot gas, in units of 10$^{9}$ \msun. For Mrk~273, the sum of the gas mass in Region 2, 3 and 4 is listed. For Mrk~231, the sum of the mass for the two gas components is listed.} 
\tablenotetext{f}{The values are not directly listed in the corresponding papers. They are estimated based on the information listed in the Table 3 of \citet{Nardini2013} and the Table 2 of \citet{Veilleux2014}.} 
\label{tab:3ulirg}
\end{deluxetable}

\subsection{The $\alpha$/Fe ratios in (U)LIRGs and Early-type Galaxies} 

Super-solar $\alpha$/Fe ratios are typically measured in the old stellar populations of early-type galaxies \citep[e.g.][]{Worthey1998,Graves2010,Conroy2014,Kriek2016} and the hot interstellar medium of some post-merger early-type galaxies with relatively young stellar populations \citep[e.g.][]{abundanceETG2012}. These results are most likely due to either a short timescale of star formation (efficient quenching before the onset of Type Ia SNe) or variations in the initial mass function (IMF). Given the mass outflow rate and gas content of Mrk~273, the implied gas depletion timescale is only $\sim$ 10 Myr \citep{Cicone2014}. The outflow in Mrk~273 might thus be able to quench star formation on this short timescale, if the ejected gas does not return to the center to form stars. Therefore, the super-solar $\alpha$/Fe ratio measured in the Nuclear Region (Region 1) of Mrk~273 may be the result of the on-going rapid starburst/quenching process.

\section{Conclusions}

We have combined a deep, 200 ksec {\em Chandra} ACIS-S observation of Mrk~273 with 44 ksec archival data acquired with the same instrument and setup. The main results are summarized as follows.

\begin{itemize}

\item The AGN associated with the SW nucleus is confirmed by the new data. A secondary peak exists in the hard X-ray images (especially 4-6 keV), which is associated with the NE nucleus at a projected distance of 0.75 kpc from the SW nucleus. 

\item The hard X-ray (3-8 keV) spectrum of the SW nucleus is well explained by a heavily absorbed AGN (N$_{H}$ = 2.79$^{+0.30}_{-0.22} \times$10$^{23}$  cm$^{-2}$. {In this statement and those below, the errors from the spectral fits correspond to a confidence range of 90\%, or $\sim$$\pm{1.6}$$\sigma$}). The hard X-ray flux of the SW nucleus has dropped by $\sim$ 60\% from the year 2000 to the year 2016. The decrease seems to be caused by the fading of the intrinsic luminosity of the central engine, although an increase in the column density of the absorbing material cannot be formally ruled out. 

\item The hard X-ray (3-8 keV) spectrum of the NE nucleus is best explained by a combination of a heavily obscured AGN {(N$_{H}$ = 6.78$^{+4.99}_{-3.31} \times$10$^{23}$ cm$^{-2}$)} and a hot gas component (kT $\sim$ 3 keV). A harder X-ray continuum compared to that of the SW nucleus is detected in the NE nucleus at 6.5-8 keV. An Fe {\sc xxv} line (6.7 keV rest frame) is found in the spectrum of the NE nucleus, while the strength of Fe K$\alpha$ line is not well constrained by the current data.  

\item A single power-law component cannot describe the hard X-ray spectrum of the Nuclear Region (r$<$3$\arcsec$, $\sim$ 2 kpc). Significant residuals above 6.5 keV in the observer's frame still exists after a hot (kT $\sim$ 3 keV) gas component is added to the model. Adding a second power-law component improves the fit (although reduced {\chitwo $\sim$ 0.8}). These results favor the existence of a dual AGN. 

\item Spatially extended excesses of 1-3 keV emission and Si {\sc xiii} 1.85 keV (also Mg {\sc xii} 1.47 keV) line emission are found to be coincident with the optical [O {\sc iii}] $\lambda$5007 emission tracing the ionization cones and outflowing gas out to $\sim$ 5 kpc. {An $\alpha$/Fe ratio of 3.3$^{+4.2}_{-2.6}$ is measured in this hot, X-ray emitting gas.} 

\item The temperature of the cooler gas component in the Nuclear Region {(kT=0.75$^{+0.07}_{-0.08}$ keV)} is similar to that of the gas in the Host Galaxy Region {(kT=0.73$^{+0.08}_{-0.10}$ keV)}. They are both higher than that of the Southern Nebula {(kT=0.57$^{+0.06}_{-0.07}$ keV)} and that of the NE-extended Nebula {(kT=0.31$^{+0.07}_{-0.01}$ keV)}. No clear spatial variation is seen within the Southern Nebula itself.

\item {Super-solar $\alpha$/Fe ratios are measured in the Nuclear Region (6.6$^{+3.5}_{-3.7}$) and the Southern Nebula (4.5$^{+3.1}_{-1.9}$). However, lower $\alpha$/Fe ratios are derived in the Host Galaxy Region (1.7$^{+1.6}_{-1.6}$, different at the $\sim$2--$\sigma$ level) and perhaps also in the NE-extended Nebula (2.1$^{+5.9}_{-1.8}$, although the uncertainties are large).}

\item These data suggest that the hot gas in the Southern Nebula has been heated and enriched by multiple outflows originated from the Nuclear Region, on a time scale of $\lesssim$0.1 Gyr. A similar origin has been suggested for the large-scale nebulae around Mrk~231 and NGC~6240. In contrast, the NE-extended Nebula is likely the pre-existing gas in the halo, which has not been affected by the outflows yet.
\\
\end{itemize}

\acknowledgements 

We thank the anonymous referee for thoughtful and constructive comments that improved this paper. W. L. thanks R. Mushotzky and E. Kara for inspiring and helpful discussions and suggestions. Support for this work was provided by NASA through {\em Chandra} contract G06-17090X  (W. L., S. V.). K. I. acknowledges support by the Spanish MINECO under grant AYA2016-76012-C3-1-P and MDM-2014-0369 of ICCUB (Unidad de Excelencia `Mar\'ia de Maeztu'). F.T. acknowledges support by the Programma per Giovani Ricercatori - anno 2014 ``Rita Levi Montalcini''. This work was conducted in part at the Aspen Center for Physics, which is supported by NSF grant PHY­1607611; V. U. thanks the Center for its hospitality during the Astrophysics of Massive Black Holes Mergers workshop in June and July 2018. This work has made use of NASA's Astrophysics Data System Abstract Service and the NASA/IPAC Extragalactic Database (NED), which is operated by the Jet Propulsion Laboratory, California Institute of Technology, under contract with the National Aeronautics and Space Administration. The HST data presented in this paper were obtained from the Mikulski Archive for Space Telescopes (MAST). STScI is operated by the Association of Universities for Research in Astronomy, Inc., under NASA contract NAS5-26555. 

\appendix
\section{Spectral Analysis of Mrk~273x}

\begin{figure}[!htb]  
\centering
\epsscale{1.0}
\plotone{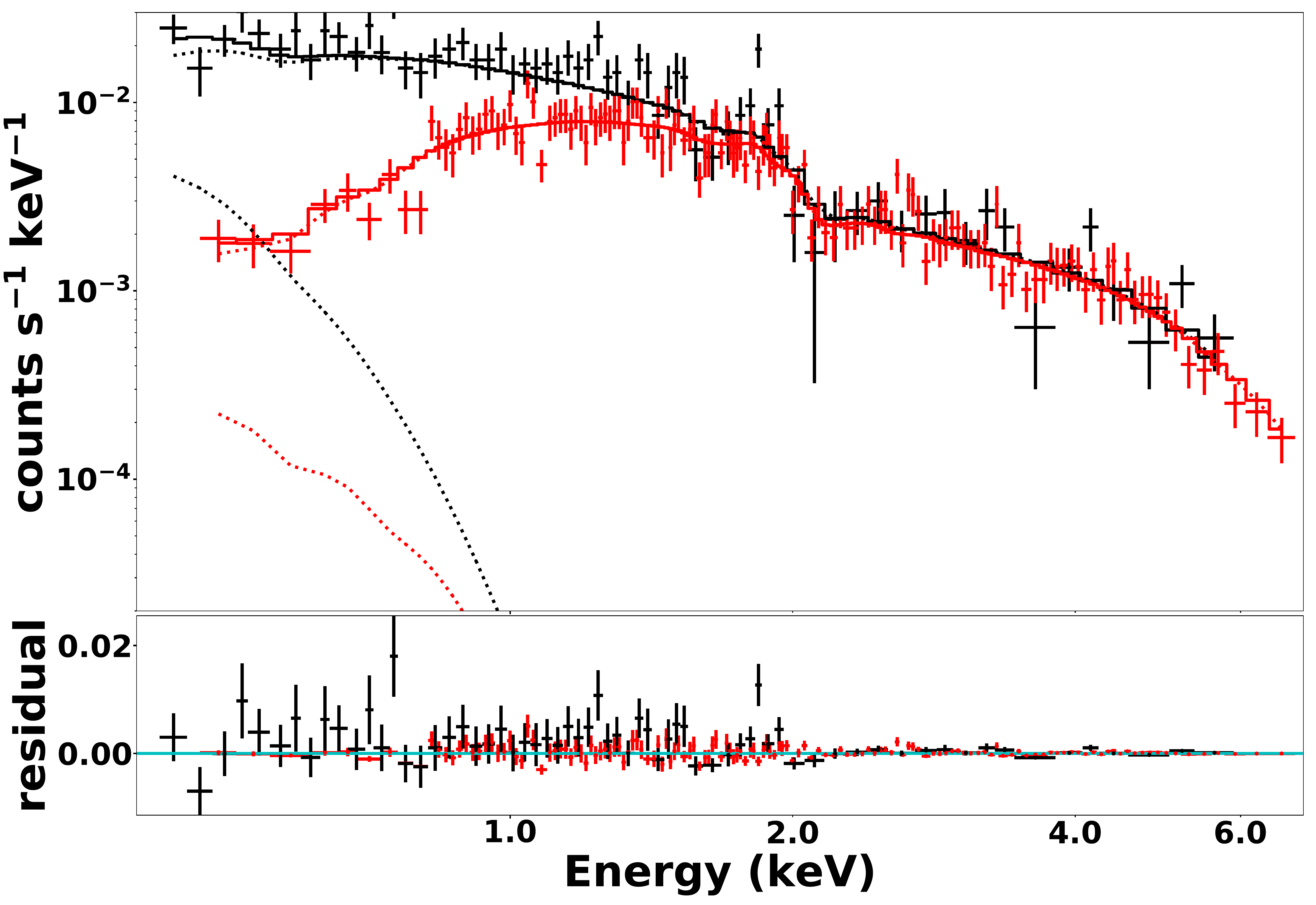}
\caption{Top: the spectra of Mrk 273x of 2000 data (black) and 2016 data (red) are shown separately. The apparent difference of the two spectra is mainly due to the drop of the instrument response in the soft X-ray from year 2000 to year 2016, but not a change in the flux of the source. Simultaneous fitting was applied to the data and the model was an absorbed power-law component {plus a blackbody component with Galactic absorption}. The model is shown in solid lines and the components of it are shown in dotted lines. Bottom: residuals (data minus model).
\label{fig:273x}}
\end{figure}

Mrk~273x is at a projected angular separation of only 1$\arcmin$.3 from Mrk~273 and it is observed simultaneously in our {\em Chandra} program. Mrk~273x is optically classified as a z = 0.46 Type 2 AGN, but its X-ray properties \citep[e.g., lack of obvious absorption at low energies, high hard X-ray luminosity, absence of Fe K$\alpha$; ][]{Xia2002} are typical for an unabsorbed Type 1 AGN. By comparing the X-ray spectrum from the 2000 and 2016 data, constraints can be obtained on the variability of this source in X-ray. The spectra of Mrk~273x from the 2000 data and the 2016 data are shown in Figure \ref{fig:273x}. The two spectra are consistent with each other and no clear variability is seen. 

{The spectra were fitted well ({reduced \chitwo = 1.2}) using a model with a power-law component ($\Gamma$=1.53$^{+0.15}_{-0.06}$) without much absorption (N$_{H}$ $\sim$ 1.0$^{+0.2}_{-0.2}$$\times$10$^{21}$ cm$^{-2}$), plus a blackbody component (kT$<$0.26 keV). {The blackbody component was used to model possible residuals at $\sim$ 0.4-0.8 keV when only a power-law component was fitted to the data, although the p-value of the F-test between the two fits is only 0.6. The current data are thus not conclusive on the existence of the blackbody component, which might be a soft X-ray excess usually seen in Type 1 AGN \citep[e.g.][]{Crummy2006}}. Potential Fe K$\alpha$ line was fitted and an upper limit of {$\sim$56 eV} on the equivalent width (EW) in the observer's frame was obtained.} The X-ray spectra over 16 years agree with each other, which show typical features for unabsorbed Type 1 AGN. This suggests that the Type 2 optical spectrum of Mrk~273x is intrinsic, i.e. due to the lack of a broad line region. This type of sources, where the X-ray spectrum shows all the characteristics of an unabsorbed Type 1 AGN while the optical spectrum points to a Type 2 AGN, remains a challenge to the standard AGN unification model.

\bibliographystyle{aasjournal}
\bibliography{mrk273}

\end{document}